\documentclass[11pt,a4paper]{article}
\usepackage[utf8]{inputenc}
\usepackage[english]{babel}
\usepackage{natbib,hyperref,graphicx}
\usepackage{amsmath,amsfonts,amssymb,amsthm}
\usepackage{geometry}
\usepackage{mathtools}
\usepackage{array}
\usepackage{booktabs,tabularx}
\usepackage{xcolor}
\usepackage{bm}
\usepackage{float}
\usepackage{comment}
\usepackage{authblk}
\input{alphabet}
\usepackage{ulem}

\newtheorem{proposition}{Proposition}
\newtheorem{remark}{Remark}
\newtheorem{lemma}{Lemma}

\title{Portfolio Decarbonization under Incomplete Carbon and Transition Information}
\author[1]{Olivier Féron\thanks{\texttt{olivier-2.feron@edf.fr}}}
\author[2]{Ruben Haalebos\thanks{\texttt{ruben.haalebos@ensae.fr}}}
\affil[1]{Électricité de France (EDF)}
\affil[2]{CREST/IP Paris - Électricité de France (EDF)}
\date{\today}

\providecommand{\keywords}[1]{%
    \small\textbf{Keywords:} #1
}

\begin{document}

\maketitle

\section*{Abstract}
We study dynamic portfolio decarbonization under incomplete information about firms' carbon intensities and the underlying transition state. The investor maximizes expected utility of terminal wealth net of a penalty on perceived cumulative financed emissions. Asset returns and carbon intensities depend on a stochastic transition factor, while the investor may need to infer the current transition state, its long-run mean and firms' carbon intensities from market observations and periodic noisy disclosures. The optimal strategy combines a Merton-type component with climate-dependent expected-return adjustments, a direct carbon intensity penalty and intertemporal hedging demands associated with stochastic carbon intensities and learning. In a numerical application, incomplete information has limited effects on welfare and target satisfaction when prior beliefs are well specified. By contrast, biased beliefs generate welfare losses in either direction but asymmetric environmental consequences: optimistic and pessimistic beliefs induce opposite portfolio tilts and therefore affect the probability of exceeding the cumulative emissions target differently. \\


\keywords{
Portfolio decarbonization;
Incomplete information;
Stochastic control;
Bayesian filtering;
Climate finance
}

\section{Introduction}

Portfolio decarbonization is inherently forward-looking, but the information on which it depends is incomplete. Firms increasingly announce Greenhouse Gas (GHG) emission-reduction targets and transition plans, yet these targets are not immutable: they may be revised as climate science, reporting methodologies, business structure or transition policies evolve \citep{Privato_2024, Jiang_2025}. At the same time, realized emissions are typically disclosed at discrete intervals, most commonly on an annual basis. They are imperfect measures of firms' underlying carbon footprints: they are subject to data and methodological uncertainty and are frequently revised ex post \citep{Cohen_2026}. Importantly, carbon information is also financially material: higher corporate emissions and emissions growth have been associated with higher expected stock returns, while carbon-related policy and disclosure shocks generate differential market reactions across firms according to their carbon exposure \citep{Bolton_2021,Florackis_2025}. An investor pursuing a multi-year decarbonization objective therefore faces a joint portfolio-choice and learning problem. \\

This paper asks how incomplete climate information affects the composition, welfare, and target satisfaction of dynamic portfolio decarbonization. By composition, we mean the allocation across assets with heterogeneous carbon intensities and decarbonization trajectories. Welfare effects are measured through the certainty-equivalent loss associated with pursuing a given cumulative financed-emissions objective, together with the incremental welfare cost generated by imperfect information. Target satisfaction refers to the ability of the resulting portfolio strategy to deliver the intended emissions reduction, which we assess through the probability of exceeding the associated cumulative emissions budget.\\

To answer this question, we study a continuous-time portfolio-choice problem in which the investor values terminal wealth net of cumulative financed emissions. We adopt the perspective of a small, price-taking investor whose portfolio decisions do not affect asset prices, firms' emissions, or the evolution of the transition state. The investor's decarbonization objective therefore concerns the emissions financed by the portfolio rather than the real-economy emissions induced by its investment decisions. Firms' expected returns and carbon intensities depend on a common stochastic transition state, while carbon intensities themselves evolve toward long-run levels determined by the long-run state of the transition. The investor may be uncertain both about the current transition state and about its long-run mean, and firm-level carbon intensities may only be observed through periodic noisy disclosures. We consider three information regimes. A full-information investor observes the transition state and carbon intensities and serves as the benchmark. A partially informed investor observes carbon intensities continuously but must infer the transition state and its long-run mean. Finally, a disclosure-based investor must additionally infer firms' carbon intensities from periodic noisy observations. Because this investor does not observe the latent carbon intensities continuously, her portfolio carbon objective is computed from their conditional estimates. \\

The central dynamic mechanism of the model is that climate information enters the portfolio problem through two distinct channels. First, beliefs about the transition state and firms' carbon intensities affect expected investment opportunities and therefore the financial attractiveness of the risky assets. Second, the same beliefs determine the expected rate at which portfolio choices consume the cumulative financed-emissions budget. Learning consequently affects both the financial and environmental margins of the control problem. These two effects need not be aligned: information that improves portfolio choice and investor welfare can simultaneously increase cumulative financed emissions if it induces a reallocation toward assets that are currently carbon intensive but are expected to decarbonize rapidly. \\

The numerical results deliver three main findings. First, incomplete climate information generates only modest welfare losses when prior beliefs are approximately correct and carbon disclosures are sufficiently frequent. Second, misspecified beliefs about the transition generate welfare losses in either direction. Both overly optimistic and overly pessimistic investors depart from the full-information allocation, and these losses become larger when disclosures are less frequent and mistaken beliefs persist for longer. Third, the consequences for portfolio decarbonization are asymmetric. In our numerical application, optimistic beliefs about the transition induce investors to tilt away from carbon-intensive assets and tend to produce emissions below the intended budget, whereas pessimistic beliefs increase carbon-intensive exposures and raise the probability of exceeding the target. Thus, belief distortions of similar magnitude but opposite sign can generate comparable welfare losses while having opposite implications for the environmental objective. \\

Taken together, these results highlight a distinction between the financial value of information and its implications for a dynamic environmental objective. More accurate beliefs improve the investor's ability to allocate capital conditional on the available investment opportunities, but they do not necessarily move the portfolio monotonically toward lower cumulative financed emissions. The environmental effect of learning depends on how new information changes the relative attractiveness of assets with heterogeneous current intensities and future decarbonization trajectories. The paper therefore contributes to dynamic portfolio choice under partial information by showing that, when the state variable being learned affects both expected returns and the accumulation of a second dynamic objective, the welfare and target-satisfaction effects of information can move in different directions. \\

\paragraph{Literature review}

The uncertainty surrounding the climate information available to investors is well studied in the literature. Corporate transition plans and emission-reduction targets are frequently revised, while reported emissions are subject to measurement uncertainty, disagreement across data providers, and ex post revisions \citep{Privato_2024,Jiang_2025,Cohen_2026}. At the same time, carbon information is financially material and valued by institutional investors \citep{Bolton_2021,Florackis_2025,Ilhan_2023}. We build on this literature by studying how such information frictions affect the dynamic implementation of portfolio decarbonization. \\

A broad literature studies portfolio decarbonization through low-carbon tilts, benchmark-relative constraints, and progressive exclusions \citep{Andersson_2016,Riposo_2023,Bolton_2022,Jondeau_2025}. These approaches show that substantial reductions in portfolio carbon exposure can often be achieved with limited financial deviation from conventional benchmarks. The choice of emissions metric nevertheless matters: \cite{Fraser_Fiedler_2023} show that reductions in financed emissions need not correspond to equivalent reductions in firms' physical emissions, and argue for complementing such measures with forward-looking indicators of portfolio alignment with climate targets. In line with this perspective, we focus on cumulative financed emissions at the portfolio level. Our main departure from this literature is therefore intertemporal: rather than controlling portfolio carbon exposure separately at each rebalancing date, we consider cumulative financed emissions over the investment horizon, so that current portfolio decisions affect emissions accumulated over the entire period. \\

Recent work has moved toward explicitly dynamic formulations of sustainable portfolio choice. \cite{Korn_2023} study continuous-time portfolio optimization under sustainability constraints; subsequent contributions extend this framework to alternative preference specifications and scenario-based settings \citep{Korn2_2025,Korn_2025}. \cite{Chen_2025} study expected-utility maximization subject to carbon-risk constraints and characterize how alternative carbon-risk measures affect the allocation between green and brown assets. Related work considers forward-looking portfolio decarbonization through tightening carbon budgets. In particular, \cite{Cenedese_2026} introduce a firm's distance to exit from a net-zero portfolio, which depends on its expected ability to remain compatible with a progressively tighter carbon budget. These papers make the dynamic nature of sustainable portfolio allocation explicit, but environmental objectives are typically represented through portfolio constraints, eligibility rules, or carbon-risk limits. Our formulation instead places the environmental objective directly in investor preferences. \\

Preference-based formulations provide an alternative to imposing sustainability as a hard constraint. At a broader level, \cite{Pastor_2021} model investors as deriving non-pecuniary utility from green holdings, while \cite{Pedersen_2021} characterize the trade-off between financial performance and ESG characteristics through an ESG-efficient frontier. Recent continuous-time models also introduce environmental objectives directly into the investor's objective function rather than restricting the admissible portfolio set \citep{Mancinelli_2025,Hausladen2026NetZero}. In particular, \cite{Hausladen2026NetZero} model stochastic firm emissions and derive continuous-time strategies in which emissions and net-zero target paths enter the investor's objective. We follow this preference-based approach but penalize cumulative financed emissions jointly with terminal wealth. The carbon penalty therefore governs the trade-off between financial welfare and the investor's portfolio-emissions objective and can be calibrated to a desired cumulative emissions reduction. We depart from these existing dynamic formulations in two main aspects. First, we consider ownership-based financed emissions, so that carbon exposure scales with the investor's monetary holdings, consistent with standard carbon accounting practices \citep{PCAF_2022}. Second, we assume that the relevant climate states are not necessarily observed by the investor. We allow investors to be uncertain about the current state and long-run direction of the transition and, in the most restrictive information regime, to observe firm-level carbon intensities only through periodic noisy disclosures. \\

Our analysis consequently also relates to the literature on dynamic portfolio choice under partial information. \cite{Lakner_1998} studies portfolio optimization when the drift of risky assets is an unobservable Gaussian process and shows how filtering converts the partially observed problem into one expressed in terms of the investor's conditional estimates. \cite{Brendle_2006} similarly solves a portfolio problem under incomplete information and explicitly quantifies the utility loss relative to the full-information benchmark. A related literature introduces additional noisy information signals. \cite{Frey_2012} consider portfolio optimization when investors learn about an unobservable return state from both asset prices and discrete expert opinions, while \cite{Gabih2024PowerUtility} study a hidden Gaussian mean-reverting drift with noisy signals arriving at fixed and known dates. These signals are methodologically close to the periodic disclosures in our model: at disclosure dates, the investor updates the conditional distribution of the latent state and subsequently adjusts the optimal portfolio. Our setting differs in that partial information concerns climate states that jointly affect expected asset returns and the dynamics of firm-level carbon intensities. Information therefore affects not only the financial attractiveness of the assets but also the rate at which the portfolio consumes its cumulative emissions budget. \\

Related recent work emphasizes the role of information and learning in dynamic financial and environmental decisions. \cite{Wu_2024} consider continuous-time portfolio selection when the market regime is unobservable, highlighting the interaction between portfolio choice and learning about latent investment opportunities. In an environmental-policy setting, \cite{Basei_2024} study Bayesian learning about an unobservable trend in the socioeconomic costs of pollution and show how posterior beliefs determine the timing of emissions-reduction policy. Our setting differs from these contributions in that investors learn about a latent transition process that simultaneously affects financial returns and firm-level carbon dynamics. Moreover, information arrives through both continuously observed market prices and discrete noisy carbon disclosures. Learning therefore changes not only investment opportunities but also the expected accumulation of the portfolio's cumulative environmental objective. \\

Methodologically, the full-information portfolio problem is related to continuous-time portfolio choice with mean-reverting state variables, as in \cite{KimOmberg_1996}. Under exponential utility and affine-Gaussian state dynamics, the associated Hamilton-Jacobi-Bellman (HJB) equation admits an exponential--quadratic representation related to \cite{Liu_2013}. We combine this structure with continuous-time filtering and discrete Bayesian updates to characterize portfolio choice under alternative climate-information regimes. The contribution relative to standard partial-information portfolio problems is that the latent state being learned affects two dynamic margins simultaneously. It changes expected investment opportunities through asset returns, but also changes firms' expected carbon trajectories and therefore the rate at which current portfolio decisions accumulate financed emissions. In the disclosure-based regime, the resulting control problem further combines continuous learning from market prices with discrete Bayesian revisions from noisy firm-level carbon disclosures. This structure allows us to study when the welfare value of improved information and its effect on cumulative-emissions target satisfaction are aligned, and when they instead move in opposite directions.

The remainder of the paper is organized as follows. Section 2 introduces the notation and formulates the optimization problem. Section 3 derives the optimal policy. Section 4 describes the calibration procedure for the penalty coefficient. Section 5 presents a numerical application, and Section 6 concludes.

\section{Financial market and optimization problem}

In this section, we introduce and formalize the notation useful to define the market setting as well as the financed emissions.

\subsection{Portfolio, cumulative emissions and the state of transition}

We denote by $E_t^i$ the carbon emissions of firm $i$ at time $t$, measured in $\text{tCO2-eq}/\text{year}$. We further define the firm's carbon intensity $U_t^i$ as emissions normalized by an appropriate output measure. This output can be physical (ton of steel produced, MWh generated) or monetary (revenues, market capitalization, enterprise value including cash (EVIC)). In this paper, we follow the guidance from the Technical Expert Group on Sustainable Finance \cite{TEG_2019}, and we use EVIC as the normalization metric, so that:
\begin{equation}
U_t^i = \frac{E_t^i}{\text{EVIC}_t^i}
\end{equation}

Now, let us consider a portfolio $P$ composed of $d$ assets, with wealth $X_t$ at time $t$. We define $\pi_t = (\pi^1_t, ...,\pi^d_t)$ as the portfolio process, describing the share invested in each asset. Following \cite{LeGuenedal_2022}, we define the portfolio’s financed emissions at time $t$ as the sum, across holdings, of each issuer’s emissions scaled by the investor’s ownership ratio. Specifically, for asset $i$, the ownership ratio is the share of the issuer’s enterprise value that is effectively “financed” by the portfolio. The portfolio financed emissions therefore take the form:
\begin{equation}
E^{\pi}_t = \sum_{i=1}^d \frac{\pi_t^iX_t}{\text{EVIC}_t^i}E^i_t = \sum_{i=1}^d \pi_t^iX_t U_t^i
\end{equation}
We can write the financed emissions in vector form:
\begin{equation}
\label{eq:dynamic_E}
E^\pi_t = X_t \bm{\pi}_t^\top \bm{U}_t
\end{equation}
The cumulative financed emissions for the investor are the integral of financed emissions over the investment period $[0,T]$, and are given by
\begin{equation}
    Y_T = \int_0^T E^\pi_s\,ds.
\end{equation}
We can define, at the portfolio level, a cumulative emission budget $\bar{Y}$, and we say that the portfolio is aligned with the budget whenever $Y_T \leq \bar{Y}$. \\

In addition to financed emissions, we introduce a common transition factor $M_t$, which summarizes the aggregate state of the climate transition at time $t$. This factor is meant to capture macro-level transition conditions, such as climate-policy stringency, investor preferences for green assets, technological progress in decarbonization, and the credibility of the transition pathway. We adopt the sign convention that higher values of $M_t$ correspond to a stronger or faster climate transition. Under this convention, an increase in $M_t$ may increase the expected returns of green assets, decrease those of brown assets, and lower firms' long-run carbon-intensity targets. The precise dynamics of $M_t$ and its effects on asset returns and carbon intensities are specified in the market model below.

\subsection{Market setting}

We model a financial market with one risk-free asset and $d$ risky assets. For a fixed date $T>0$, we consider a probability space $(\Omega, \mathcal{G}, \mathbb{G}, \mathbb{P})$, equipped with a right-continuous filtration $\mathbb{G} = (\mathcal{G}_t)_{t \in [0, T]}$. All processes are assumed to be $\mathbb{G}-$ adapted. 
The price dynamics of the risk-free asset are given by:
\begin{equation}
dS^0_t = rS^0_tdt, \, S^0_0 = 1
\end{equation}
The risky assets are denoted by $\Sb_t = (S_t^1, ..., S_t^d)^\top$, with the following dynamics:
\begin{equation}
    \frac{dS_t}{S_t} = \mub_t dt + \Sigmab_S^{1/2}dW_t, 
\end{equation}
where $\frac{dS_t}{S_t}$ is understood componentwise,  $\Sb_0 = (S_0^1, \dots, S_0^d) \in \Rbb^d_+$, $\Sigmab_S \in \Sbb^d_+$ and $W=(\Wb_t)_{t\in[0,T]}$ is a standard $\mathbb{G}-$ Brownian motion in $\mathbb{R}^d$. The covariance matrix $\Sigmab_S$ is assumed to be constant over time and positive definite. Moreover, we assume that the drift $\bm \mu_t$ is stochastic and depends on climate transition. In particular, we assume that the drift process is of the form 
\begin{equation}
\label{eq:mut}
    \bm \mu_t = \bm \mu + \alphab M_t + \betab (\Ub_t - \bar{\Ub}_0)
\end{equation}
with $\bm \mu \in \mathbb{R}^d$, $\alphab \in \mathbb{R}^d$ and $\betab \in \mathbb{R}^{d \times d}$, where $\beta$ is diagonal, with diagonal elements $\beta_i$. The drift therefore depends both on a common transition factor and a firm-specific decarbonization level. \\

The process $M = (M_t)_{t \in [0,T]}$ is the common transition factor, representing the aggregated state of the climate transition. It follows an Ornstein--Uhlenbeck (OU) process:
\begin{equation}
    dM_t=\kappa_M(m_{\infty}-M_t)dt+\sigma_MdW^M_t, M_0 \sim \mathcal{N}(\bar{M_0},q^M_0 )
\end{equation}
where \(\kappa_M>0\), \(m_{\infty}\in\mathbb R\), \(\sigma_M>0\), and \(W^M\) is a one-dimensional Brownian motion. We can interpret this factor as an indicator of the preference towards climate policy and green transition. Thus, a positive value of $M$ indicates a stronger preference for green assets and stricter climate policy. For interpretability, we assume that $M_t$ is centred so that $M_t=0$ corresponds to a neutral transition state. Positive values of $M_t$ then represent stronger-than-neutral climate-transition conditions. \\

We next model firms' carbon intensities denoted by $\Ub_t=(U_t^1,\ldots,U_t^d)$. For each asset $i$, the long-run carbon-intensity target is assumed to depend on the long-term value of the transition factor:
\begin{equation}
    U^i_{\infty} = \bar U_{\infty}^i + \delta_i m_{\infty},
\end{equation}
where $\bar{U}^i_{\infty}$ is the baseline long-run intensity target of asset $i$, corresponding to a neutral transition state, and $\delta_i$ measures the sensitivity of this target to the long-term value of the transition factor. If positive values of $m_{\infty}$ represent stronger climate-policy pressure, then a natural specification for most firms is $\delta_i<0$, so that the long-run intensity target decreases when the long-term value of transition becomes more stringent. \\

The carbon-intensity dynamics are given by an OU process\footnote{Although carbon intensities are nonnegative by construction, we use a Gaussian OU specification for tractability. This approximation is appropriate when the probability of negative intensities is negligible over the relevant parameter range. Alternatively, one could impose positivity through a log-OU or square-root specification.}:
\begin{equation}
\label{eq:Ut}
    d\Ub_t = \mathcal K(\Ub_{\infty}-\Ub_t)dt+\Sigmab_U^{1/2} d\Bb_t
\end{equation}

where $\mathcal{K} = \text{diag}(\kappa_1, \dots, \kappa_d)\in \Rbb^*_+$ contains the mean-reversion speeds of the carbon intensity of asset $i$, $\Ub_{\infty} = (U_{\infty}^1, \dots, U_{\infty}^d)$,  $\Sigmab_U \in \Sbb^d_+$  and
$\Bb=(B^1,\ldots,B^d)$ is a $d$-dimensional Brownian motion. The vector of initial values $\Ub_0$ is assumed to be an independent normally distributed random vector, with mean $\bar{\Ub_0} \in \mathbb{R}^d$ and covariance matrix $\Qb^U_0 \in \Rbb^{d \times d}$, assumed to be symmetric positive. \\

To allow correlation between asset returns and intensity shocks, set $\mathrm d\!\langle \bm W,\bm B\rangle_t = \bm\rho\,\mathrm dt$, for some correlation matrix $\bm{\rho}\in\mathbb R^{d\times d}$. Then the cross-covariance of return and intensity shocks is $\Sigmab_{SU} := \Sigmab_{S}^{1/2}\, \bm{\rho}\, (\Sigmab_U^{1/2})^\top$. A negative entry in $\bm \Sigma^{SU}$ indicates a negative correlation between shocks to the asset’s return and shocks to its carbon intensity. This situation can arise when an asset tends to perform better as the underlying firm decarbonizes—for example, when reductions in intensity are associated with efficiency gains or improved investor sentiment. Conversely, a positive entry represents a positive correlation, which may be observed for high-carbon firms whose expansion in production (and hence revenues) coincides with increases in emissions intensity. We have the following overall covariance matrix, which we assume to be positive definite:
\begin{equation}
\label{eq:Sigma}
\Sigmab = \begin{bmatrix}
    \Sigmab_S & \Sigmab_{SU} \\
    \Sigmab_{SU}^\top & \Sigmab_U
\end{bmatrix}
\end{equation}
Unless otherwise stated, the Brownian motion $W^M$ driving the transition factor is assumed to be independent of $(\Wb,\Bb)$.\\

The carbon intensity of the risk-free asset is deterministic and given by: 
\begin{equation}
    u_{rf}(t) = U_{rf}e^{-\kappa_{rf}t}
\end{equation}

Let $X_t$ denote the wealth at time $t$,  $\pi_t = (\pi^1_t, \dots, \pi^d_t)$, with $\pi^i_t$ the proportion at time $t$ of wealth invested in the $i^{\text{th}}$ asset. The dynamics of $X_t$ are given by
\begin{equation}
\label{eq:dynamique_x}
\frac{dX_t}{X_t} = rdt + \pi_t^\top (\mub_t - r \mathbf{1})dt + \pi_t^\top \Sigmab_S^{1/2} dW_t
\end{equation}
Define $Y_t$ as the cumulated carbon emissions, we obtain
\begin{equation}
\label{eq:dynamique_y}
dY_t = E^\pi_t dt = X_t \left[ \pi_t^\top U_t + (\mathbf{1}-\pi_t\mathbf{1})u_{rf}(t)\right]dt
\end{equation}

To illustrate the meaning of the coefficients $\alphab$, $\betab$ and $\deltab$ of equations \eqref{eq:mut} and \eqref{eq:Ut}, we take the case of two risky assets: a green and a brown asset. The green asset has a lower intensity and an ambitious decarbonization target. The brown asset has a higher initial intensity, with a less ambitious decarbonization target. As in \cite{Bolton_2021}, the brown asset has a higher initial expected return. \\

The coefficient $\alpha_i$ captures the sensitivity of asset $i$'s expected return to $M$. Therefore, a green firm is expected to have $\alpha_i > 0$, while the brown firm is expected to have $\alpha_i < 0$. This is consistent with \cite{Pastor_2021}, where positive shocks to ESG concerns boost green assets. \\

The coefficient $\beta_i$ captures the sensitivity of asset $i$'s expected return to $U^i_t - \bar{U^i}_0$. Therefore, a  firm that is decarbonizing is expected to have $\beta_i < 0$, while a firm that is not reducing its intensity is expected to have $\beta_i > 0$. Thus, we assume that decarbonization is a positive feature for a company. In reality, this might depend on the political context and $\betab$ could, for example, depend on $M_t$. We retain the simpler specification in this paper.  \\

The coefficient $\delta_i$ captures how the long-run carbon-intensity target of asset $i$ reacts to the long-term transition state. Under our sign convention, $\delta_i<0$ means that a stronger transition lowers the firm's long-run carbon-intensity target. For a green asset, this effect may be moderate because the firm is already on a relatively credible decarbonization path. For a brown asset, the effect may be stronger because the firm starts from a higher carbon intensity and faces greater regulatory pressure and larger required abatement. Hence, in the two-asset illustration, we assume
\begin{equation}
        \delta_G<0,\qquad \delta_B<0,\qquad |\delta_B|>|\delta_G|.
\end{equation}

The green firm is less sensitive because it already has a credible decarbonization path. For the brown firm, decarbonizing is more costly, so it is more influenced by the transition state.

\subsection{Optimization problem and information structure}
In the sequel, we consider the same portfolio optimization problem under different information regimes, depending on whether carbon intensities are fully observed, partially observed, or only disclosed at discrete dates. We therefore consider three types of information regimes for the investor, represented by $\Fbb^H=(\mathcal F^H_t)_{t\in[0,T]}$ with \(H\in\{F, U, Z\}\):

\begin{itemize}
    \item The \(F\)-investor has full information \((S,M,U, m_\infty)\):
    \begin{equation}
        \mathcal F^F_t = \sigma(S_s,U_s,M_s,m_\infty, \;0\leq s\leq t)
    \end{equation}
    \item The \(U\)-investor observes the asset price process \(S\) and the carbon intensity process \(U\), but neither \(M\) nor $m_{\infty}$:
     \begin{equation}
        \mathcal F^U_t=\sigma(S_s,U_s,\;0\leq s\leq t)
    \end{equation}   
    \item The \(Z\)-investor observes the asset price process \(S\) and only the noisy disclosures \(Z_n\) at deterministic dates, but neither \(M\) nor $m_{\infty}$:
     \begin{equation}
        \mathcal F^Z_t=\sigma(S_s,\;0\leq s\leq t;\; Z_n,\tau_n\leq t)
    \end{equation}
As in \cite{Gabih2024PowerUtility}, the noisy disclosures of carbon intensity are observed at deterministic disclosure dates $0=\tau_0<\tau_1<\dots<\tau_N<T$ with the observation equation
\begin{equation}
    Z_n=U_{\tau_n}+\epsilon_n,
    \qquad
    \epsilon_n\sim\mathcal N(0,\Sigma^\varepsilon_n),
\end{equation}
with \(\Sigma^\varepsilon_n\in\mathbb R^{d\times d}\) positive definite. 
\end{itemize}
All investors know the model parameters (except for the realized value of $m_{\infty}$ for the $U$- and $Z$-investors) and the distribution of the initial values of $M$ and $U$. \\

We assume that the $H-$investor starts with initial capital $x_0^H$ and aims to maximize terminal wealth while limiting cumulative financed emissions. 
Let $\lambda \geq 0$ denote the wealth-equivalent valuation assigned to cumulative financed emissions. It represents the strength of the investor's decarbonization mandate by converting one additional unit of cumulative financed emissions into the same units as terminal wealth. The parameter should therefore be interpreted as a shadow valuation within the portfolio objective, rather than as a market carbon price or a monetary payment on realized emissions. Under a strategy $\pi$, the wealth and cumulative-emissions processes evolve according to \eqref{eq:dynamique_x}-\eqref{eq:dynamique_y}.

The investor considers the optimization problem
\begin{equation}
    V^H(x_0^H)
    =
    \sup_{\pi\in\Ac^H}
    \Ebb_0
    \left[
        g\left(X_T^\pi-\lambda Y_T^{H,\pi}\right)
    \right],
\end{equation}
where $\Ac^H$ denotes the set of $\Fbb^H$-progressively measurable strategies satisfying \citep{Hu_2005}
\begin{equation}
    \Ebb
    \left[
        \int_0^T
        X_t^2\pi_t^\top\Sigma_S\pi_t\,dt
    \right]
    <\infty.
\end{equation}
In addition, we have the uniform-integrability condition used in the verification argument: for the classical HJB solution $v^H$ characterized below, the process
\begin{equation}
    V_t^{H,\pi}
    :=
    v^H
    \left(
        t,X_t^\pi,Y_t^{H,\pi},\ell_t^H
    \right)
\end{equation}
is required to be of class $D$, i.e.
\begin{equation}
    \left\{
        V_\tau^{H,\pi}:
        \tau\leq T,\;
        \tau\ \text{an }\Fbb^H\text{-stopping time}
    \right\}
\end{equation}
is uniformly integrable. Finally,
\begin{equation}
    g(x)=-e^{-\eta x}
\end{equation}
is the CARA utility function.

\begin{remark}[Impact of short-selling]
We do not impose short-selling constraints, so risky-asset positions $\pi_t^i$ may take negative values. Under the financed-emissions definition in \eqref{eq:dynamic_E}, such positions mechanically generate negative contributions to portfolio financed emissions. In practice, however, the accounting standard of \cite{PCAF_2022} does not treat short positions as generating negative financed emissions and instead attributes financed emissions only to long positions. To preserve the tractability of the model, we abstract from this asymmetry and leave its explicit treatment for future research. We assess the quantitative importance of this modelling choice in Appendix~\ref{app:supplementary figures}.
\end{remark}

\section{Partial observation and filtering}

Because all coefficients are linear and all noises are Gaussian, the conditional law remains Gaussian. Therefore, the filters for both the $U-$ and $Z-$ investor is fully characterized by the conditional mean and the conditional covariance matrix. We assume that $m_{\infty}$ is an unknown Gaussian random variable, constant over time, with prior distribution
\begin{equation}
    m_{\infty} \sim \mathcal N(m_{\infty,0},q^{m_\infty}_0).
\end{equation}

\paragraph{Filtering for the \(U\)-investor with unknown long-run transition mean}

The \(U\)-investor observes the asset-price process \(\Sb\) and the carbon-intensity
process \(\Ub\) continuously, but does not observe either the transition factor
\(M_t\) or its long-run mean \(m_\infty\). We define the augmented state $L^U$, as follows:
\begin{equation}
    L^U_t
    :=
    \begin{pmatrix}
        M_t\\
        m_{\infty}
    \end{pmatrix}.
\end{equation}
Since \(m_{\infty}\) is constant over time, we have \(dm_{\infty}=0\). Therefore, the
hidden state satisfies the linear dynamics
\begin{equation}
\label{eq:hidden_state_U_unknown_barm}
    dL^U_t= A_U L^U_t\,dt+G_U\,dW^M_t,
\end{equation}
where
\begin{equation}
    A_U = \begin{pmatrix}
        -\kappa_M & \kappa_M\\
        0 & 0
    \end{pmatrix},
    \qquad
    G_U = \begin{pmatrix}
        \sigma_M\\
        0
    \end{pmatrix}.
\end{equation}
Indeed, the first component of \eqref{eq:hidden_state_U_unknown_barm} gives
\begin{equation}
    dM_t=\kappa_M(m_\infty-M_t)dt+\sigma_MdW^M_t,
\end{equation}
while the second component states that \(m_\infty\) is an unknown but fixed
parameter.

The \(U\)-investor observes \(S\) and \(\Ub\), and therefore observes both return
and carbon-intensity residuals. Since \(\Ub_t\) is observed, the term
\(\betab(\Ub_t-\bar{\Ub}_0)\) in the return drift is known and should be subtracted from
the return observation. Define
\begin{align}
    \label{eq:return_residual_U_unknown_barm}  
    dR_t &= \frac{dS_t}{S_t}- \left[\mub+\betab(\Ub_t-\bar \Ub_0)\right]dt =\alphab M_t\,dt+\Sigmab_S^{1/2}dW_t \\
    dJ_t &=d\Ub_t-\mathcal K(\bar  \Ub_\infty-\Ub_t)dt = \mathcal K\deltab\,m_\infty\,dt +\Sigmab_U^{1/2}d\Bb_t 
\end{align}

Stacking the return and carbon-intensity residuals gives
\begin{equation}
\label{eq:observation_U_unknown_barm}
    dO_t
    =
    \begin{pmatrix}
        dR_t\\
        dJ_t
    \end{pmatrix}
    =
    H_U L^U_t\,dt+dN_t,
\end{equation}
where
\begin{equation}
\label{eq:observation_matrix_U_unknown_barm}
    H_U
    =
    \begin{pmatrix}
        \alphab & 0_d\\
        0_d & \mathcal K \deltab
    \end{pmatrix}
    \in\mathbb R^{2d\times 2},
\end{equation}
and
\begin{equation}
    dN_t
    :=
    \begin{pmatrix}
        \Sigmab_S^{1/2}dW_t\\
        \Sigmab_U^{1/2}d\Bb_t
    \end{pmatrix}.
\end{equation}
The observation-noise covariance is
\begin{equation}
    d\langle N\rangle_t
    =
    \Sigmab\,dt,
    \qquad
    \Sigmab
    =
    \begin{pmatrix}
        \Sigmab_S & \Sigmab_{SU}\\
        \Sigmab_{SU}^\top & \Sigmab_U
    \end{pmatrix},
\end{equation}
which is assumed to be positive definite. Unless otherwise stated, we assume
that the Brownian motion \(W^M\) driving the hidden transition factor is
independent of the observation noise \((W,\Bb)\).

The optimal estimator of the hidden state \(L^U_t=(M_t,m_\infty)^\top\), conditional on the \(U\)-investor's information \(\mathcal F^U_t\), is
\begin{equation}
    \widehat L^U_t
    :=
    \mathbb E[L_t\mid\mathcal F^U_t]
    =
    \begin{pmatrix}
        \widehat M^U_t\\
        \widehat{m}_{\infty,t}^U
    \end{pmatrix}.
\end{equation}
The associated conditional covariance matrix is
\begin{equation}
    Q^U_t =
    \mathbb E\left[
        (L^U_t-\widehat L^U_t)(L^U_t-\widehat L^U_t)^\top
        \mid \mathcal F^U_t
    \right]
    \in\mathbb S^2_+.
\end{equation}

\begin{proposition}[Kalman--Bucy filter for the \(U\)-investor with unknown \(m_\infty\)]
\label{prop:filter_U_unknown_barm}

Assume that the initial hidden state is Gaussian:
\begin{equation}
    L^U_0
    =
    \begin{pmatrix}
        M_0\\
        m_\infty
    \end{pmatrix}
    \sim
    \mathcal N(\widehat L^U_0,Q^U_0),
\end{equation}
with
\begin{equation}
    \widehat L^U_0
    =
    \begin{pmatrix}
        \bar M_0\\
        m_{\infty,0}
    \end{pmatrix},
    \qquad
    Q^U_0
    =
    \begin{pmatrix}
        q^M_0 & q^{M,m_\infty}_0\\
        q^{M,m_\infty}_0 & q^{m_\infty}_0
    \end{pmatrix}.
\end{equation}

Then the conditional mean \(\widehat L^U_t\) satisfies
\begin{equation}
\label{eq:KB_mean_U_unknown_barm}
    d\widehat L^U_t
    =
    A_U\widehat L^U_t\,dt
    +
    K_t^U
    \left(
        dO_t-H_U\widehat L^U_t\,dt
    \right)
\end{equation}
where the Kalman gain is
\begin{equation}
    K_t^U  = Q_t^U H_U^\top \Sigmab^{-1}.
\end{equation}

The conditional covariance matrix \(Q^U_t\) is the unique symmetric
nonnegative solution of the Riccati equation
\begin{equation}
\label{eq:KB_covariance_U_unknown_barm}
    \dot Q^U_t
    =
    A_UQ^U_t+Q^U_tA_U^\top+G_UG_U^\top
    -
    K_t^UH_UQ^U_t,
\end{equation}

\end{proposition}

\begin{proof}
See Appendix~\ref{app:proof_filter_U_unknown_barm}.
\end{proof}

Under this specification, the return residual \(dR_t\) provides information about
the current transition state \(M_t\), while the carbon-intensity residual \(dJ_t\)
provides information about the unknown long-run transition mean \(m_\infty\).
The two estimates remain coupled through the state equation for \(M_t\), since
\(M_t\) mean-reverts toward \(m_\infty\). Hence, even though \(m_\infty\) is constant,
its posterior estimate \(\widehat{m}^U_{\infty,t}\) is updated dynamically through
the innovation process.

\paragraph{Filtering for the \(Z\)-investor with unknown long-run transition mean}

Since the \(Z\)-investor does not observe \(\Ub_t\) continuously, the hidden state
must include the carbon-intensity vector. We therefore define the augmented
hidden state
\begin{equation}
    L_t^Z = \begin{pmatrix}  
    M_t\\
    m_\infty\\
    \Ub_t
    \end{pmatrix}
    \in\mathbb R^{d+2}.
\end{equation}
The hidden state satisfies the affine linear dynamics
\begin{equation}
\label{eq:hidden_state_Z_unknown_barm}
    dL_t^Z
    =
    \left(A_ZL_t^Z+c_Z\right)dt
    +
    G_Zd\widetilde W_t,
\end{equation}
where
\[
    \widetilde W_t =
    \begin{pmatrix}
        W_t^M\\
        \Bb_t
    \end{pmatrix}
    \in\mathbb R^{d+1},
\]
and
\begin{equation}
\label{eq:AZ_Z_unknown_barm}
    A_Z
    =
    \begin{pmatrix}
        -\kappa_M & \kappa_M & 0_{1\times d}\\
        0 & 0 & 0_{1\times d}\\
        0_{d\times 1} & \mathcal K \deltab & -\mathcal K
    \end{pmatrix},
    \qquad
    c_Z
    =
    \begin{pmatrix}
        0\\
        0\\
        \mathcal K\bar \Ub_\infty
    \end{pmatrix},
\end{equation}
and
\begin{equation}
\label{eq:GZ_Z_unknown_barm}
    G_Z
    =
    \begin{pmatrix}
        \sigma_M & 0_{1\times d}\\
        0 & 0_{1\times d}\\
        0_{d\times 1} & \Sigmab_U^{1/2}
    \end{pmatrix}.
\end{equation}
The first component of \eqref{eq:hidden_state_Z_unknown_barm} is
\begin{equation}
    dM_t=\kappa_M(m_\infty-M_t)dt+\sigma_MdW^M_t,
\end{equation}
the second component is \(dm_\infty=0\), and the last \(d\) components correspond
to the carbon-intensity dynamics. We treat \(\bar{\Ub}_0\) as a known reference carbon-intensity level. Hence the continuous return observation can be written as
\begin{equation}
    dR_t^Z = \frac{dS_t}{S_t} - \left(\mub-\mathcal \betab \bar \Ub_0\right)dt.
\end{equation}
Then
\begin{equation}
    dR_t^Z = H_ZL_t^Z\,dt+\Sigmab_S^{1/2}dW_t,
\end{equation}
where
\begin{equation}
    H_Z= \begin{pmatrix}
        \alphab & 0_d & \betab
    \end{pmatrix}
    \in\mathbb R^{d\times(d+2)}.
\end{equation}
Here \(\alphab\) and \(0_d\) are understood as \(d\times 1\) columns, while
\(\betab\in\mathbb R^{d\times d}\).

The disclosure equation can be written as:
\begin{equation}
    Z_n  = D_ZL_{\tau_n}^Z+\epsilon_n, \quad  D_Z
    :=
    \begin{pmatrix}
        0_{d\times 1} & 0_{d\times 1} & I_d
    \end{pmatrix}
    \in\mathbb R^{d\times(d+2)}.
\end{equation}

The state noise $G_Zd\widetilde W_t$ and the return-observation noise $\Sigmab_S^{1/2}dW_t$ are correlated. We define the correlation matrix:
\begin{equation}
\Sigma_{SL}\,dt := d\left\langle G_Z\widetilde W, \Sigmab_S^{1/2}W \right\rangle_t = \begin{pmatrix} 0_{1\times d}\\ 0_{1\times d}\\ \Sigmab_{SU}^\top \end{pmatrix} \in\mathbb R^{(d+2)\times d}
\end{equation} 

The conditional mean of the hidden state is
\begin{equation}
    \widehat L_t^Z = \mathbb E[L_t^Z\mid\mathcal F_t^Z]  =
    \begin{pmatrix}
        \widehat M_t^Z\\
        \widehat{m}_{\infty,t}^Z\\
        \widehat \Ub_t^Z
    \end{pmatrix}
\end{equation}
and the conditional covariance matrix is
\begin{equation}
    Q_t^Z
    =
    \mathbb E\left[
        (L_t^Z-\widehat L_t^Z)(L_t^Z-\widehat L_t^Z)^\top
        \mid \mathcal F_t^Z
    \right]
\end{equation}

Recall that the $Z-$investor observes $Z_n$ at disclosure dates $\tau_n$ for $n=0, \dots, N$. In the next proposition,  $\widehat L_t^Z$ (resp. $Q_t^Z$) denotes the conditional mean (resp. covariance) between disclosure dates, while $\bar L_n^Z$ (resp. $\bar Q_n^Z$) represents the posterior mean (resp. posterior covariance matrix) after observing the disclosure \(Z_n\). This proposition characterizes the filtered process $(\widehat{L}^Z, Q^Z)$ over the whole interval $[0~;~T]$.

\begin{proposition}[Filtering equations for the \(Z\)-investor with unknown \(m_\infty\)]
\label{prop:filter_Z_unknown_barm}

Between two disclosure dates, namely for \(t\in(\tau_n,\tau_{n+1})\), the conditional mean satisfies
\begin{equation}
\label{eq:KB_mean_Z_unknown_barm}
    d\widehat L_t^Z  = \left(A_Z\widehat L_t^Z+c_Z\right)dt  +
    \left(Q_t^ZH_Z^\top+\Sigma_{SL}\right)\Sigmab_S^{-1} \left( dR_t^Z-H_Z\widehat L_t^Zdt \right), \quad \hat{L}^Z_{\tau_n} = \bar{L}^Z_n
\end{equation}
The associated conditional covariance matrix satisfies the Riccati equation
\begin{equation}
\label{eq:KB_covariance_Z_unknown_barm}
    \dot Q_t^Z
    =
    A_ZQ_t^Z+Q_t^ZA_Z^\top+G_ZG_Z^\top
    -
    \left(Q_t^ZH_Z^\top+\Sigma_{SL}\right)
    \Sigmab_S^{-1}
    \left(Q_t^ZH_Z^\top+\Sigma_{SL}\right)^\top,  \quad Q^Z_{\tau_n} = \bar{Q}^Z_n
\end{equation}
At each disclosure date $(\tau_n)_{n > 0}$, the filter is updated according to: 
\begin{equation}
\label{eq:disclosure_update_mean_Z_unknown_barm}
    \widehat L_{\tau_n}^Z  = \widehat L_{\tau_n^-}^Z + K_n^Z\nu_n^Z,
\end{equation}
with
\begin{align}
    \nu_n^Z & =Z_n-D_Z\widehat L_{\tau_n^-}^Z, \\
    K_n^Z & = Q_{\tau_n^-}^ZD_Z^\top( D_ZQ_{\tau_n^-}^ZD_Z^\top+\Sigmab^\epsilon_n)^{-1}
\end{align}
the disclosure innovation and gain, respectively, where
\begin{equation}
Q_{\tau_n}^Z  = Q_{\tau_n^-}^Z  - Q_{\tau_n^-}^ZD_Z^\top \left(D_ZQ_{\tau_n^-}^ZD_Z^\top+\Sigmab_n^\epsilon \right)^{-1} D_ZQ_{\tau_n^-}^Z
\end{equation}

\end{proposition}

\begin{proof}
See Appendix~\ref{app:proof_filter_Z_unkown_barm}.
\end{proof}

The filtering problem for the $Z$-investor is therefore hybrid. Between
disclosure dates, learning comes only from continuously observed returns. These
returns contain information about \(M_t\) and \(\Ub_t\) through the drift term
\begin{equation*}
    \alphab M_t+\mathcal \betab\Ub_t.
\end{equation*}
At disclosure dates, the investor receives a noisy direct signal about
$\Ub_{\tau_n}$. This disclosure also updates beliefs about $m_\infty$, because
$m_\infty$ determines the long-run target of $\Ub_t$, and it updates beliefs
about $M_t$ through the covariance structure of the posterior state. Hence the
discrete disclosures affect the whole filtered state
\begin{equation}
    \widehat L_t^Z
    =
    \left(
        \widehat M_t^Z,
        \widehat{m}^Z_{\infty,t},
        \widehat \Ub_t^Z
    \right)^\top.
\end{equation}

\section{The optimization problem}

In this section, we will solve this problem for the different types of investors. We define a generic state vector: 
\begin{equation}
\label{eq:state_vector}
    \ell=
    \begin{pmatrix}
        m\\
        \xi\\
        \ub
    \end{pmatrix}
    \in\mathbb R^{d+2}.
\end{equation}
Define the selection matrices
\begin{equation}
    E_m=
    \begin{pmatrix}
        1 & 0 & 0_{1\times d}
    \end{pmatrix},
    \qquad
    E_{\xi}=
    \begin{pmatrix}
        0 & 1 & 0_{1\times d}
    \end{pmatrix}, 
    \qquad
    E_u=
    \begin{pmatrix}
        0_{d\times1} & 0_{d\times1} & I_d
    \end{pmatrix}
\end{equation}
Thus,
\begin{equation}
    m=E_m \ell,
    \qquad
    \xi=E_{\xi}\ell,
    \qquad
    \ub=E_u \ell.
\end{equation}
The excess-return vector is
\begin{equation}
\label{eq:theta_F}
    \begin{aligned}
    \theta(\ell) &= \mub-r\mathbf 1 +\alphab E_m\ell + \betab(E_u\ell-\bar \Ub_0) \\
    &= \theta_0 + \Theta\ell
    \end{aligned}
\end{equation}
where
\begin{equation}
    \theta_0:=\mu-r\mathbf 1-\betab \bar U_0,
    \qquad
    \Theta:=\alphab E_m+\betab E_u.
\end{equation}
Finally, we define: 
\begin{equation}
    \gamma(t)=e^{r(T-t)}  - \lambda \int_{t}^{T}e^{r(s-t)}u_{rf}(s)ds
\end{equation}

\begin{remark}
We assume that $\inf_{t \in [0,T]} \gamma(t) > 0$. This condition ensures that the marginal coefficient of current wealth in the transformed terminal objective remains strictly positive and, in particular, that the feedback controls derived below are well defined. Since $u_{\mathrm{rf}}(t)=U_{\mathrm{rf}}e^{-\kappa_{\mathrm{rf}}t}$, this is equivalent, for $r\neq\kappa_{\mathrm{rf}}$, to
\begin{equation}
    0\leq\lambda<\bar\lambda,
    \qquad
    \bar\lambda
    :=
    \left[
    U_{\mathrm{rf}}
    \frac{
    e^{-\kappa_{\mathrm{rf}}T}-e^{-rT}
    }{
    r-\kappa_{\mathrm{rf}}
    }
    \right]^{-1}.
\end{equation}

\end{remark}

\subsection{Optimal strategy for the \(F\)-investor}

The $F-$investor has access to full information and solves the following problem: 
\begin{equation}\label{problem-investor-F}
    V^F(x_0^F) = \sup_{\pi \in \Ac^F}
    \mathbb E_0\left[
    g(X_T-\lambda Y_T)
    \right],
\end{equation}
Note that, because the $F$-investor knows the value of $m_\infty$, it will be treated as a parameter of the model. The control variables are \(\pi_i\), \(i=1,\dots,d\), and the state variables are
$$ 
(X_t,Y_t, \mathcal{L}_t^F)
$$
where:
\begin{equation}
\mathcal{L}_t^F = \begin{pmatrix}
        M_t\\
        \Ub_t
        \end{pmatrix}
\end{equation}
The dynamics of $\mathcal{L}_t^F$ is given by: 
\begin{equation}
\label{eq:state_U_control}
    d\mathcal L_t^F
    =
    (A_F\mathcal{L}_t^F + c_F)dt
    +
    D_F d\widetilde W_t,
\end{equation}
where
\begin{equation}
    A_F
    =
    \begin{pmatrix}
        -\kappa_M & 0_{1\times d}\\
        0_{d\times 1} & -\mathcal K
    \end{pmatrix},
    \quad
    c_F
    =
    \begin{pmatrix}
        \kappa_M m_\infty\\
        \mathcal K (\bar\Ub_\infty +\deltab m_\infty)
    \end{pmatrix},
    \quad    
    D_F =
    \begin{pmatrix}
        \sigma_M \quad 0_{1\times d}\\
        0_{d\times d} \quad \Sigma_U^{1/2}
    \end{pmatrix}
    \in\mathbb R^{(d+1)\times (d+1)}.
\end{equation}
and
\begin{equation}
    \widetilde W_t =
    \begin{pmatrix}
        W^M_t\\
        \Bb_t
    \end{pmatrix}
    \in\mathbb R^{d+1}.
\end{equation}

The state noise $D_Fd\widetilde W_t$ and the return-observation noise $\Sigmab_S^{1/2}dW_t$ are correlated. We define the correlation matrix:
\begin{equation}
\Gamma_F\,dt = 
d\left\langle D_F\widetilde W, \Sigmab_S^{1/2}W \right\rangle_t = 
\begin{pmatrix}  
0_{1\times d}\\ 
\Sigmab_{SU}^\top 
\end{pmatrix} 
\in\mathbb R^{(d+1)\times d}
\end{equation} 

We use the simplified state vector for the $F$-investor:
\begin{equation}
\ell^F :=\begin{pmatrix}
        m\\
        \ub
        \end{pmatrix}
\in \mathbb R^{d+1} 
\end{equation}
and, for the case of the $F$-investor, we adapt the selection matrices as follows:
\begin{equation}
    E_m=
    \begin{pmatrix}
        1  & 0_{1\times d}
    \end{pmatrix},
    \qquad
    \qquad
    E_u=
    \begin{pmatrix}
        0_{d\times1}  & I_d
    \end{pmatrix}
\end{equation}
leading to: 
\begin{equation}
    \theta^F(\ell^F)=\theta_0+\Theta^F\ell^F, \qquad \Theta^F:=\alphab E_m^F+\betab E_u^F,
\end{equation}

\begin{proposition}[Optimal control and value function for the \(F\)-investor]
\label{prop:control-F}

For \(t\in[0,T]\), the value function of the \(F\)-investor is denoted by $v^F(t,x,y,\ell)$ and is given by
\begin{equation}
\label{eq:value_F}
    v^F(t,x,y,\ell^F)
    =
    -\exp
    \left\{
        -\eta
        \left[
            \gamma(t)x-\lambda y+\phi^F(t,\ell^F)
        \right]
    \right\}
\end{equation}
where $\phi^F$ admits the quadratic representation
\begin{equation}
\phi^F(t,\ell^F)=p^F_0(t)+p^F_1(t)^{\top}\ell^F+\frac12(\ell^F)^{\top}P^F(t)\ell^F,
\end{equation}
with $p^F_0(t) \in\mathbb R$, $p^F_1(t) \in\mathbb R^{d+1}$ and $P^F(t)\in\mathbb R^{(d+1)\times(d+1)}$ satisfying the backward system

\begin{equation}
    p^F_0(T)=0,\qquad p^F_1(T)=0,\qquad P^F(T)=0,
\end{equation}
and
\begin{equation}
\begin{aligned}
        0 = & \dot p^F_0(t) + c_F^\top p^{F}_1 +\frac12\operatorname{Tr} \left( D_FD_F^{\top}P^F(t) \right) -\frac{\eta}{2}(p_1^{F})^\top D_FD_F^{\top}p^F_1(t) +\frac{1}{2\eta} b^F_0(t)^{\top}\Sigma_S^{-1}b^F_0(t) \\
        0= & \dot p^F_1(t) + A_F^{\top}p^F_1(t)+P^F(t)c_F-\eta P^F(t)D_FD_F^{\top}p^F_1(t)+\frac{1}{\eta}B^F(t)^{\top}\Sigma_S^{-1}b^F_0(t) \\
        0=&\dot P^F(t)+P^F(t)A_F+A_F^{\top}P^F(t)-\eta P^F(t)D_FD_F^{\top}P^F(t)+\frac{1}{\eta}B^F(t)^{\top}\Sigma_S^{-1}B^F(t),
\end{aligned}
\end{equation}
where
\begin{equation}
    b_0^F(t)
    :=
    \theta_0-\eta\Gamma_F^\top p_1^F(t)
    + \frac{\lambda}{\gamma(t)}u_{rf}(t)\mathbf{1}
    \qquad
    B^F(t)
    :=
    \Theta^F-\frac{\lambda}{\gamma(t)}E_u^F
    -\eta\Gamma_F^\top P^F(t).
\end{equation}

Moreover, the optimal control is
\begin{equation}
    \label{eq:optimal_pi_F_general}
    \pi^{F,*}(t,x,\ell^F)=\frac{1}{\eta\gamma(t)x}\Sigma_S^{-1}\left[\theta(\ell^F)-\frac{\lambda}{\gamma(t)}(E_u\ell^F - u_{rf}(t)\mathbf{1})-\eta\Gamma_F^\top\nabla_{\ell}\phi^F(t,\ell^F)\right]
\end{equation}
\end{proposition}

\begin{proof}
See Appendix~\ref{app:proof-control-F}.
\end{proof}

\begin{remark}[Interpretation in the original variables]

We use the fact that:
\begin{equation}
    \Gamma_F^\top\nabla_\ell\phi^F
    =
    \Sigmab_{SU}\nabla_u\phi^F
\end{equation}

The optimal strategy can be written in the original variables as
\begin{equation}
\begin{aligned}
\label{eq:pi_F_original_variables}
    \pi^{F,*}(t,x,m,\ub)
    &=
    \frac{1}{\eta\gamma(t)x}
    \Sigmab_S^{-1}
    \Bigg[
        \mub-r\mathbf 1
        +
        \alphab m
        +
        \betab(\ub-\bar{\Ub}_0) \\
        & - 
        \frac{\lambda}{\gamma(t)}(\ub - u_{rf}(t)\mathbf{1})
        -
        \eta\Sigmab_{SU}\nabla_u\phi^F(t,m,\ub)
    \Bigg]
\end{aligned}
\end{equation}

The optimal allocation has four components. The term
\begin{equation}
    \mub-r\mathbf 1
\end{equation}
is the standard financial excess return. The term
\begin{equation}
    \alphab m
\end{equation}
captures the impact of the transition factor on expected returns. The term
\begin{equation}
    \betab(\ub-\bar\Ub_0)
\end{equation}
captures the direct effect of firms' observed carbon intensities on expected
returns. The term
\begin{equation}
    -\frac{\lambda}{\gamma(t)}(\ub - u_{rf}(t)\mathbf{1})
\end{equation}
is the direct carbon-penalty component: assets with higher current carbon
intensities are penalized more strongly. Finally,
\begin{equation}
    -\eta\Sigmab_{SU}\nabla_u\phi^F(t,m,\ub)
\end{equation}
is an intertemporal hedging demand induced by the instantaneous correlation
between return shocks and carbon-intensity shocks.
\end{remark}

\subsection{Optimal strategy for the \(U\)-investor}

The \(U\)-investor observes the price process \(S\) and the carbon-intensity process \(\Ub\) continuously, but does not observe either the transition factor \(M_t\) or the long-run transition mean \(m_\infty\). Because \(\Ub_t\) is observed continuously and enters both the return drift and the emission penalty, the relevant Markov state for the control problem of the $U$-investor is :
\begin{equation}
    \hat{\mathcal L}_t^U
    =
    \begin{pmatrix}
        \hat{L}^U_t\\
        \Ub_t
    \end{pmatrix}
    \in\mathbb R^{d+2}.
\end{equation}
Here, $\hat{L}_t^U$ is the stochastic Kalman filtered mean and the conditional covariance matrix \(Q_t^U\) is deterministic and satisfies the Kalman--Bucy Riccati equation from Proposition~\ref{prop:filter_U_unknown_barm}. 
We define the innovation process $I_t^U$:
\begin{equation}
    dI_t^U = dO_t - H_U\hat{L}_t^Udt 
\end{equation}
The dynamics of $\hat{\mathcal L}^U$ is:
\begin{equation}
\label{eq:state_U_control}
    d\hat{\mathcal L}_t^U
    =
    (A_U\hat{\mathcal L}_t^U+c_U)dt
    +
    D_t^U dI_t^U,
\end{equation}
where
\begin{equation}
    A_U
    =
    \begin{pmatrix}
        -\kappa_M & \kappa_M & 0_{1\times d}\\
        0 & 0 & 0_{1\times d}\\
        0_{d\times 1} & \mathcal K\deltab & -\mathcal K
    \end{pmatrix},
    \qquad
    c_U
    =
    \begin{pmatrix}
        0\\
        0\\
        \mathcal K\bar\Ub_\infty
    \end{pmatrix},
\end{equation}
and
\begin{equation}
    D_t^U =
    \begin{pmatrix}
        K_t^U\\
        0_{d\times d} \quad I_d
    \end{pmatrix}
    \in\mathbb R^{(d+2)\times 2d}.
\end{equation}
The matrix \(D_t^U\) should be read blockwise: the first two rows correspond to
the Kalman--Bucy update of \((\widehat M_t^U,\widehat{m}_{\infty,t}^U)\), while the
last \(d\) rows extract the intensity innovation \(dJ_t\).

The covariance matrices of the return innovation and filtered states are:
\begin{equation}
    d\langle \widehat R^U\rangle_t=\Sigmab_Sdt,
    \qquad
    d\langle \hat{\mathcal{L}}^U\rangle_t=\Lambda_t^Udt \quad, \Lambda_t^U =  D_t^U\Sigmab(D_t^U)^\top  \in\mathbb R^{(d+2)\times (d+2)}
\end{equation}

\begin{lemma}[Covariance between the filtered state and the return innovation]
\label{lem:cov_RL_U}
Define
\begin{equation}
    \Gamma_t^U = D_t^U\Sigmab E_R^\top  
    \in\mathbb R^{(d+2)\times d}, \quad     E_R =
    \begin{pmatrix}
        I_d & 0_{d\times d}
    \end{pmatrix}
\end{equation}
The instantaneous covariance between the filtered state and the return innovation is
\begin{equation}
d\langle L^U,\widehat R^U\rangle_t
=
\Gamma^U_tdt,
\end{equation}
The conditional covariance matrix of the hidden state $\hat{L}^U_t$ can be written blockwise as
\begin{equation}
    Q_t^U
    =
    \begin{pmatrix}
        q^U_M(t) & q_{M,m_\infty}^U(t)\\
        q_{M,m_\infty}^U(t) & q_{m_\infty}^U(t)
    \end{pmatrix}.
\end{equation}
and
\begin{equation}
    \Gamma_t^U
    =
    \begin{pmatrix}
        q^U_M(t)\alphab^\top\\
        q_{M,m_\infty}^U(t)\alphab^\top\\
        \Sigmab_{SU}^\top
    \end{pmatrix}
    \in\mathbb R^{(d+2)\times d}.
\end{equation}

\end{lemma}

\begin{proof}
See Appendix~\ref{app:proof_lemma_cov_RL_U}.
\end{proof}

The \(U\)-investor therefore solves
\begin{equation}
    V^U(x_0^U) =    \sup_{\pi\in\mathcal A^U}
    \mathbb E_0
    \left[
        -\exp\left(
            -\eta(X_T-\lambda Y_T)
        \right)
    \right].
\end{equation}

\begin{proposition}[Optimal control and value function for the \(U\)-investor]
\label{prop:optimal-U}
For the \(U\)-investor, the generic state-space variable is
\begin{equation}
        \ell=     \begin{pmatrix}
        m\\
        \xi\\
        \ub
    \end{pmatrix}\in\mathbb R^{d+2},
\end{equation}
where \(m\) represents the filtered estimate of the transition factor \(M_t\), \(\xi\) represents the filtered estimate of the long-run transition mean \(m_\infty\), and \(\ub\) is the observed carbon-intensity vector. The value function of the \(U\)-investor is 
\begin{equation}
    v^U(t,x,y,\ell)
    =
    -\exp
    \left\{
        -\eta
        \left(
            \gamma(t)x-\lambda y+\phi^U(t,\ell)
        \right)
    \right\}.
\end{equation}
where \(\phi^U\) admits the quadratic representation
\begin{equation}
    \phi^U(t,\ell)
    =
    p_0^U(t)
    +
    p_1^U(t)^\top\ell
    +
    \frac12\ell^\top P^U(t)\ell,
\end{equation}
with $p_0^U(t)\in\mathbb R, \, p_1^U(t)\in\mathbb R^{d+2}$ and $P^U(t)\in\mathbb R^{(d+2)\times(d+2)}$.
Then the coefficient functions solve the backward system with the terminal conditions
\begin{equation}
    p_0^U(T)=0, \qquad p_1^U(T)=0,  \qquad  P^U(T)=0
\end{equation}
and
\begin{equation}
\begin{aligned}
    0={}& \dot p_0^U(t) + p_1^U(t)^\top c_U + \frac12
    \operatorname{Tr} \left(
        \Lambda_t^U P^U(t)
    \right)
    -  \frac{\eta}{2} p_1^U(t)^\top\Lambda_t^U p_1^U(t)\\
    &+ \frac{1}{2\eta} b_0^U(t)^\top \Sigmab_S^{-1} b_0^U(t), \\
    0={}&\dot p_1^U(t) + A_U^\top p_1^U(t) + P^U(t)c_U
    - \eta P^U(t) \Lambda_t^U p_1^U(t)\\
    &+ \frac{1}{\eta} B^U(t)^\top \Sigmab_S^{-1}b_0^U(t), \\
    0={}& \dot P^U(t) + P^U(t)A_U + A_U^\top P^U(t)
    - \eta P^U(t) \Lambda_t^U P^U(t)\\
    &+ \frac{1}{\eta} B^U(t)^\top \Sigmab_S^{-1} B^U(t).
\end{aligned}
\end{equation}
where
\begin{equation}
    b_0^U(t)
    :=
    \theta_0-\eta(\Gamma_t^U)^\top p_1^U(t)
    + \frac{\lambda}{\gamma(t)}u_{rf}(t)\mathbf{1})
    \qquad
    B^U(t)
    :=
    \Theta-\frac{\lambda}{\gamma(t)}E_u
    -\eta(\Gamma_t^U)^\top P^U(t).
\end{equation}

The optimal control is
\begin{equation}
\label{eq:optimal_pi_U_general}
    \pi_t^{U,*}
    =
    \frac{1}{\eta\gamma(t)X_t}
    \Sigmab_S^{-1}
    \left[
        \theta^U(\ell_t^U)
        -
        \frac{\lambda}{\gamma(t)}(E_u\ell_t^U- u_{rf}(t)\mathbf{1})
        -
        \eta(\Gamma_t^U)^\top
        \nabla_\ell\phi^U(t,\ell_t^U)
    \right].
\end{equation}

\end{proposition}

\begin{proof}
See Appendix~\ref{app:proof-control-U}.
\end{proof}

\begin{remark}[Interpretation in the original variables for the \(U\)-investor]

For the \(U\)-investor, the generic state-space variable is
\begin{equation}
        \ell^U=     \begin{pmatrix}
        m\\
        \xi\\
        \ub
    \end{pmatrix}\in\mathbb R^{d+2},
\end{equation}
where \(m\) represents the filtered estimate of the transition factor \(M_t\), \(\xi\) represents the filtered estimate of the long-run transition mean \(m_\infty\), and \(\ub\) is the observed carbon-intensity vector.

Recall that
\begin{equation}
    \Gamma_t^U
    =
    \begin{pmatrix}
        q^U_M(t)\alphab^\top\\
        q_{M,m_\infty}^U(t)\alphab^\top\\
        \Sigmab_{SU}^\top
    \end{pmatrix}
    \in\mathbb R^{(d+2)\times d}.
\end{equation}
Therefore, writing
\begin{equation}
    \nabla_\ell\phi^U
    =
    \begin{pmatrix}
        \partial_m\phi^U\\
        \partial_\xi\phi^U\\
        \nabla_u\phi^U
    \end{pmatrix},
\end{equation}
we obtain
\begin{equation}
\label{eq:Gamma_U_original_variables}
    (\Gamma_t^U)^\top\nabla_\ell\phi^U
    =
    \alphab
    \left[
        q^U_M(t)\partial_m\phi^U
        +
        q_{M,m_\infty}^U(t)\partial_\xi\phi^U
    \right]
    +
    \Sigmab_{SU}\nabla_u\phi^U.
\end{equation}

The optimal strategy can therefore be written in the original variables as
\begin{equation}
\label{eq:pi_U_original_variables}
\begin{split}
    \pi^{U,*}(t,x,m,\xi,\ub)
    =
    \frac{1}{\eta\gamma(t)x}
    \Sigmab_S^{-1}
    \Bigg[
        &\mub-r\mathbf 1
        +
        \alphab m
        +
        \betab(\ub-\bar{\Ub}_0)
        -
        \frac{\lambda}{\gamma(t)}(\ub - u_{rf}(t)\mathbf{1})\\
        &-
        \eta\alphab
        \left(
            q^U_M(t)\partial_m\phi^U(t,m,\xi,\ub)+
            q_{M,m_\infty}^U(t)\partial_\xi\phi^U(t,m,\xi,\ub)
        \right) \\
        &-
        \eta\Sigmab_{SU}
        \nabla_u\phi^U(t,m,\xi,\ub)
    \Bigg].
\end{split}
\end{equation}
The first four terms are identical to those for the $F$-investor. The remaining terms are intertemporal hedging demands. The term
\begin{equation}
    -\eta\alphab q^U_M(t)
    \partial_m\phi^U(t,m,\xi,\ub)
\end{equation}
hedges uncertainty about the current transition factor \(M_t\). It appears
because return innovations contain information about \(M_t\), and therefore
affect the investor's estimate of the transition state.
The term
\begin{equation}
    -\eta\alphab q_{M,m_\infty}^U(t)
    \partial_\xi\phi^U(t,m,\xi,\ub)
\end{equation}
captures the fact that learning about the current transition factor also affects
beliefs about the long-run transition mean \(m_\infty\), through the posterior
covariance \(q_{M,m_\infty}^U(t)\).
Finally,
\begin{equation}
    -\eta\Sigmab_{SU}\nabla_u\phi^U(t,m,\xi,\ub)
\end{equation}
is the intertemporal hedging demand induced by the instantaneous correlation
between return shocks and carbon-intensity shocks. This term is already present
under full information, because carbon intensities are observed but stochastic.

\end{remark}

\subsection{Optimal strategy for the \(Z\)-investor}

The \(Z\)-investor observes the price process continuously and receives noisy carbon-intensity disclosures at deterministic dates
\begin{equation}
    0=\tau_0<\tau_1<\cdots<\tau_N<T.
\end{equation}
Portfolio decisions and the associated carbon objective are based on the investor's filtered estimate. Perceived cumulative financed emissions are then defined by:
\begin{equation}
    dY_t^Z
    =
    X_t \left[\pi_t^\top \widehat \Ub_t^Z + (\mathbf{1}-\pi_t\mathbf{1})u_{rf}(t)\right]dt,
    \qquad
    \widehat \Ub_t^Z
    :=
    \mathbb E[\Ub_t\mid\mathcal F_t^Z].
\end{equation}
This specification reflects a portfolio manager whose decarbonization mandate is implemented using the carbon data and estimates available at the time portfolio decisions are made. The investor therefore penalizes its assessed financed emissions rather than an unobservable latent carbon footprint. \\

The \(Z\)-investor therefore penalizes perceived cumulative financed emissions and solves
\begin{equation}
    V^Z(x_0^Z)
    =
    \sup_{\pi\in\mathcal A^Z}
    \mathbb E_0
    \left[
        -\exp\left(
            -\eta(X_T-\lambda Y_T^Z)
        \right)
    \right].
\end{equation}

The filtered state is
\begin{equation}
    \widehat L_t^Z =
    \begin{pmatrix}
        \widehat M_t^Z\\
        \widehat{m}_{\infty,t}^Z\\
        \widehat \Ub_t^Z
    \end{pmatrix}
    =
    \mathbb E[L_t^Z\mid\mathcal F_t^Z]
    \in\mathbb R^{d+2},
\end{equation}
and satisfies, between the disclosure dates:
\begin{equation}
    d\widehat L_t^Z = (A_Z\widehat L_t^Z+c_Z)dt +  \Gamma_t^Z\Sigmab_S^{-1}d\widehat R_t^Z,
\end{equation}
where
\begin{equation}
    d\widehat R_t^Z = dR_t^Z-H_Z\widehat L_t^Zdt,
\end{equation}
and
\begin{equation}
    \Gamma_t^Z = Q_t^ZH_Z^\top+\Sigma_{SL}\in\mathbb R^{(d+2)\times d}.
\end{equation}
Moreover,
\begin{equation}
    d\langle \widehat R^Z\rangle_t = \Sigmab_Sdt, \qquad d\langle \widehat L^Z,\widehat R^Z\rangle_t =\Gamma_t^Zdt,
\end{equation}
and
\begin{equation}
    d\langle \widehat L^Z\rangle_t = \Lambda_t^Zdt, \qquad
    \Lambda_t^Z = \Gamma_t^Z\Sigmab_S^{-1}(\Gamma_t^Z)^\top.
\end{equation}

The observable wealth and filtered-emission dynamics are therefore
\begin{equation}
    dX_t
    =
    X_t
    \left[
        r+\pi_t^\top \theta^Z(\widehat L_t^Z)
    \right]dt
    +
    X_t\pi_t^\top d\widehat R_t^Z,
\end{equation}
and
\begin{equation}
    dY_t^Z
    =
    X_t\left[u_{rf}(t) + \pi_t^\top \left( E_u\widehat L_t^Z - u_{rf}(t) \mathbf 1  \right)\right]dt.
\end{equation}

For \(n=0,\ldots,N\), set \(\tau_{N+1}:=T\). Denote by
\[
    v_n^Z(t,x,y,\ell)
\]
the value function on the interval \(t\in[\tau_n,\tau_{n+1})\), with the
convention that \(v_n^Z(\tau_n,\cdot)\) is the value immediately after the
disclosure at \(\tau_n\), for \(n\geq1\).

\begin{remark}
The process $Y_t^Z$ should not be confused with the posterior estimate $\mathbb{E}[Y_t|\mathcal{F}_t^Z]$ of realized cumulative emissions. Our formulation corresponds to an investor whose portfolio mandate is defined in terms of available carbon estimates.
\end{remark}

\begin{proposition}[Optimal control and value function for the \(Z\)-investor between disclosure dates]
\label{prop:optimal-Z}
For the \(Z\)-investor, the generic state-space variable is
\begin{equation}
        \ell=     \begin{pmatrix}
        m\\
        \xi\\
        \ub
    \end{pmatrix}\in\mathbb R^{d+2},
\end{equation}
where \(m\) represents the filtered estimate of the transition factor \(M_t\), \(\xi\) represents the filtered estimate of the long-run transition mean \(m_\infty\), and \(\ub\) is the filtered estimate of the carbon- intensity vector. 
For \(t\in[\tau_n,\tau_{n+1})\), the value function of the \(Z\)-investor is
given by
\begin{equation}
    v_n^Z(t,x,y,\ell)  =  -\exp \left\{ -\eta \left( \gamma(t)x-\lambda y+\phi_n^Z(t,\ell)\right) \right\},
\end{equation}
where \(\phi_n^Z\) admits the quadratic representation
\begin{equation}
    \phi_n^Z(t,\ell) = p_{0,n}^Z(t) + p_{1,n}^Z(t)^\top\ell + \frac12\ell^\top P_{n}^Z(t)\ell,
\end{equation}
where
\begin{equation}
    p_{0,n}^Z(t)\in\mathbb R,
    \qquad
    p_{1,n}^Z(t)\in\mathbb R^{d+2},
    \qquad
    P_{n}^Z(t)\in\mathbb R^{(d+2)\times(d+2)}.
\end{equation}

The terminal condition is imposed on the last interval \((\tau_N,T]\)
\begin{equation}
    p_{0,N}^Z(T)=0,\qquad
    p_{1,N}^Z(T)=0,\qquad
    P_N^Z(T)=0.
\end{equation}
For \(n<N\), the terminal values at \(\tau_{n+1}^{-}\) are obtained from the disclosure-date jump
conditions applied at \(\tau_{n+1}\).

Define
\begin{equation}
    b_{0,n}^Z(t)
    :=
    \theta_0-\eta(\Gamma_t^Z)^\top p_{1,n}^Z(t)
    +\frac{\lambda}{\gamma(t)}u_{rf}(t)\mathbf{1}
    \qquad
    B_n^Z(t)
    :=
    \Theta-\frac{\lambda}{\gamma(t)}E_u 
    -\eta(\Gamma_t^Z)^\top P_n^Z(t).
\end{equation}
Then the coefficient functions solve, on each interval
\((\tau_n,\tau_{n+1})\), the backward system
\begin{equation}
\begin{aligned}
    0={}& \dot p_{0,n}^Z(t)  +  p_{1,n}^Z(t)^\top c_Z + \frac12\operatorname{Tr} \left(
        \Lambda_t^Z P_{n}^Z(t)\right) -\frac{\eta}{2} p_{1,n}^Z(t)^\top \Lambda_t^Z
    p_{1,n}^Z(t) \\
    &+\frac{1}{2\eta}b_{0,n}^Z(t)^\top \Sigmab_S^{-1} b_{0,n}^Z(t) \\
    0={}& \dot p_{1,n}^Z(t) + A_Z^\top p_{1,n}^Z(t) + P_{n}^Z(t)c_Z  -  \eta P_{n}^Z(t)
    \Lambda_t^Z p_{1,n}^Z(t) \\
    &+ \frac{1}{\eta} B_n^Z(t)^\top \Sigmab_S^{-1} b_{0,n}^Z(t) \\
    0={}& \dot P_{n}^Z(t) + P_{n}^Z(t)A_Z + A_Z^\top P_{n}^Z(t)- \eta P_{n}^Z(t)
    \Lambda_t^Z P_{n}^Z(t) \\
    &+ \frac{1}{\eta} B_n^Z(t)^\top \Sigmab_S^{-1} B_n^Z(t)
\end{aligned}
\end{equation}

The optimal control is
\begin{equation}
\label{eq:optimal_pi_Z_general}
    \pi_t^{Z,*} = \frac{1}{\eta\gamma(t)X_t} \Sigmab_S^{-1} \left[ \theta^Z(\widehat L_t^Z)
        -  \frac{\lambda}{\gamma(t)}(E_u\widehat L_t^Z- u_{rf}(t)\mathbf{1})
        -\eta(\Gamma_t^Z)^\top
        \nabla_\ell\phi_n^Z(t,\widehat L_t^Z)
    \right].
\end{equation}
In quadratic form, the optimal strategy is
\begin{equation}
\label{eq:optimal_pi_Z_general}
    \pi_t^{Z,*}
    =
    \frac{1}{\eta\gamma(t)X_t}
    \Sigmab_S^{-1}
    \left[
        b_{0,n}^Z(t)
        +
        B_n^Z(t)\widehat L_t^Z
    \right],
    \qquad
    t\in[\tau_n,\tau_{n+1}).
\end{equation}

\end{proposition}

\begin{proof}
See Appendix~\ref{app:proof-control-Z}.
\end{proof}

\begin{proposition}[Jump conditions at disclosure dates]
\label{prop:optimal-Z-jumps}

At a disclosure date \(\tau_n\), the filtered state jumps because the investor
incorporates the new noisy disclosure. By contrast, the wealth process \(X_t\)
and the filtered cumulative-emission process \(Y_t^Z\) are continuous at
\(\tau_n\).

The disclosure equation is
\begin{equation}
    Z_n=D_ZL_{\tau_n}^Z+\epsilon_n,
    \qquad
    D_Z
    =
    \begin{pmatrix}
        0_{d\times1} & 0_{d\times1} & I_d
    \end{pmatrix}.
\end{equation}
The filtered state update is
\begin{equation}
    \widehat L_{\tau_n}^Z  = \widehat L_{\tau_n^-}^Z + K_n^Z( Z_n-D_Z\widehat L_{\tau_n^-}^Z) =\widehat L_{\tau_n^-}^Z + K_n^Z\nu_n^Z,
\end{equation}
where conditionally on \(\mathcal F_{\tau_n^-}^Z\),
\begin{equation}
    \nu_n^Z  \sim \mathcal N(0,S_n^Z), \qquad
    S_n^Z = D_ZQ_{\tau_n^-}^ZD_Z^\top+\Sigmab_n^\epsilon.
\end{equation}

The dynamic programming principle gives
\begin{equation}
\label{eq:DPP_jump_Z}
    v_{n-1}^Z(\tau_n^-,x,y,\ell)  = \mathbb E \left[ v_n^Z \left( \tau_n,  x,
            y,\ell+ K_n^Z\nu_n^Z \right) \mid \mathcal F_{\tau_n^-}^Z \right].
\end{equation}

Let
\begin{equation}
    \begin{aligned}
    p_i^+ &= p_{i,n}^Z(\tau_n), \qquad p_i^- = p_{i,n-1}^Z(\tau_n^-), \qquad i=0,1. \\
    P^+ &= P_{n}^Z(\tau_n), \qquad P^- = P_{n-1}^Z(\tau_n^-),
    \end{aligned}
\end{equation}
and
\begin{equation}
    A_n^Z  =  (K_n^Z)^\top P^+K_n^Z
    \in\mathbb R^{d\times d},
\end{equation}

The solution is constructed backward. First solve the ODE system on \((\tau_N,T)\) with
terminal condition
\begin{equation}
p_{0,N}^Z(T)=p_{1,N}^Z(T)=P_{N}^Z(T)=0.
\end{equation}
Then apply the disclosure jump condition at \(\tau_N\), which gives terminal values for the
ODE system on \((\tau_{N-1},\tau_N)\). Repeating this procedure recursively yields the value
function and the optimal control on all intervals. More explicitly, the disclosure-date jump conditions for the quadratic coefficients are
\begin{equation}
\begin{split}
\label{eq:jump_K2_Z}
    P^- &= P^+ - \eta P^+K_n^Z ((S_n^Z)^{-1} + \eta A_n^Z)^{-1} ( K_n^Z)^\top  P^+ \\
    p^-_1  &=  p^+_1 -  \eta P^+ K_n^Z ((S_n^Z)^{-1} + \eta A_n^Z)^{-1} ( K_n^Z)^\top p^+_1 \\
    p^-_0 &=  p^+_0 + \frac{1}{2\eta} \log\det \left(  I_d+\eta S_n^ZA_n^Z\right) \\
    &- \frac{\eta}{2}(p^+_1)^\top  K_n^Z ((S_n^Z)^{-1} + \eta A_n^Z)^{-1} ( K_n^Z)^\top  p^+_1
\end{split}
\end{equation}

\end{proposition}

\begin{proof}
See Appendix~\ref{app:proof-control-Z-jumps}.
\end{proof}

\begin{remark}[Interpretation in the original variables for the \(Z\)-investor]

The conditional covariance matrix of the hidden state $L^Z_t$ can be written blockwise as
\begin{equation}
\label{eq:block_covariance_Z_original_variables}
    Q_t^Z
    =
    \begin{pmatrix}
        q_M^Z(t) & q_{M m_\infty}^Z(t) & q_{MU}^Z(t)^\top\\
        q_{M m_\infty}^Z(t) & q_{m_\infty}^Z(t) & q_{m_\infty U}^Z(t)^\top\\
        q_{MU}^Z(t) & q_{m_\infty U}^Z(t) & q_U^Z(t)
    \end{pmatrix},
\end{equation}
Thus:
\begin{equation}
    \Gamma_t^Z = Q_t^ZH_Z^\top+\Sigma_{SL} =\begin{pmatrix}
        q_M^Z(t)\alphab^\top
        +
        q_{MU}^Z(t)^\top\betab^\top\\[0.15cm]
        q_{Mm_\infty}^Z(t)\alphab^\top
        +
        q_{m_\infty U}^Z(t)^\top\betab^\top\\[0.15cm]
        q_{MU}^Z(t)\alphab^\top
        +
        q_U^Z(t)\betab^\top
        +
        \Sigmab_{SU}^\top
    \end{pmatrix}.
\end{equation}

Writing
\begin{equation}
    \nabla_\ell\phi_n^Z
    =
    \begin{pmatrix}
        \partial_m\phi_n^Z\\
        \partial_{m_\infty}\phi_n^Z\\
        \nabla_u\phi_n^Z
    \end{pmatrix},
\end{equation}
the optimal strategy can be written in the original variables as
\begin{equation}
\label{eq:pi_Z_original_variables}
\begin{split}
    \pi_t^{Z,*}  = \frac{1}{\eta\gamma(t)X_t}\Sigmab_S^{-1}
    \Bigg[
        &\mub-r\mathbf 1 + \alphab\widehat M_t^Z +\betab \left( \widehat\Ub_t^Z-\bar{\Ub}_0
        \right) -\frac{\lambda}{\gamma(t)}(\widehat\Ub_t^Z- u_{rf}(t)\mathbf{1})\\
        &-\eta \Big( \alphab q_M^Z(t) + \betab q_{MU}^Z(t)\Big)
        \partial_m\phi_n^Z \left( t,\widehat M_t^Z,\widehat{m}_{\infty,t}^Z,\widehat\Ub_t^Z
        \right)\\
        &-\eta
        \Big(
            \alphab q_{M m_\infty}^Z(t) + \betab q_{m_\infty U}^Z(t) \Big)
        \partial_{m_\infty}\phi_n^Z \left(
            t,\widehat M_t^Z,\widehat{m}_{\infty,t}^Z,\widehat\Ub_t^Z
        \right)\\
        &-\eta \Big(\alphab q_{MU}^Z(t)^\top + \betab q_U^Z(t) +\Sigmab_{SU} \Big)
        \nabla_u\phi_n^Z \left(t,\widehat M_t^Z,\widehat{m}_{\infty,t}^Z,\widehat\Ub_t^Z
        \right)
    \Bigg].
\end{split}
\end{equation}

The first line in \eqref{eq:pi_Z_original_variables} is the myopic component.
It contains the filtered financial excess return and the direct carbon-emission
penalty. The remaining terms are intertemporal hedging demands induced by
partial observation and by the correlation between financial and carbon-intensity
shocks.

The term involving \(q_M^Z(t)\) hedges uncertainty about the current transition
factor \(M_t\). The term involving \(q_{Mm_\infty}^Z(t)\) captures the fact that
learning about the long-run transition mean \(m_\infty\) also changes beliefs
about the current transition state. The terms involving \(q_{MU}^Z(t)\) and
\(q_{m_\infty U}^Z(t)\) capture the covariance between beliefs about transition
conditions and beliefs about firms' carbon intensities. The term involving
\(q_U^Z(t)\) reflects uncertainty about the unobserved carbon intensities
themselves. Finally, the term involving \(\Sigmab_{SU}\) hedges the instantaneous
correlation between return shocks and carbon-intensity shocks.

\end{remark}

\section{Numerical application}

In this section, we use a calibrated numerical application to illustrate how incomplete climate information affects dynamic portfolio decarbonization. We compare the different information regimes along three dimensions: portfolio composition, welfare, and the target satisfaction of the cumulative financed-emissions objective. We first characterize the benchmark allocation and calibrate the carbon penalty $\lambda$ to a common emissions-reduction target. The calibration procedure is presented in Appendix \ref{app:lambda_calibration}. We then study how disclosure frequency and misspecified beliefs about the current and long-run transition states affect optimal portfolio decisions. Particular attention is paid to the asymmetry between welfare and environmental outcomes: belief errors of similar magnitude but opposite sign can generate comparable welfare losses while having different effects on cumulative financed emissions. \\

To compare the different strategies, we consider three complementary outcomes: the optimal portfolio allocation, the welfare consequences of the strategy, and the target satisfaction of the decarbonization objective. Welfare is measured using the carbon-adjusted certainty equivalent, while target satisfaction is measured by the probability that cumulative financed emissions exceed the intended emissions budget. For a given strategy $\pi$, the carbon-adjusted certainty equivalent $CE_{X-\lambda Y}$ is the sure amount of carbon-penalized terminal wealth that makes the investor indifferent to the risky payoff $X_T^\pi-\lambda Y_T^\pi$. It is defined by
\begin{equation}
    g\!\left(CE_{X-\lambda Y}(\pi)\right)
    =
    \mathbb{E}\left[
        g\!\left(X_T^\pi-\lambda Y_T^\pi\right)
    \right].
\end{equation}
Under CARA utility,
\begin{equation}
    CE_{X-\lambda Y}(\pi)
    =
    -\frac{1}{\eta}
    \log
    \left(
        \mathbb{E}\left[
            e^{-\eta(X_T^\pi-\lambda Y_T^\pi)}
        \right]
    \right).
\end{equation}

The quantity $CE_{X-\lambda Y}$ therefore provides a monetary valuation of the investor's complete objective, incorporating both terminal wealth and the wealth-equivalent valuation assigned to cumulative financed emissions through $\lambda$. Holding $\lambda$ fixed across information regimes allows us to compare investors under the same decarbonization mandate and isolates the effect of differences in information. For the $Z$-investor, the optimization problem is defined in terms of perceived cumulative financed emissions $Y_T^Z$, consistently with the information available to the investor. Ex post target satisfaction can, however, also be evaluated using realized cumulative financed emissions
computed from the latent carbon-intensity process $U_t$. In the baseline calibration, this distinction is quantitatively small. Figure \ref{fig:actual_vs_perceived_emissions} in Appendix~\ref{app:supplementary figures} shows that perceived and realized cumulative financed emissions remain very close over the investment horizon, both in terms of their mean trajectories and their Monte Carlo dispersion. \\

Welfare, however, does not directly measure whether the decarbonization objective is achieved. We therefore complement the certainty-equivalent comparison with the probability of exceeding this budget,
\begin{equation}
    p^{\mathrm{exc}}(\pi)
    :=
    \mathbb{P}\left(
        Y_T^\pi > \bar Y
    \right).
\end{equation}
This target satisfaction measure allows us to distinguish the welfare consequences of imperfect information from its consequences for the delivery of the decarbonization objective. Two information structures may therefore generate similar certainty-equivalent values while implying substantially different probabilities of meeting the emissions target. \\

In the numerical analysis, the penalty is calibrated using the full-information investor. For a target reduction, we choose $\lambda_F^\star$ such that the probability that the full-information investor exceeds the associated emissions budget
$\bar Y$ is equal to the tolerance level $\epsilon$:
\begin{equation}
    \mathbb P
    \left(
        Y_T^{F,\star}(\lambda_F^\star)>\bar Y
    \right)
    =\epsilon.
\end{equation}
The parameter $\epsilon$ determines the quantile of the full-information emissions distribution used to calibrate the common carbon valuation. In the baseline analysis, we set $\epsilon=0.5$. This choice should be interpreted as a median-based calibration convention rather than as a normative tolerance for missing the emissions target: $\lambda_F^\star$ is chosen so that the target budget coincides approximately with the median of cumulative financed emissions under the fully informed optimal strategy. This provides a stable reference point around which the effects of incomplete information can be measured in either direction. The resulting penalty is then held fixed across information regimes, so that
\begin{equation}
    \lambda_F=\lambda_U=\lambda_Z=\lambda_F^\star(R).
\end{equation}

This specification represents a common decarbonization mandate: the valuation assigned to financed emissions is fixed independently of the investor's information structure. The resulting differences in portfolio composition and cumulative emissions therefore reflect differences in information and beliefs rather than differences in the strength of the environmental preference. We use a target reduction of 50\% as the upper bound on our baseline decarbonization target for two reasons. It is approximately consistent with the reduction in cumulative emissions required under the Network for Greening the Financial System (NGFS) Below~$2^\circ$C scenario over the corresponding horizon. Moreover, beyond this level, short positions begin to contribute more materially to the measured reduction in financed emissions, as shown in Figure~\ref{fig:negative_emissions} in Appendix~\ref{app:supplementary figures}. Restricting the baseline analysis to reductions of up to 50\% therefore limits the quantitative importance of this modelling feature. Holding $\lambda$ fixed across investors then allows the climate performance of the different information structures to vary under a common valuation of financed emissions.

\subsection{Illustrative example}

In the subsequent illustrative examples, we examine three representative asset profiles:
\begin{itemize}
    \item \textbf{Green asset}: characterized by a low initial carbon intensity and an ambitious decarbonization target that is not affected by the transition factor process;
    \item \textbf{Decarbonizing asset}: characterized by a high initial carbon intensity, but a low target under a positive transition factor process;
    \item \textbf{Brown asset}: characterized by a high initial carbon intensity and a medium target under a positive transition factor process.
\end{itemize}

Table \ref{tab:numerical_parameters} summarizes the parameter values used in the numerical application. In the baseline calibration, we assume a neutral long-run transition state, with $m_\infty = 0$, and annual disclosures of firm-level carbon intensities. The model is simulated at a monthly frequency over a 25-year investment horizon. \\

\begin{table}[!htbp]
\centering
\small
\setlength{\tabcolsep}{4pt}
\renewcommand{\arraystretch}{1.04}
\begin{tabularx}{\linewidth}{@{}l>{\raggedright\arraybackslash}Xrrr@{}}
\toprule
Parameter & Description & Brown & Decarbonizing & Green \\
\midrule
\multicolumn{5}{@{}l}{\textit{A. Common economic and transition-state parameters}} \\
\addlinespace[2pt]
$T$ & Investment horizon (years) & \multicolumn{3}{c}{25} \\
$r$ & Risk-free rate & \multicolumn{3}{c}{0.02} \\
$\eta$ & Absolute risk aversion & \multicolumn{3}{c}{2} \\
$X_0$ & Initial wealth & \multicolumn{3}{c}{1} \\
$u_{\mathrm{rf},0}$ & Initial risk-free carbon intensity & \multicolumn{3}{c}{25} \\
$\kappa_{\mathrm{rf}}$ & Risk-free intensity decay rate & \multicolumn{3}{c}{0.1} \\
$M_0$ & Initial transition state & \multicolumn{3}{c}{0} \\
$\widehat M_0$ & Initial estimate of the transition state & \multicolumn{3}{c}{0} \\
$m_\infty$ & True long-run transition mean & \multicolumn{3}{c}{0} \\
$\widehat m_{\infty,0}$ & Initial estimate of the long-run mean & \multicolumn{3}{c}{0} \\
$\kappa_M$ & Transition-state mean reversion & \multicolumn{3}{c}{0.15} \\
$\sigma_M$ & Transition-state volatility & \multicolumn{3}{c}{0.02} \\
\addlinespace[5pt]
\multicolumn{5}{@{}l}{\textit{B. Asset-specific parameters}} \\
\addlinespace[2pt]
$\mu_i$ & Expected return (calibrated) & 0.0584 & 0.0524 & 0.0411 \\
$\alpha_i$ & Return exposure to the transition state & $-0.05$ & 0.04 & 0.02 \\
$\beta_{ii}$ & Return loading on carbon intensity & $-1.5\times10^{-4}$ & $-3\times10^{-4}$ & $-7.5\times10^{-5}$ \\
$\sigma_{ii}$ & Return-shock volatility & 0.18 & 0.17 & 0.16 \\
$(\sigma_B)_{ii}$ & Carbon-intensity shock volatility & 12 & 7 & 1 \\
$\rho_{ii}$ & Return--intensity shock correlation & 0.15 & $-0.15$ & $-0.075$ \\
$S_{i,0}$ & Initial asset price & 1 & 1 & 1 \\
$U_{i,0}$ & Initial true carbon intensity & 100 & 75 & 20 \\
$\overline U_{i,0}$ & Initial carbon-intensity estimate & 100 & 75 & 20 \\
$U_{\mathrm{ref},i}$ & Reference carbon intensity & 100 & 75 & 20 \\
$\overline U_{\infty,i}$ & Baseline long-run carbon intensity & 70 & 35 & 10 \\
$\delta_i$ & Long-run intensity sensitivity to $m_\infty$ & $-60$ & $-300$ & 0 \\
$(K_U)_{ii}$ & Carbon-intensity mean reversion & 0.02 & 0.075 & 0.02 \\
\addlinespace[5pt]
\multicolumn{5}{@{}l}{\textit{C. Disclosure and filtering parameters}} \\
\addlinespace[2pt]
$q_{M,0}$ & Initial transition-state variance & \multicolumn{3}{c}{0.01} \\
$q_{m_\infty,0}$ & Initial long-run-mean variance & \multicolumn{3}{c}{0.01} \\
$q_{M,m_\infty,0}$ & Initial state--long-run-mean covariance & \multicolumn{3}{c}{0} \\
$(Q_{U,0})_{ii}$ & Initial carbon-intensity variance & 1 & 1 & 1 \\
$\Delta\tau$ & Disclosure interval (years) & \multicolumn{3}{c}{1} \\
$N_D$ & Disclosures before maturity & \multicolumn{3}{c}{24} \\
$(\Sigma_\varepsilon)_{ii}$ & Disclosure-noise variance & 20 & 20 & 20 \\
\bottomrule
\end{tabularx}
\par\vspace{4pt}
\caption{Baseline numerical parameters}
\par\vspace{4pt}
\begin{minipage}{\linewidth}
\footnotesize
\textit{Notes:} Time is measured in years; rates and volatilities use this time scale. Wealth, prices, and carbon intensities follow the model's normalization. Matrix rows show diagonal entries; omitted off-diagonal
entries are zero. The return vector $\mu$ is calibrated to initial unpenalized risky-asset weights $(0.36,\,0.34,\,0.25)$ and rounded to four decimal places. The model is simulated at a monthly frequency over the $25$-year horizon using $N_{\mathrm{MC}}=\text{5000}$ Monte Carlo paths. The Gaussian state processes are simulated using an Euler-Maruyama scheme, while the deterministic filtering Riccati equations and the backward coefficient equations are solved numerically on the same time grid. For comparisons across information regimes, disclosure frequencies, and prior biases, we use common underlying simulated paths so that differences across specifications reflect changes in the investor's information and strategy rather than simulation noise. Disclosures occur at $t=1,\ldots,24$. The carbon penalty is calibrated for the full-information investor for each emissions-reduction target and is held fixed across information regimes.
\end{minipage}
\label{tab:numerical_parameters}
\end{table}

\begin{figure}[H]
    \centering
    \includegraphics[width=0.9\linewidth]{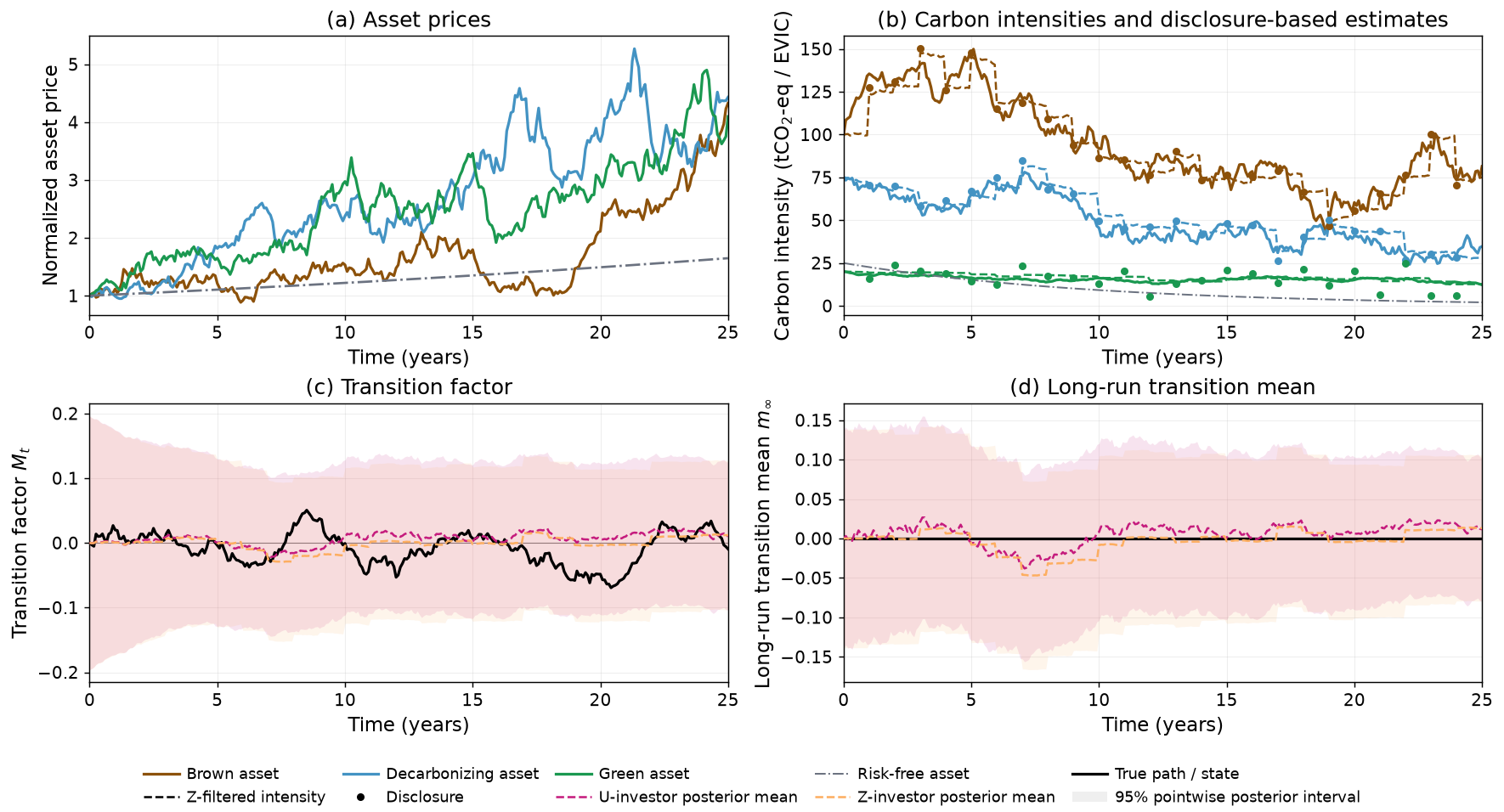}
    \caption{Simulated market paths and filtered estimates of the transition state (one representative path).}
    \par\vspace{4pt}
    \begin{minipage}{\linewidth}
    \footnotesize
    \textit{Notes:} Panel (a): risky-asset and risk-free prices. Panel (b): true carbon-intensity paths (solid), Z-investor filtered estimates (dashed), and discrete noisy disclosures (dots) for the brown, decarbonizing, and green assets. Panel (c): true transition factor $M_t$ against its $U-$ and $Z-$investor filtered estimates $\hat{M}_t^U,\, \hat{M}_t^Z$. Panel (d): true long-run transition mean $m_\infty$ against its filtered estimates $\hat{m}_{\infty,t}^U,\,\hat{m}_{\infty,t}^Z$. Both investors update their estimates continuously from observed returns; the \(Z\)-investor additionally updates its estimates discretely at carbon-intensity disclosure dates.
    \end{minipage}
    \label{fig:single_path}
\end{figure}

Figure \ref{fig:single_path} illustrates the temporal evolution of the principal state variables along a representative sample path, with the aim of providing an intuitive visualization of the model’s core mechanisms. In particular, the filtered estimates of the transition factor exhibit only minor differences under continuous versus discrete observation of carbon intensity; however, disclosure at predetermined times may induce discontinuities (i.e., jump responses) in the corresponding trajectories. Second, the figure highlights the role of uncertainty and the associated filtering procedure. Over the period from year 5 to year 10, both the filtered transition factor and the long-run transition mean remain below their corresponding true values. This discrepancy can be attributed to a sequence of disclosures from decarbonising and brown assets indicating limited progress in decarbonisation. Within the model interpretation, these observations are consistent with a “high-carbon” state of the world, preceding a subsequent phase of gradual adjustment. A decomposition of the effect of the different terms in the optimal portfolio solution is provided in Figure \ref{fig:decomposition} in Appendix \ref{app:supplementary figures}. \\

\begin{figure}[H]
    \centering
    \includegraphics[width=0.9\linewidth]{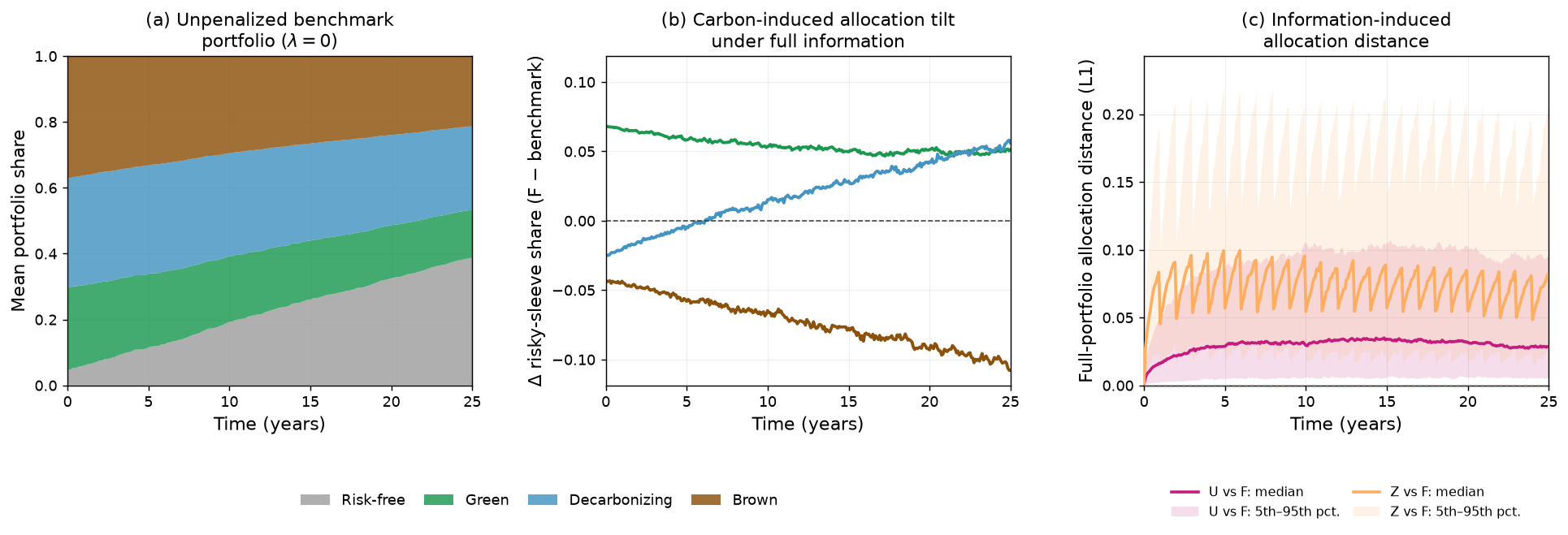}
    \caption{Portfolio allocation effects of the carbon objective and incomplete climate information.}
    \par\vspace{4pt}
    \begin{minipage}{\linewidth}
    \footnotesize
    \textit{Notes:}  Panel (a) reports the mean composition of the unpenalized benchmark portfolio, including the risk-free asset and the three risky assets. Panel (b) shows, for each risky asset, the change in the risky-sleeve share induced by the carbon objective under full information, measured relative to the unpenalized benchmark. Positive values therefore indicate an overweight relative to the benchmark and negative values an underweight. Panel (c) measures the additional portfolio distortion generated by incomplete information using the $L^1$ distance $\|w_t^{H}-w_t^{F}\|_1$, for $H\in\{U,Z\}$.
    \end{minipage}
    \label{fig:comparison}
\end{figure}

We now analyse the impact of introducing a carbon penalty on the optimal portfolio allocation, as illustrated in Figure \ref{fig:comparison}. Specifically, we compare the unpenalized benchmark ($\lambda = 0$) with the $F$-investor's penalized allocation, and we further examine how partial observation affects the resulting strategies. Relative to the unpenalized benchmark, the introduction of a penalization term initially reduces the portfolio weights allocated to brown and decarbonizing assets. However, because the decarbonizing asset reduces its carbon intensity, the allocation to the decarbonizing asset increases over time, eventually exceeding the weight of the green asset. This initial divestment from the decarbonizing asset is consistent with the model’s mechanism: the penalty induces an early reduction in exposure, which is gradually attenuated as the carbon intensity of the decarbonizing asset declines. Consequently, the proposed framework is well suited to passive strategies whose objective is to reduce the portfolio footprint. By contrast, if the objective is to actively induce real-economy change by investing in firms to accelerate reductions in their carbon intensity, an alternative modelling framework would be necessary. \\

Limited information can generate substantial deviations from the full-information portfolio allocation. Panel (c) of Figure \ref{fig:comparison} illustrates this by reporting, along the same simulated path, the differences in portfolio weights between the $U$- and $F$-investors and between the $Z$- and $F$-investors. As expected, deviations from the full-information allocation are larger when the investor has access to less information. In addition, discrete disclosures can generate sharp adjustments in portfolio weights as the investor updates beliefs and rebalances positions. Although transaction costs are not included in the model, such rebalancing episodes could have additional wealth consequences in practice.\\

\subsection{Impact of partial information on welfare and target satisfaction}

\begin{figure}[H]
    \centering
    \includegraphics[width=0.9\linewidth]{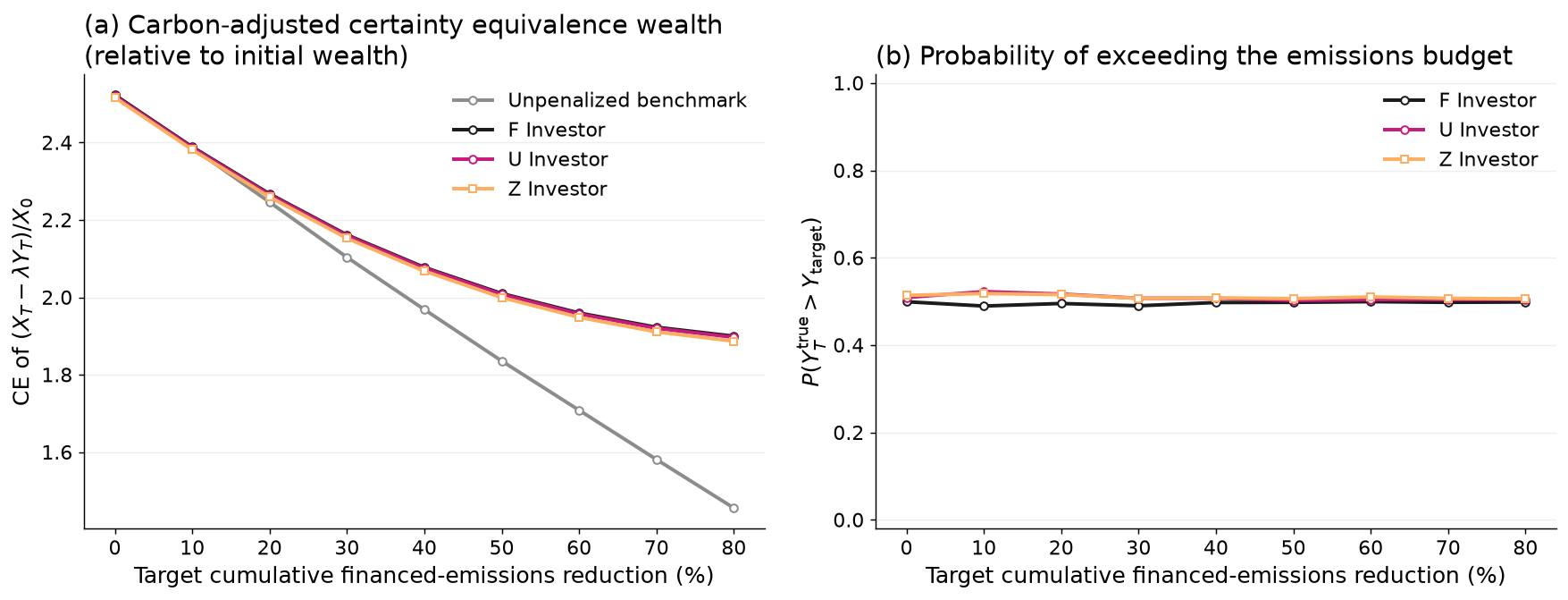}
    \caption{Welfare and probability of exceeding the emissions budget. }
    \par\vspace{4pt}
    \begin{minipage}{\linewidth}
    \footnotesize
    \textit{Notes:}  Panel (a) reports the certainty equivalent of terminal wealth net of the cumulative financed-emissions penalty,
    $CE_{X-\lambda Y}$, as a function of the targeted reduction in cumulative financed emissions. For each target, the carbon penalty $\lambda=\lambda_F^\star$ is calibrated for the full-information investor and the same penalty is then used to evaluate the $U$-investor, the $Z$-investor, and the unpenalized strategy. The gap between the unpenalized benchmark and the $F$-investor captures the value of incorporating the emissions objective directly into portfolio optimization, whereas the gaps between the $F$-, $U$-, and $Z$-investors isolate the welfare consequences of incomplete climate information. Panel (b) reports the probability that realized cumulative financed emissions exceed the corresponding budget $\bar Y$ under the same common penalty. By construction, the $F$-investor has an exceedance probability of approximately $50\%$ at each target; departures of the
    $U$- and $Z$-investors from this reference therefore measure the effect of incomplete information on target satisfaction. 
    \end{minipage}
    \label{fig:no_bias}
\end{figure}

As shown by the optimal policies of the $F$-investor \eqref{eq:optimal_pi_F_general}, the $U$-investor \eqref{eq:optimal_pi_U_general}, and the $Z$-investor \eqref{eq:optimal_pi_Z_general}, introducing a penalty on cumulative financed emissions alters the optimal portfolio relative to the unpenalized solution. We now examine the welfare implications of this penalization in Figure \ref{fig:no_bias}. Panel (a) reports the carbon-adjusted certainty equivalent. For the unpenalized strategy, this quantity is evaluated ex post: cumulative financed emissions are penalized using the same value of $\lambda^\star$ as that calibrated for the $F$-investor. \\

We observe that an investor with a preference for lower cumulative financed emissions is better off incorporating this preference directly into the optimization problem rather than applying the same penalty ex post to the unpenalized strategy. The carbon-adjusted certainty equivalent also decreases as the targeted emissions reduction becomes more stringent, as deeper decarbonization requires larger departures from the unconstrained allocation. The partially informed investor attains only slightly lower welfare than the fully informed investor. With correctly specified initial priors, partial information therefore has only a limited effect on welfare in this baseline calibration.

The effect of partial information is also limited in terms of target satisfaction. The probability of exceeding the emissions budget increases only slightly relative to the full-information benchmark. In the baseline calibration, this suggests that, when initial priors are correctly specified and disclosures are sufficiently informative, the investor is able to filter both carbon intensities and the transition state with enough precision to keep the resulting strategy close to the full-information allocation. \\

\subsection{Initial bias and portfolio composition}

In practice, investors may hold different, and potentially biased, beliefs about both the current state of the climate transition and its long-run direction. When information is limited, for example because disclosures are infrequent, such belief distortions can lead to persistent deviations from the full-information allocation. Our framework allows us to incorporate these effects and study their consequences for portfolio choice. We therefore analyse the $Z$-investor under alternative carbon-intensity disclosure frequencies and misspecified initial beliefs about the climate-transition process. The scalar bias $b$ jointly shifts the initial prior means of the current transition state and its long-run mean:
\begin{equation}
    \mathbb{E}^{Z}_{0}[M_0]
    =
    \widehat M_0 + b,
    \qquad
    \mathbb{E}^{Z}_{0}[m_\infty]
    =
    \widehat m_{\infty,0} + b,
\end{equation}
while the true data-generating process and the prior variances are held fixed. When $b>0$ (resp. $b<0$), we refer to the investor as having an optimistic (resp. pessimistic) bias about the transition. \\

\begin{figure}[H]
    \centering
    \includegraphics[width=0.9\linewidth]{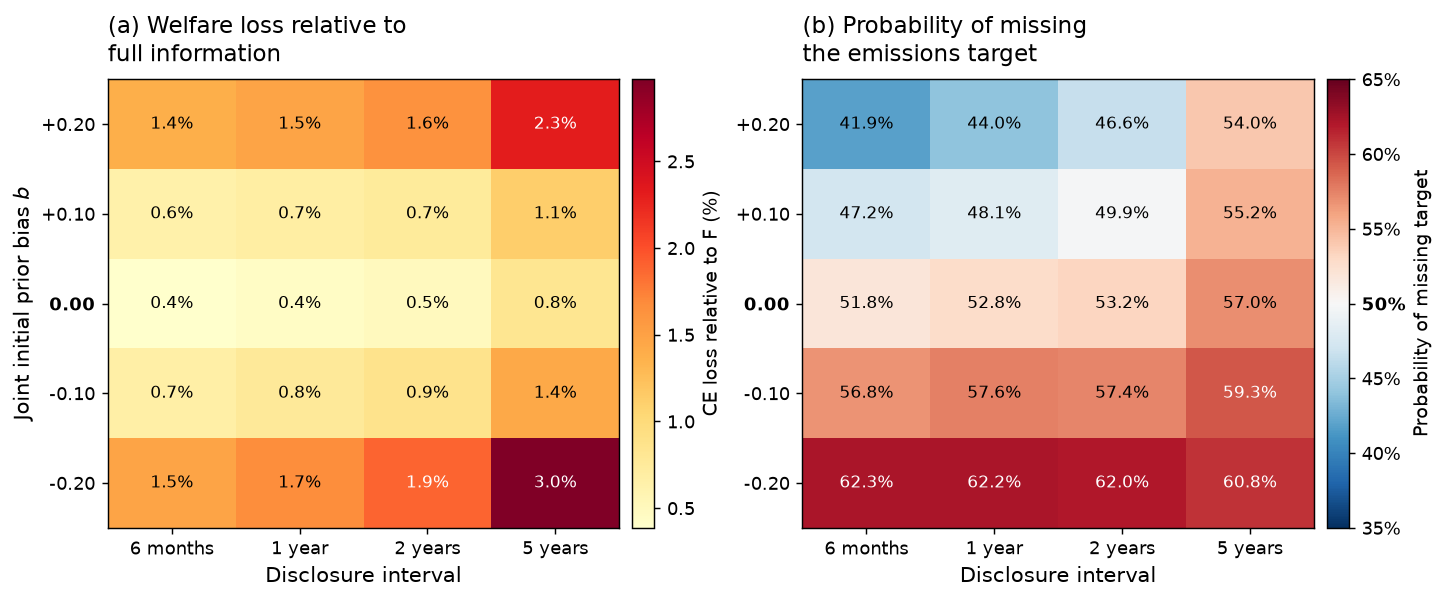}
    \caption{Effects of disclosure frequency and joint initial prior bias on welfare and target satisfaction.}
    \par\vspace{4pt}
    \begin{minipage}{\linewidth}
    \footnotesize
    \textit{Notes:} Panel (a) reports the welfare loss of the $Z$-investor relative to the full-information benchmark, measured from the carbon-adjusted certainty equivalent $CE_{X-\lambda Y}$. Panel (b) reports the probability that realized cumulative financed emissions exceed the target. The carbon penalty $\lambda=\lambda_F^\star$ is calibrated once for the $F$-investor to a $50\%$ cumulative-emissions-reduction target and is held fixed across all disclosure frequencies and prior-bias configurations. Thus, variation across cells reflects differences in information quality and initial beliefs rather than differences in the strength of the carbon incentive.
    \end{minipage}
    \label{fig:welfare_reliability}
\end{figure}

Figure \ref{fig:welfare_reliability} reports the effects of the initial belief bias and disclosure frequency on two outcomes: the carbon-adjusted certainty-equivalent loss relative to the $F$-investor and the probability of exceeding the cumulative emissions budget. The results are computed for a carbon penalty $\lambda^\star$, calibrated for a $50\%$ reduction target for cumulative emissions with $\epsilon = 0.5$. Under the baseline calibration, the welfare effect is approximately U-shaped in the bias. Large deviations from the true transition state in either direction reduce welfare, with the certainty-equivalent loss reaching approximately $3.0\%$ in the most adverse case. Welfare losses also increase as disclosures become less frequent, since erroneous beliefs and uncertainty about carbon intensities persist for longer.

The effect on target satisfaction is instead strongly related to the direction of the bias. An optimistic investor tends to tilt the portfolio toward lower-carbon assets and consequently faces a lower probability of exceeding the emissions budget, despite not attaining the full-information welfare level. Conversely, a pessimistic investor allocates relatively more to carbon-intensive assets and experiences both a welfare loss and a deterioration in target satisfaction, with the probability of exceeding the emissions budget reaching $62.3\%$ in the most adverse case. Thus, belief distortions of similar magnitude but opposite sign can generate comparable welfare losses while having opposite implications for the decarbonization objective.\\

\begin{figure}[H]
    \centering
    \includegraphics[width=0.9\linewidth]{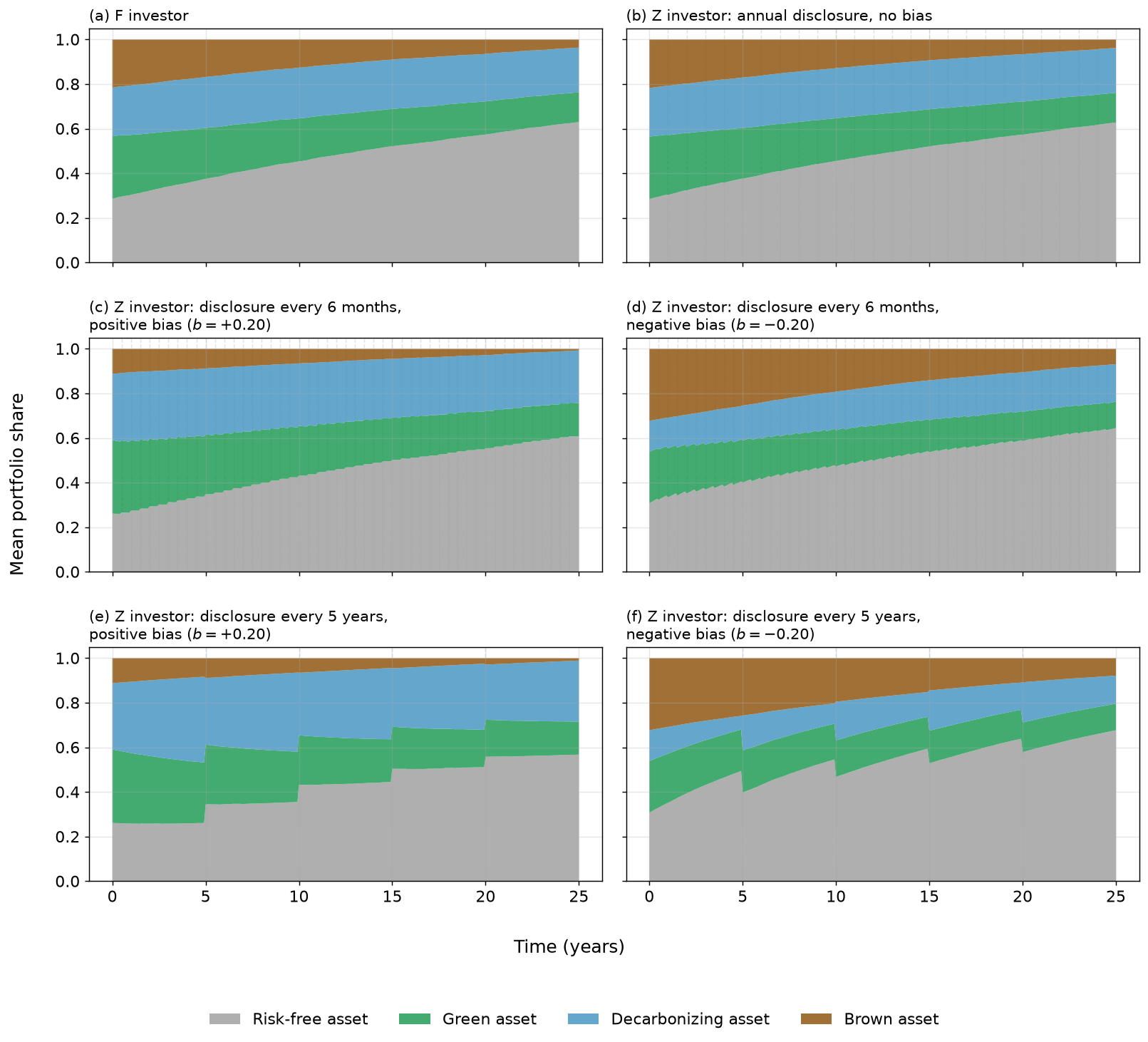}
    \caption{Portfolio composition under different disclosure frequencies and joint prior biases.}
    \par\vspace{4pt}
    \begin{minipage}{\linewidth}
    \footnotesize
    \textit{Notes:} Panel (a) provides the full-information benchmark, while the remaining panels illustrate how the $Z$-investor's allocation changes with the frequency of carbon-intensity disclosures and with the joint initial prior bias $b$ in $M_0$ and $m_\infty$. 
    Portfolio shares are averaged across simulated paths at each date and include the risk-free, green, decarbonizing, and brown assets. The figure illustrates the portfolio-composition channel underlying the welfare and target satisfaction effects reported in Figure~\ref{fig:welfare_reliability}.
    \end{minipage}
    \label{fig:allocation}
\end{figure}

Interestingly, more frequent disclosures do not uniformly improve target satisfaction when initial beliefs are biased. Instead, disclosure frequency interacts with the direction of the initial bias: more frequent disclosures reduce the probability of exceeding the emissions budget for optimistically biased investors, but increase it for pessimistically biased investors. To shed light on this asymmetry, Figure~\ref{fig:allocation} examines the corresponding portfolio-allocation dynamics. Panel~(a) reports the allocation of the fully informed $F$-investor, while Panel~(b) shows the baseline $Z$-investor with annual disclosures and unbiased initial beliefs. Panels~(c)-(f) illustrate how portfolio composition changes under positive and negative initial biases and across disclosure frequencies. A positive bias about the transition leads the $Z$-investor to allocate substantially less to the brown asset and more to the decarbonizing asset relative to the full-information benchmark. This pattern is visible under both semi-annual disclosures in Panel~(c) and five-year disclosures in Panel~(e). Conversely, a negative bias induces a markedly larger exposure to the brown asset and a lower exposure to the decarbonizing asset, particularly at the beginning of the investment horizon, as shown in Panels~(d) and~(f).

Disclosure frequency also affects the dynamics of portfolio adjustment. With semi-annual disclosures, new information arrives frequently, so revisions to estimated carbon intensities and the resulting portfolio reallocations are relatively small. This generates a comparatively smooth allocation path. When disclosures occur only every five years, carbon-intensity estimates are corrected less frequently, leading to much larger portfolio adjustments when new information is eventually released. This effect is particularly pronounced under a negative initial bias, as Panel~(f) shows sizeable reallocations at each disclosure date. Under pessimistic beliefs about the transition, the investor initially assigns a less favourable emissions trajectory to the decarbonizing asset and therefore holds substantially less of it. At each disclosure date, the observed emissions profile is better than anticipated, prompting the investor to revise its beliefs and increase its exposure to the decarbonizing asset. With sparse disclosures, these corrections occur only infrequently, so the investor maintains a lower exposure to the decarbonizing asset for longer periods, most visibly during the first five years. \\

This mechanism helps explain why more information need not translate into lower cumulative financed emissions. In the calibration, the decarbonizing asset combines relatively high initial intensity with a favourable subsequent decarbonization trajectory. A pessimistically biased investor therefore reduces its exposure to this asset, which mechanically lowers financed emissions in the short run. More frequent disclosures correct this pessimism more rapidly and induce the investor to increase its exposure to the decarbonizing asset. Although this improves the investor's information set and welfare, it can raise cumulative financed emissions over the investment horizon and hence increase the probability of exceeding the emissions budget. The results therefore highlight an important distinction between the welfare value of information and its environmental consequences: more accurate and timely information is beneficial for portfolio choice, but does not necessarily lead to lower realized financed emissions. \\

\section{Conclusion}

Global temperature outcomes are closely related to cumulative \(\mathrm{CO}_2\) emissions \citep{Rogelj_2019}, making cumulative emissions a meaningful criterion for studying portfolio decarbonization. In this paper, we study portfolio decarbonization through a cumulative financed-emissions objective. This objective is dynamic and forward-looking: portfolio decisions taken today affect the emissions accumulated over the entire investment horizon and therefore depend on expectations about firms' future carbon trajectories. In practice, however, climate information is incomplete. This paper therefore asks how incomplete climate information affects the composition, welfare, and target satisfaction of a dynamically decarbonized portfolio. Our main contribution is to develop a framework that combines cumulative financed emissions with partial information about both the climate-transition state and firm-level carbon intensities. \\

The numerical application highlights two distinct consequences of incomplete information. When initial beliefs are correctly specified, partial information has only a limited effect on welfare and target satisfaction in our baseline calibration. When beliefs are misspecified, errors of similar magnitude but opposite sign generate welfare losses of comparable order, but their consequences for the emissions objective are asymmetric. Optimistic and pessimistic beliefs induce different portfolio tilts and therefore affect the probability of exceeding the cumulative emissions budget in opposite directions. Disclosure frequency determines how quickly these belief errors are corrected and, consequently, how strongly they affect portfolio allocations. The results therefore show that the welfare and decarbonization consequences of incomplete climate information do not move together: inaccurate beliefs reduce the welfare of the investor in either direction, while their effect on cumulative financed emissions depends critically on the direction of the bias. \\

Several limitations remain. First, the numerical results are conditional on the chosen calibration. Although the parameter values are selected to be economically plausible, a fully empirical estimation and calibration exercise remains an important extension. Second, as in \cite{Hausladen2026NetZero}, the treatment of short positions is simplified. In the current specification, negative portfolio weights can mechanically generate negative financed emissions and therefore offset emissions elsewhere in the portfolio. From a carbon-accounting perspective, however, short positions should not necessarily be interpreted as generating negative financed emissions; under accounting conventions such as \cite{PCAF_2022}, a more appropriate treatment may be to truncate their contribution to financed emissions at zero. Incorporating such asymmetric treatment of long and short positions would introduce additional nonlinearities into the control problem. Third, we model the disclosure-based investor as managing a carbon objective constructed from contemporaneously available estimates of firm-level carbon intensities. This reflects an information-based portfolio mandate, but differs from preferences defined directly over latent realized emissions. Studying an investor who is explicitly exposed to uncertainty about unobserved realized cumulative emissions would require a risk-sensitive filtering formulation. Developing this joint filtering formulation constitutes another extension of the framework. \\

The analysis treats the investor as small and non-consequentialist. She takes firms' emissions trajectories as given and seeks only to reduce the financed emissions associated with its own portfolio. Extending the framework to multiple investors whose portfolio decisions may affect asset prices, firms' financing conditions, and ultimately endogenous carbon-intensity dynamics would provide a natural direction for future research. \\

\begin{center}
\begin{minipage}{\textwidth}
\textbf{Use of generative artificial intelligence.}
GPT5.6 Sol was used during the preparation of this manuscript to assist with language editing and text correction, consistency and verification checks, and the development and debugging of numerical code and figures. All AI-assisted outputs, including mathematical expressions, code, numerical results, and graphical representations, were reviewed and verified by the authors. The authors remain fully responsible for the scientific content, analysis, and conclusions of the paper.
\end{minipage}
\end{center}


\newpage
\appendix

\section{Proofs}
\label{app:proofs}

\renewcommand{\theequation}{A.\arabic{equation}}
\renewcommand{\thefigure}{A.\arabic{figure}}
\renewcommand{\thetable}{A.\arabic{table}}

\setcounter{equation}{0}
\setcounter{figure}{0}
\setcounter{table}{0}

\subsection{Filtering equations for the $U$-investor (Proposition~\ref{prop:filter_U_unknown_barm})}
\label{app:proof_filter_U_unknown_barm}

\begin{proof}

The \(U\)-investor observes the return and carbon-intensity residuals. Stacking these two observations gives
\begin{equation}
        dO_t
    :=
    \begin{pmatrix}
        dR_t\\
        dJ_t
    \end{pmatrix}
    =
    H_U L^U_t\,dt+dN_t,
\end{equation}
where
\begin{equation}
    H_U=
    \begin{pmatrix}
        \alphab & 0_d\\
        0_d & \mathcal K\deltab
    \end{pmatrix}
    \in\mathbb R^{2d\times 2},
\end{equation}
and
\begin{equation}
        dN_t
    =
    \begin{pmatrix}
        \Sigmab_S^{1/2}dW_t\\
        \Sigmab_U^{1/2}d\Bb_t
    \end{pmatrix}.
\end{equation}
The covariance matrix of the observation noise is
\begin{equation}
    d\langle N\rangle_t=\Sigmab\,dt,
    \qquad
    \Sigmab
    =
    \begin{pmatrix}
        \Sigmab_S & \Sigmab_{SU}\\
        \Sigmab_{SU}^{\top} & \Sigmab_U
    \end{pmatrix}.
\end{equation}
where \(\Sigmab\) is positive definite and \(W^M\) is independent of the observation noise \((W,\Bb)\). The state equation and the observation equation are linear, and the initial
hidden state is assumed to be Gaussian. Hence, by the Kalman--Bucy filtering
theorem, the conditional distribution of \(L_t\) given \(\mathcal F_t^U\)
remains Gaussian. It is therefore fully characterized by its conditional mean
\begin{equation}
    \widehat L_t^U = \mathbb E[L^U_t\mid\mathcal F_t^U] =
    \begin{pmatrix}
        \widehat M_t^U\\
        \widehat{m}_{\infty,t}^U
    \end{pmatrix},
\end{equation}
and its conditional covariance matrix
\begin{equation}
    Q_t^U = \mathbb E\left[(L^U_t-\widehat L_t^U)(L^U_t-\widehat L_t^U)^\top\mid \mathcal F_t^U
    \right]
\end{equation}

The Kalman--Bucy filter for the conditional mean is
\begin{equation}
    d\widehat L_t^U = A_U\widehat L_t^U\,dt + K_t^U \left(  dO_t-H\widehat L_t^Udt \right)
\end{equation}
where the Kalman gain is
\begin{equation}
    K_t^U  = Q_t^U H_U^\top \Sigmab^{-1}.
\end{equation}
Therefore,
\begin{equation}
    d\widehat L_t^U =A_U\widehat L_t^U\,dt + Q_t^U H_U^\top \Sigmab^{-1}\left(
        dO_t-H_U\widehat L_t^Udt\right)
\end{equation}
In components, this gives
\begin{equation}
\begin{split}
    d\widehat M_t^U &= \kappa_M \left( \widehat{m}_{\infty,t}^U-\widehat M_t^U\right)dt
    + e_1^\top K_t^U\left(dO_t-H_U\widehat L_t^Udt\right),\\
    d\widehat{m}_{\infty,t}^U &= e_2^\top K_t^U\left(dO_t-H_U\widehat L_t^Udt\right),
\end{split}
\end{equation}
where \(e_1=(1,0)^\top\) and \(e_2=(0,1)^\top\). This proves the first statement.

We now derive the Riccati equation for the conditional covariance matrix.
Define the filtering error
\begin{equation}
        e_t=L^U_t-\widehat L_t^U
\end{equation}

Using the observation equation, the innovation process satisfies
\begin{equation}
    dI_t^U = dO_t-H_U\widehat L_t^Udt =  H_U(L^U_t-\widehat L_t^U)dt+dN_t
    =
    H_Ue_tdt+dN_t.
\end{equation}
Moreover,
\begin{equation}
\begin{split}
    de_t    &=  dL^U_t-d\widehat L_t^U\\
    &= A_U L^U_tdt+G_UdW_t^M - A_U\widehat L^U_t dt - K_t^U\left(dO_t-H_U\widehat L_t^Udt\right)\\
    &= A_U e_tdt+G_UdW_t^M - K_t^U \left( H_Ue_tdt+dN_t \right)\\
    &=\left(A_U-K_t^UH_U\right)e_tdt + G_UdW_t^M - K_t^UdN_t
\end{split}
\end{equation}
Applying Itô's formula to \(e_te_t^\top\), we obtain
\begin{equation}
    \begin{split}
    d(e_te_t^\top)  &= (de_t)e_t^\top +e_t(de_t)^\top +(de_t)(de_t)^\top
\end{split}
\end{equation}

The drift terms coming from the first two terms are
\begin{equation}
    \left(A_U-K_t^UH_U\right)e_te_t^\top dt + e_te_t^\top
    \left(A_U-K_t^UH_U\right)^\top dt.
\end{equation}
Since \(W^M\) is independent of \(N\), the quadratic-variation term is
\begin{equation}
    (de_t)(de_t)^\top = \left( G_UG_U^\top  + K_t^U\Sigmab (K_t^U)^\top\right)dt
\end{equation}
Taking conditional expectations given \(\mathcal F_t^U\), and using
\begin{equation}
    Q_t^U = \mathbb E[e_te_t^\top\mid\mathcal F_t^U]
\end{equation}
we get
\begin{equation}
    \dot Q_t^U = \left(A_U-K_t^UH_U\right)Q_t^U + Q_t^U\left(A_U-K_t^UH_U\right)^\top + G_UG_U^\top +
    K_t^U\Sigmab (K_t^U)^\top
\end{equation}
Substituting
\[
    K_t^U=Q_t^U H_U^\top\Sigma^{-1}
\]
into the covariance equation yields
\begin{equation}
\begin{split}
    \dot Q_t^U
    &=
    A_UQ_t^U+Q_t^U A_U^\top
    +G_UG_U^\top
    -Q_t^U H_U^\top\Sigma^{-1}H_UQ_t^U.
\end{split}
\end{equation}
Equivalently,
\begin{equation}
    \dot Q_t^U
    =
    A_UQ_t^U+Q_t^U A_U^\top
    +G_UG_U^\top
    -K_t^U\Sigma(K_t^U)^\top.
\end{equation}
The initial conditions are determined by the prior distribution of
\begin{equation}
    L_0^U=
    \begin{pmatrix}
        M_0\\
        m_\infty
    \end{pmatrix}.
\end{equation}
Hence
\begin{equation}
    \widehat L_0^U=
    \begin{pmatrix}
        \overline M_0\\
        \overline m_\infty
    \end{pmatrix},
    \qquad
    Q_0^U=
    \begin{pmatrix}
        q_0^M & q_0^{M,m_\infty}\\
        q_0^{M,m_\infty} & q_0^{m_\infty}
    \end{pmatrix}.
\end{equation}
Since the system is linear Gaussian, \(\Sigma\) is positive definite and \(Q_0^U\) is symmetric
nonnegative, the Kalman--Bucy covariance equation admits a unique symmetric nonnegative
solution on \([0,T]\). This completes the proof.

\end{proof}

\subsection{Filtering equations and disclosure updates for the $Z$-investor (Proposition~\ref{prop:filter_Z_unknown_barm})}
\label{app:proof_filter_Z_unkown_barm}

\begin{proof}

Recall that, for the \(Z\)-investor, the hidden state is
\begin{equation}
        L_t^Z  = \begin{pmatrix}
        M_t\\
        m_\infty\\
        \Ub_t
    \end{pmatrix}
    \in\mathbb R^{d+2}
\end{equation}

The \(Z\)-investor observes the asset-price process continuously and receives
noisy disclosures of \(\Ub_t\) only at deterministic dates
\begin{equation}
        0=\tau_0<\tau_1<\dots<\tau_N<T
\end{equation}

Between two disclosure dates, no new direct observation of carbon intensities is
received. Hence, on each interval \((\tau_n,\tau_{n+1})\), the only continuous
observation available to the \(Z\)-investor is the return process. Between disclosure dates, the conditional law of \(L_t^Z\) given the \(Z\)-investor filtration remains
Gaussian. It is fully characterized by its conditional mean
\begin{equation}
    \widehat L_t^Z
    :=
    \mathbb E[L_t^Z\mid\mathcal F_t^Z],
\end{equation}
and its conditional covariance matrix
\begin{equation}
    Q_t^Z
    :=
    \mathbb E\left[
        (L_t^Z-\widehat L_t^Z)
        (L_t^Z-\widehat L_t^Z)^\top
        \mid\mathcal F_t^Z
    \right].
\end{equation}

By the Kalman-Bucy filtering theorem with correlated state and observation
noises, the conditional mean satisfies, for
\(t\in(\tau_n,\tau_{n+1})\),
\begin{equation}
    d\widehat L_t^Z
    =
    \left(A_Z\widehat L_t^Z+c_Z\right)dt
    +
    \left(Q_t^ZH_Z^\top+\Sigma_{SL}\right)
    \Sigmab_S^{-1}
    \left(
        dR_t^Z-H_Z\widehat L_t^Zdt
    \right).
\end{equation}
This proves the mean equation between disclosure dates.

We now derive the covariance equation. The Kalman--Bucy covariance equation with
correlated state and observation noises is:
\begin{equation}
    \dot Q_t^Z  =  A_ZQ_t^Z+Q_t^ZA_Z^\top+G_ZG_Z^\top - \left(Q_t^ZH_Z^\top+\Sigma_{SL}\right)
    \Sigmab_S^{-1} \left(H_ZQ_t^Z+\Sigma_{SL}^\top\right).
\end{equation}

Equivalently,
\begin{equation}
    \begin{aligned}
    \dot Q_t^Z  &= \left(A_Z-\Sigma_{SL}\Sigmab_S^{-1}H_Z\right)Q_t^Z + Q_t^Z \left(A_Z-\Sigma_{SL}\Sigmab_S^{-1}H_Z\right)^\top  \\
    &\quad + \left(  G_ZG_Z^\top-\Sigma_{SL}\Sigmab_S^{-1}\Sigma_{SL}^\top \right)
    -  Q_t^ZH_Z^\top\Sigmab_S^{-1}H_ZQ_t^Z .
\end{aligned}
\end{equation}

Moreover,
\begin{equation}
     G_ZG_Z^\top =
    \begin{pmatrix}
        \sigma_M^2 & 0 & 0_{1\times d}\\
        0 & 0 & 0_{1\times d}\\
        0_{d\times 1} & 0_{d\times 1} & \Sigmab_U
    \end{pmatrix}.
\end{equation}

Since
\begin{equation}
    \Sigma_{SL} =
    \begin{pmatrix}
        0_{1\times d}\\
        0_{1\times d}\\
        \Sigmab_{SU}^\top
    \end{pmatrix},
\end{equation}
we obtain
\begin{equation}
        \Sigma_{SL}\Sigmab_S^{-1}\Sigma_{SL}^\top
    =
    \begin{pmatrix}
        0 & 0 & 0_{1\times d}\\
        0 & 0 & 0_{1\times d}\\
        0_{d\times 1} & 0_{d\times 1}
        & \Sigmab_{SU}^\top\Sigmab_S^{-1}\Sigmab_{SU}
    \end{pmatrix}.
\end{equation}

Therefore,
\begin{equation}
    G_ZG_Z^\top-\Sigma_{SL}\Sigmab_S^{-1}\Sigma_{SL}^\top
    =
    \begin{pmatrix}
        \sigma_M^2 & 0 & 0_{1\times d}\\
        0 & 0 & 0_{1\times d}\\
        0_{d\times 1} & 0_{d\times 1}
        & \Sigmab_U-\Sigmab_{SU}^\top\Sigmab_S^{-1}\Sigmab_{SU}
    \end{pmatrix}.
\end{equation}
The lower-right block
\begin{equation}
    \Sigmab_U-\Sigmab_{SU}^\top\Sigmab_S^{-1}\Sigmab_{SU}
\end{equation}
is the Schur complement of \(\Sigmab_S\) in the joint covariance matrix
\begin{equation}
    \Sigmab
    =
    \begin{pmatrix}
        \Sigmab_S & \Sigmab_{SU}\\
        \Sigmab_{SU}^\top & \Sigmab_U
    \end{pmatrix}.
\end{equation}
Since \(\Sigmab\) is positive definite and \(\Sigmab_S\) is positive definite,
this Schur complement is positive semidefinite. Hence
\begin{equation}
    G_ZG_Z^\top-\Sigma_{SL}\Sigmab_S^{-1}\Sigma_{SL}^\top
    \in\mathbb S^{d+2}_+.
\end{equation}
The right-hand side of the Riccati equation is locally Lipschitz in \(Q_t^Z\).
Since the equation is the covariance equation associated with a linear Gaussian
filtering problem, it admits a unique symmetric nonnegative solution on each
interval \((\tau_n,\tau_{n+1})\). This proves the covariance equation between
disclosure dates. \\

It remains to prove the discrete update at disclosure dates. At time \(\tau_n\),
the \(Z\)-investor observes
\begin{equation}
    Z_n=D_ZL_{\tau_n}^Z+\epsilon_n.
\end{equation}
where $\epsilon_n$ is independent of \(L_{\tau_n}^Z\) conditionally on \(\mathcal F_{\tau_n^-}^Z\). Just before the disclosure, the conditional distribution is
\begin{equation}
    L_{\tau_n}^Z\mid\mathcal F_{\tau_n^-}^Z
    \sim
    \mathcal N
    \left(
        \widehat L_{\tau_n^-}^Z,
        Q_{\tau_n^-}^Z
    \right).
\end{equation}
Define the disclosure innovation by
\begin{equation}
    \nu_n^Z=Z_n-D_Z\widehat L_{\tau_n^-}^Z.
\end{equation}
Conditionally on \(\mathcal F_{\tau_n^-}^Z\), this innovation is Gaussian with
mean zero and covariance
\begin{equation}
    \operatorname{Var} \left(\nu_n^Z\mid\mathcal F_{\tau_n^-}^Z \right)  =
    D_ZQ_{\tau_n^-}^ZD_Z^\top+\Sigmab_n^\epsilon
\end{equation}
Moreover,
\begin{equation}
    \operatorname{Cov}
    \left( L_{\tau_n}^Z,\nu_n^Z \mid \mathcal F_{\tau_n^-}^Z \right) =Q_{\tau_n^-}^ZD_Z^\top.
\end{equation}

By the Gaussian conditioning formula,
\begin{equation}
\begin{aligned}
    \widehat L_{\tau_n}^Z &=\mathbb E\left[ L_{\tau_n}^Z \mid \mathcal F_{\tau_n}^Z \right] \\
    &=  \mathbb E\left[ L_{\tau_n}^Z \mid \mathcal F_{\tau_n^-}^Z,Z_n
    \right] \\
    &= \widehat L_{\tau_n^-}^Z + Q_{\tau_n^-}^ZD_Z^\top \left( D_ZQ_{\tau_n^-}^ZD_Z^\top+\Sigmab_n^\epsilon
    \right)^{-1} \left( Z_n-D_Z\widehat L_{\tau_n^-}^Z \right).
\end{aligned}
\end{equation}

Define the discrete Kalman gain by
\begin{equation}
    K_n^Z= Q_{\tau_n^-}^ZD_Z^\top \left( D_ZQ_{\tau_n^-}^ZD_Z^\top+\Sigmab_n^\epsilon
    \right)^{-1}.
\end{equation}
Then
\begin{equation}
    \widehat L_{\tau_n}^Z  = \widehat L_{\tau_n^-}^Z + K_n^Z\left(  Z_n-D_Z\widehat L_{\tau_n^-}^Z
    \right)
\end{equation}
This proves the stated mean update at disclosure dates.

The Gaussian conditioning formula also gives the posterior covariance matrix:
\begin{equation}
    Q_{\tau_n}^Z  = Q_{\tau_n^-}^Z  - Q_{\tau_n^-}^ZD_Z^\top \left(D_ZQ_{\tau_n^-}^ZD_Z^\top+\Sigmab_n^\epsilon \right)^{-1} D_ZQ_{\tau_n^-}^Z.
\end{equation}
Equivalently,
\begin{equation}
    Q_{\tau_n}^Z = Q_{\tau_n^-}^Z -  K_n^ZD_ZQ_{\tau_n^-}^Z.
\end{equation}
Since \(Q_{\tau_n}^Z\) is the conditional covariance matrix of
\(L_{\tau_n}^Z\) given \(\mathcal F_{\tau_n}^Z\), it is symmetric positive
semidefinite. This completes the proof.
\end{proof}

\subsection{HJB solution for the $F$-investor (Proposition~\ref{prop:control-F})}
\label{app:proof-control-F}

\begin{proof}
For the F-investor, the state variable is
\[
    \ell^F=(m,u)^\top\in\mathbb R^{d+1}.
\]
The HJB equation is
\begin{equation}
\begin{split}
0
&=
\partial_t v^F
+rxv_x^F
+xv_yu_{rf}
+\nabla_\ell v^{F\top}(A_F\ell^F+c_F)
+\frac12
\operatorname{Tr}\left(D_FD_F^\top D_{\ell\ell}^2v^F\right) \\
&\quad
+\sup_{\pi\in\mathbb R^d}
\Bigg\{
    xv_x^F\pi^\top\theta^F(\ell^F)
    +xv_y^F\pi^\top (E_u^F\ell^F - u_{rf}\mathbf 1)
    +\frac12x^2v_{xx}^F\pi^\top\Sigma_S\pi
    +x\pi^\top\Gamma_F^\top\nabla_{x\ell}v^F
\Bigg\}.
\end{split}
\end{equation}
The terminal condition is
\begin{equation}
    v^F(T,x,y,\ell^F)=-\exp\{-\eta(x-\lambda y)\}.
\end{equation}
Following \cite{Liu_2013}, we look for a solution of the form
\begin{equation}
    v^F(t,x,y,\ell^F)
    =
    -\exp\left\{
        -\eta\left(
            \gamma(t)x-\lambda y+\phi^F(t,\ell^F)
        \right)
    \right\},
    \qquad
    \gamma(t)=e^{r(T-t)}- \lambda \int_{t}^{T}e^{r(s-t)}u_{rf}(s)ds.
\end{equation}
Then \(\gamma'(t)+r\gamma(t) - \lambda u_{rf}=0\), \(\gamma(T)=1\), and the terminal condition implies
\begin{equation}
    \phi^F(T,\ell^F)=0.
\end{equation}
The derivatives are
\begin{equation}
    \begin{aligned}
    \partial_t v^F
    &=
    -\eta v^F\left(\gamma'(t)x+\partial_t\phi^F\right),\\
    v_x^F
    &=
    -\eta\gamma(t)v^F,\\
    v_y^F
    &=
    \eta\lambda v^F,\\
    v_{xx}^F
    &=
    \eta^2\gamma(t)^2v^F,\\
    \nabla_\ell v^F
    &=
    -\eta v^F\nabla_\ell\phi^F,\\
    D_{\ell\ell}^2v^F
    &=
    \eta^2v^F\nabla_\ell\phi^F\nabla_\ell\phi^{F\top}
    -\eta v^FD_{\ell\ell}^2\phi^F,\\
    \nabla_{x\ell}v^F
    &=
    \eta^2\gamma(t)v^F\nabla_\ell\phi^F.
    \end{aligned}
\end{equation}

Substituting into the HJB equation and dividing by \(-\eta v^F>0\), the terms proportional to \(x\) outside the supremum cancel because \(\gamma'(t)+r\gamma(t) - \lambda u_{rf} = 0\). We obtain
\begin{equation}
\begin{split}
0
&=
\partial_t\phi^F
+\nabla_\ell\phi^{F\top}(A_F\ell^F+c_F)
+\frac12\operatorname{Tr}\left(D_FD_F^\top D_{\ell\ell}^2\phi^F\right)
-\frac{\eta}{2}\nabla_\ell\phi^{F\top}D_FD_F^\top\nabla_\ell\phi^F \\
&\quad
+\sup_{\pi\in\mathbb R^d}
\Bigg\{
    \gamma(t)x\pi^\top\theta^F(\ell^F)
    -\lambda x\pi^\top (E_u^F\ell^F - u_{rf}\mathbf 1)
    -\frac{\eta}{2}\gamma(t)^2x^2\pi^\top\Sigma_S\pi
    -\eta\gamma(t)x\pi^\top\Gamma_F^\top\nabla_\ell\phi^F
\Bigg\}.
\end{split}
\end{equation}
The expression inside the supremum is strictly concave in \(\pi\), since \(\Sigma_S\) is positive
definite. The first-order condition gives
\begin{equation}
    \pi^{F,*}(t,x,\ell^F)
    =
    \frac{1}{\eta\gamma(t)x}
    \Sigma_S^{-1}
    \left[
        \theta^F(\ell^F)
        -\frac{\lambda}{\gamma(t)}(E_u^F\ell^F - u_{rf}(t)\mathbf 1)
        -\eta\Gamma_F^\top\nabla_\ell\phi^F(t,\ell^F)
    \right].
\end{equation}
Substituting the maximized value back into the HJB gives the PDE
\begin{equation}
\begin{split}
0
&=
\partial_t\phi^F
+\nabla_\ell\phi^{F\top}(A_F\ell^F+c_F)
+\frac12\operatorname{Tr}\left(D_FD_F^\top D_{\ell\ell}^2\phi^F\right)
-\frac{\eta}{2}\nabla_\ell\phi^{F\top}D_FD_F^\top\nabla_\ell\phi^F \\
&\quad
+\frac{1}{2\eta}
\left[
    \theta^F(\ell^F)
    -\frac{\lambda}{\gamma(t)}(E_u^F\ell^F - u_{rf}(t)\mathbf 1)
    -\eta\Gamma_F^\top\nabla_\ell\phi^F
\right]^\top
\Sigma_S^{-1}
\left[
    \theta^F(\ell^F)
    -\frac{\lambda}{\gamma(t)}(E_u^F\ell^F - u_{rf}(t)\mathbf 1)
    -\eta\Gamma_F^\top\nabla_\ell\phi^F
\right].
\end{split}
\end{equation}
Now impose the quadratic ansatz
\begin{equation}
    \phi^F(t,\ell^F)
    =
    p_0^F(t)+p_1^F(t)^\top\ell^F
    +\frac12(\ell^F)^\top P^F(t)\ell^F,
\end{equation}
where \(P^F(t)\) is symmetric. Then
\begin{equation}
    \nabla_\ell\phi^F=p_1^F+P^F\ell^F,
    \qquad
    D_{\ell\ell}^2\phi^F=P^F.
\end{equation}
Define
\begin{equation}
    b_0^F(t)
    :=
    \theta_0-\eta\Gamma_F^\top p_1^F(t)
    + \frac{\lambda}{\gamma(t)}u_{rf}(t)\mathbf 1)
    \qquad
    B^F(t)
    :=
    \Theta^F-\frac{\lambda}{\gamma(t)}E_u^F 
    -\eta\Gamma_F^\top P^F(t).
\end{equation}
Identifying the constant, linear and quadratic terms in \(\ell^F\) gives
\begin{equation}
\begin{aligned}
0
&=
\dot p_0^F
+p_1^{F\top}c_F
+\frac12\operatorname{Tr}\left(D_FD_F^\top P^F\right)
-\frac{\eta}{2}p_1^{F\top}D_FD_F^\top p_1^F
+\frac{1}{2\eta}b_0^{F\top}\Sigma_S^{-1}b_0^F,
\\
0
&=
\dot p_1^F
+A_F^\top p_1^F
+P^F c_F
-\eta P^FD_FD_F^\top p_1^F
+\frac{1}{\eta}B^{F\top}\Sigma_S^{-1}b_0^F,
\\
0
&=
\dot P^F
+P^FA_F+A_F^\top P^F
-\eta P^FD_FD_F^\top P^F
+\frac{1}{\eta}B^{F\top}\Sigma_S^{-1}B^F.
\end{aligned}
\end{equation}
The terminal conditions are
\begin{equation}
    p_0^F(T)=0,\qquad p_1^F(T)=0,\qquad P^F(T)=0.
\end{equation}
Finally, let
\begin{equation}
    C^F(t):=\Theta^F-\frac{\lambda}{\gamma(t)}E_u^F.
\end{equation}
The Riccati equation for \(P^F\) can be rewritten as
\begin{equation}
    0=
    \dot P^F
    +P^F\widetilde A_F(t)
    +\widetilde A_F(t)^\top P^F
    +R_F(t)
    -P^F M_F P^F,
\end{equation}
where
\begin{equation}
\begin{aligned}
    \widetilde A_F(t)
    &:=
    A_F-\Gamma_F\Sigma_S^{-1}C^F(t),\\
    R_F(t)
    &:=
    \frac{1}{\eta}C^F(t)^\top\Sigma_S^{-1}C^F(t),\\
    M_F
    &:=
    \eta\left(
        D_FD_F^\top-\Gamma_F\Sigma_S^{-1}\Gamma_F^\top
    \right).
\end{aligned}
\end{equation}
The matrix \(R_F(t)\) is nonnegative definite. Moreover,
\begin{equation}
    D_FD_F^\top-\Gamma_F\Sigma_S^{-1}\Gamma_F^\top
\end{equation}
is the Schur complement of \(\Sigma_S\) in the joint covariance matrix of the return noise and
the state noise,
\begin{equation}
    \begin{pmatrix}
        \Sigma_S & \Gamma_F^\top\\
        \Gamma_F & D_FD_F^\top
    \end{pmatrix}.
\end{equation}
Hence \(M_F\in\mathbb S_+^{d+1}\). Therefore, the finite-horizon Riccati equation admits a unique symmetric solution on \([0,T]\) (see Theorem 4.1.6 of \cite{Aboukandil2003}). The equations for \(p_1^F\) and \(p_0^F\) are then linear
backward equations once \(P^F\) is known. This proves the result. Admissibility and verification are provided in Appendix~\ref{app:verification}.
\end{proof}

\subsection{Covariance structure of the $U$-investor’s filtered state (Lemma ~\ref{lem:cov_RL_U})}
\label{app:proof_lemma_cov_RL_U}

\begin{proof}
Let
\begin{equation}
    E_R=
    \begin{pmatrix}
        I_d & 0_{d\times d}
    \end{pmatrix}.
\end{equation}

The return innovation is the first block of the observation innovation:
\begin{equation}
    d\widehat R_t^U=E_R\,dI_t^U.
\end{equation}
Since
\begin{equation}
    d\langle I^U\rangle_t=\Sigma\,dt,
\end{equation}
and
\begin{equation}
    d\widehat{\mathcal L}_t^U
    =
    (A_U\widehat{\mathcal L}_t^U+c_U)dt
    +D_t^U\,dI_t^U,
\end{equation}
we have
\begin{equation}
    d\left\langle
        \widehat{\mathcal L}^U,\widehat R^U
    \right\rangle_t
    =
    D_t^U\Sigma E_R^\top dt.
\end{equation}
Therefore
\begin{equation}
    \Gamma_t^U
    =
    D_t^U\Sigma E_R^\top.
\end{equation}
Using the block representation
\begin{equation}
    D_t^U=
    \begin{pmatrix}
        K_t^U\\
        0_{d\times d} & I_d
    \end{pmatrix},
    \qquad
    K_t^U=Q_t^U H_U^\top\Sigma^{-1},
\end{equation}
we get
\begin{equation}
    \Gamma_t^U
    =
    \begin{pmatrix}
        Q_t^UH_U^\top\Sigma^{-1}\\
        0_{d\times d} & I_d
    \end{pmatrix}
    \Sigma E_R^\top.
\end{equation}
Since
\begin{equation}
    \Sigma E_R^\top
    =
    \begin{pmatrix}
        \Sigma_S\\
        \Sigma_{SU}^\top
    \end{pmatrix},
\end{equation}
we obtain
\begin{equation}
    \Gamma_t^U
    =
    \begin{pmatrix}
        Q_t^UH_U^\top E_R^\top\\
        \Sigma_{SU}^\top
    \end{pmatrix}.
\end{equation}
Moreover,
\begin{equation}
    H_U^\top E_R^\top
    =
    \begin{pmatrix}
        \alpha^\top\\
        0_{1\times d}
    \end{pmatrix}.
\end{equation}
Writing
\begin{equation}
    Q_t^U=
    \begin{pmatrix}
        q_M^U(t) & q_{M,m_\infty}^U(t)\\
        q_{M,m_\infty}^U(t) & q_{m_\infty}^U(t)
    \end{pmatrix},
\end{equation}
we finally get
\begin{equation}
    \Gamma_t^U
    =
    \begin{pmatrix}
        q_M^U(t)\alpha^\top\\
        q_{M,m_\infty}^U(t)\alpha^\top\\
        \Sigma_{SU}^\top
    \end{pmatrix}
    \in\mathbb R^{(d+2)\times d}.
\end{equation}
This proves the lemma.
\end{proof}

\subsection{HJB solution for the $U$-investor (Proposition~\ref{prop:optimal-U})}
\label{app:proof-control-U}

\begin{proof}
For the U-investor, the Markov state is
\begin{equation}
    \ell^U=
    \begin{pmatrix}
        m\\
        \xi\\
        u
    \end{pmatrix}
    \in\mathbb R^{d+2},
\end{equation}
where \(m\) is the filtered estimate of \(M_t\), \(\xi\) is the filtered estimate of \(m_\infty\),
and \(u\) is the observed carbon-intensity vector. The HJB equation is therefore
\begin{equation}
\begin{split}
0
&=
\partial_t v^U
+rxv_x^U
+xv_y^Uu_{rf}
+\nabla_\ell v^{U\top}(A_U\ell^U+c_U)
+\frac12\operatorname{Tr}\left(\Lambda_t^U D_{\ell\ell}^2v^U\right)\\
&\quad
+\sup_{\pi\in\mathbb R^d}
\Bigg\{
    xv_x^U\pi^\top\theta(\ell^U)
    +xv_y^U\pi^\top (E_u^U\ell^U - u_{rf}(t)\mathbf 1)
    +\frac12x^2v_{xx}^U\pi^\top\Sigma_S\pi
    +x\pi^\top(\Gamma_t^U)^\top\nabla_{x\ell}v^U
\Bigg\}.
\end{split}
\end{equation}
The terminal condition is
\[
    v^U(T,x,y,\ell^U)=-\exp\{-\eta(x-\lambda y)\}.
\]
We use the ansatz
\begin{equation}
    v^U(t,x,y,\ell^U)
    =
    -\exp\left\{
        -\eta\left(
            \gamma(t)x-\lambda y+\phi^U(t,\ell^U)
        \right)
    \right\}.
\end{equation}
Proceeding as in the proof of Proposition~3, the terms proportional to \(x\) outside the supremum cancel because \(\gamma'(t)+r\gamma(t) - \lambda u_{rf} = 0\). The reduced PDE for \(\phi^U\) is
\begin{equation}
\begin{split}
0
&=
\partial_t\phi^U
+\nabla_\ell\phi^{U\top}(A_U\ell^U+c_U)
+\frac12\operatorname{Tr}\left(\Lambda_t^U D_{\ell\ell}^2\phi^U\right)
-\frac{\eta}{2}\nabla_\ell\phi^{U\top}\Lambda_t^U\nabla_\ell\phi^U\\
&\quad
+\frac{1}{2\eta}
\left[
    \theta(\ell^U)
    -\frac{\lambda}{\gamma(t)}(E_u^U\ell^U - u_{rf}(t)\mathbf 1)
    -\eta(\Gamma_t^U)^\top\nabla_\ell\phi^U
\right]^\top
\Sigma_S^{-1}
\left[
    \theta(\ell^U)
    -\frac{\lambda}{\gamma(t)}(E_u^U\ell^U - u_{rf}(t)\mathbf 1)
    -\eta(\Gamma_t^U)^\top\nabla_\ell\phi^U
\right].
\end{split}
\end{equation}
The maximizer is
\begin{equation}
    \pi^{U,*}(t,x,\ell^U)
    =
    \frac{1}{\eta\gamma(t)x}
    \Sigma_S^{-1}
    \left[
        \theta(\ell^U)
        -\frac{\lambda}{\gamma(t)}(E_u^U\ell^U - u_{rf}(t)\mathbf 1)
        -\eta(\Gamma_t^U)^\top\nabla_\ell\phi^U(t,\ell^U)
    \right].
\end{equation}

Now impose the quadratic ansatz
\begin{equation}
    \phi^U(t,\ell^U)
    =
    p_0^U(t)+p_1^U(t)^\top\ell^U
    +\frac12(\ell^U)^\top P^U(t)\ell^U,
\end{equation}
with \(P^U(t)\) symmetric. Define
\begin{equation}
    b_0^U(t)
    :=
    \theta_0-\eta(\Gamma_t^U)^\top p_1^U(t)
     + \frac{\lambda}{\gamma(t)}u_{rf}(t)\mathbf 1)
    \qquad
    B^U(t)
    :=
    \Theta-\frac{\lambda}{\gamma(t)}E_u^U
    -\eta(\Gamma_t^U)^\top P^U(t).
\end{equation}
Identifying the constant, linear and quadratic terms gives
\begin{equation}
\begin{aligned}
0
&=
\dot p_0^U
+p_1^{U\top}c_U
+\frac12\operatorname{Tr}\left(\Lambda_t^UP^U\right)
-\frac{\eta}{2}p_1^{U\top}\Lambda_t^Up_1^U
+\frac{1}{2\eta}b_0^{U\top}\Sigma_S^{-1}b_0^U,
\\
0
&=
\dot p_1^U
+A_U^\top p_1^U
+P^Uc_U
-\eta P^U\Lambda_t^Up_1^U
+\frac{1}{\eta}B^{U\top}\Sigma_S^{-1}b_0^U,
\\
0
&=
\dot P^U
+P^UA_U+A_U^\top P^U
-\eta P^U\Lambda_t^UP^U
+\frac{1}{\eta}B^{U\top}\Sigma_S^{-1}B^U.
\end{aligned}
\end{equation}
The terminal conditions are
\begin{equation}
    p_0^U(T)=0,\qquad p_1^U(T)=0,\qquad P^U(T)=0.
\end{equation}
Finally, let
\begin{equation}
    C^U(t):=\Theta-\frac{\lambda}{\gamma(t)}E_u.
\end{equation}
The Riccati equation can be rewritten as
\begin{equation}
    0
    =
    \dot P^U
    +P^U\widetilde A_U(t)
    +\widetilde A_U(t)^\top P^U
    +R_U(t)
    -P^U M_U(t)P^U,
\end{equation}
where
\begin{equation}
\begin{aligned}
    \widetilde A_U(t)
    &:=
    A_U-\Gamma_t^U\Sigma_S^{-1}C^U(t),\\
    R_U(t)
    &:=
    \frac{1}{\eta}C^U(t)^\top\Sigma_S^{-1}C^U(t),\\
    M_U(t)
    &:=
    \eta\left(
        \Lambda_t^U-\Gamma_t^U\Sigma_S^{-1}(\Gamma_t^U)^\top
    \right).
\end{aligned}
\end{equation}
The matrix \(R_U(t)\) is nonnegative definite because \(\Sigma_S\) is positive definite.

It remains to check that \(M_U(t)\) is nonnegative definite. Recall that
\begin{equation}
    \Lambda_t^U=D_t^U\Sigma(D_t^U)^\top,
    \qquad
    \Gamma_t^U=D_t^U\Sigma E_R^\top,
    \qquad
    E_R=
    \begin{pmatrix}
        I_d & 0_{d\times d}
    \end{pmatrix}.
\end{equation}
Moreover,
\begin{equation}
    E_R\Sigma E_R^\top=\Sigma_S.
\end{equation}
Therefore,
\begin{equation}
\begin{split}
    \Lambda_t^U-\Gamma_t^U\Sigma_S^{-1}(\Gamma_t^U)^\top
    &=
    D_t^U
    \left[
        \Sigma
        -\Sigma E_R^\top
        \left(E_R\Sigma E_R^\top\right)^{-1}
        E_R\Sigma
    \right]
    (D_t^U)^\top .
\end{split}
\end{equation}
The matrix
\begin{equation}
    \Sigma
    -\Sigma E_R^\top
    \left(E_R\Sigma E_R^\top\right)^{-1}
    E_R\Sigma
\end{equation}
is the Schur complement corresponding to the component of the observation noise which is
orthogonal to the return innovation. Since \(\Sigma\) is positive definite and
\(E_R\Sigma E_R^\top=\Sigma_S\) is positive definite, this matrix is nonnegative definite. Hence
\begin{equation}
    \Lambda_t^U-\Gamma_t^U\Sigma_S^{-1}(\Gamma_t^U)^\top
    \in\mathbb S_+^{d+2},
\end{equation}
and therefore
\begin{equation}
    M_U(t)\in\mathbb S_+^{d+2}.
\end{equation}
Thus the Riccati equation for \(P^U\) is a finite-horizon Riccati equation with
\(R_U(t)\succeq 0\) and \(M_U(t)\succeq 0\). Under the usual boundedness assumptions on the
deterministic coefficient functions, it admits a unique symmetric solution on \([0,T]\)
(see Theorem~4.1.6 of \cite{Aboukandil2003}). Once \(P^U\) is known, the equations for
\(p_1^U\) and \(p_0^U\) are linear backward equations. This proves the result. Admissibility and verification are provided in Appendix~\ref{app:verification}.
\end{proof}

\subsection{HJB solution for the $Z$-investor (Proposition \ref{prop:optimal-Z})}
\label{app:proof-control-Z}

\begin{proof}
Fix \(n\in\{0,\ldots,N\}\). On the interval \((\tau_n,\tau_{n+1})\), the Z-investor receives
no new disclosure. The filtered state satisfies
\begin{equation}
    d\widehat L_t^Z
    =
    (A_Z\widehat L_t^Z+c_Z)dt
    +\Gamma_t^Z\Sigma_S^{-1}d\widehat R_t^Z,
\end{equation}
where
\begin{equation}
    d\langle \widehat R^Z\rangle_t=\Sigma_Sdt,
    \qquad
    d\langle \widehat L^Z,\widehat R^Z\rangle_t=\Gamma_t^Zdt,
\end{equation}
and
\begin{equation}
    d\langle \widehat L^Z\rangle_t
    =
    \Lambda_t^Zdt,
    \qquad
    \Lambda_t^Z:=\Gamma_t^Z\Sigma_S^{-1}(\Gamma_t^Z)^\top.
\end{equation}
The HJB equation on \((\tau_n,\tau_{n+1})\) is
\begin{equation}
\begin{split}
0
&=
\partial_t v_n^Z
+rxv_{n,x}^Z
+xv_{n,y}^Zu_{rf}
+\nabla_\ell v_n^{Z\top}(A_Z\ell+c_Z)
+\frac12\operatorname{Tr}\left(\Lambda_t^Z D_{\ell\ell}^2v_n^Z\right)\\
&\quad
+\sup_{\pi\in\mathbb R^d}
\Bigg\{
    xv_{n,x}^Z\pi^\top\theta(\ell)
    +xv_{n,y}^Z\pi^\top (E_u\ell - u_{rf}(t)\mathbf 1)
    +\frac12x^2v_{n,xx}^Z\pi^\top\Sigma_S\pi
    +x\pi^\top(\Gamma_t^Z)^\top\nabla_{x\ell}v_n^Z
\Bigg\}.
\end{split}
\end{equation}
We use the exponential ansatz
\begin{equation}
    v_n^Z(t,x,y,\ell)
    =
    -\exp\left\{
        -\eta\left(
            \gamma(t)x-\lambda y+\phi_n^Z(t,\ell)
        \right)
    \right\}.
\end{equation}
The same calculation as in Proposition~3 gives the reduced PDE
\begin{equation}
\begin{split}
0
&=
\partial_t\phi_n^Z
+\nabla_\ell\phi_n^{Z\top}(A_Z\ell+c_Z)
+\frac12\operatorname{Tr}\left(\Lambda_t^ZD_{\ell\ell}^2\phi_n^Z\right)
-\frac{\eta}{2}\nabla_\ell\phi_n^{Z\top}\Lambda_t^Z\nabla_\ell\phi_n^Z\\
&\quad
+\frac{1}{2\eta}
\left[
    \theta(\ell)
    -\frac{\lambda}{\gamma(t)}(E_u\ell - u_{rf}(t)\mathbf 1)
    -\eta(\Gamma_t^Z)^\top\nabla_\ell\phi_n^Z
\right]^\top
\Sigma_S^{-1}
\left[
    \theta(\ell)
    -\frac{\lambda}{\gamma(t)}(E_u\ell - u_{rf}(t)\mathbf 1)
    -\eta(\Gamma_t^Z)^\top\nabla_\ell\phi_n^Z
\right].
\end{split}
\end{equation}
The optimal control on \((\tau_n,\tau_{n+1})\) is
\begin{equation}
    \pi_t^{Z,*}
    =
    \frac{1}{\eta\gamma(t)X_t}
    \Sigma_S^{-1}
    \left[
        \theta(\widehat L_t^Z)
        -\frac{\lambda}{\gamma(t)}(E_u\widehat L_t^Z - u_{rf}(t)\mathbf 1)
        -\eta(\Gamma_t^Z)^\top
        \nabla_\ell\phi_n^Z(t,\widehat L_t^Z)
    \right].
\end{equation}

Now impose the quadratic ansatz
\begin{equation}
    \phi_n^Z(t,\ell)
    =
    p_{0,n}^Z(t)+p_{1,n}^Z(t)^\top\ell
    +\frac12\ell^\top P_n^Z(t)\ell,
\end{equation}
where \(P_n^Z(t)\) is symmetric. Define
\begin{equation}
    b_{0,n}^Z(t)
    :=
    \theta_0-\eta(\Gamma_t^Z)^\top p_{1,n}^Z(t) 
    + \frac{\lambda}{\gamma(t)}u_{rf}(t)\mathbf 1
    \qquad
    B_n^Z(t)
    :=
    \Theta-\frac{\lambda}{\gamma(t)}E_u
    -\eta(\Gamma_t^Z)^\top P_n^Z(t).
\end{equation}
Identifying constant, linear and quadratic terms gives
\begin{equation}
\begin{aligned}
0
&=
\dot p_{0,n}^Z
+p_{1,n}^{Z\top}c_Z
+\frac12\operatorname{Tr}\left(\Lambda_t^ZP_n^Z\right)
-\frac{\eta}{2}p_{1,n}^{Z\top}\Lambda_t^Zp_{1,n}^Z
+\frac{1}{2\eta}b_{0,n}^{Z\top}\Sigma_S^{-1}b_{0,n}^Z,
\\
0
&=
\dot p_{1,n}^Z
+A_Z^\top p_{1,n}^Z
+P_n^Zc_Z
-\eta P_n^Z\Lambda_t^Zp_{1,n}^Z
+\frac{1}{\eta}B_n^{Z\top}\Sigma_S^{-1}b_{0,n}^Z,
\\
0
&=
\dot P_n^Z
+P_n^ZA_Z+A_Z^\top P_n^Z
-\eta P_n^Z\Lambda_t^ZP_n^Z
+\frac{1}{\eta}B_n^{Z\top}\Sigma_S^{-1}B_n^Z.
\end{aligned}
\end{equation}
On the last interval, the terminal condition is
\begin{equation}
    p_{0,N}^Z(T)=0,\qquad
    p_{1,N}^Z(T)=0,\qquad
    P_N^Z(T)=0.
\end{equation}
For \(n<N\), the boundary values at \(\tau_{n+1}^-\) are obtained from the disclosure-date jump
conditions.

Finally, fix \(n\in\{0,\ldots,N\}\) and let
\begin{equation}
    C^Z(t):=\Theta-\frac{\lambda}{\gamma(t)}E_u.
\end{equation}
The Riccati equation can be rewritten as
\begin{equation}
    0
    =
    \dot P_n^Z
    +P_n^Z\widetilde A_Z(t)
    +\widetilde A_Z(t)^\top P_n^Z
    +R_Z(t)
    -P_n^Z M_Z(t)P_n^Z,
\end{equation}
where
\begin{equation}
\begin{aligned}
    \widetilde A_Z(t)
    &:=
    A_Z-\Gamma_t^Z\Sigma_S^{-1}C^Z(t),\\
    R_Z(t)
    &:=
    \frac{1}{\eta}C^Z(t)^\top\Sigma_S^{-1}C^Z(t),\\
    M_Z(t)
    &:=
    \eta\left(
        \Lambda_t^Z-\Gamma_t^Z\Sigma_S^{-1}(\Gamma_t^Z)^\top
    \right).
\end{aligned}
\end{equation}
The matrix \(R_Z(t)\) is nonnegative definite because \(\Sigma_S\) is positive definite.

Using the fact that
\begin{equation}
    \Lambda_t^Z
    =
    \Gamma_t^Z\Sigma_S^{-1}(\Gamma_t^Z)^\top.
\end{equation}
the $M_Z$term vanishes and the Riccati equation for \(P_n^Z\) reduces, on each interval
\((\tau_n,\tau_{n+1})\), to the linear matrix equation
\begin{equation}
    0
    =
    \dot P_n^Z
    +P_n^Z\widetilde A_Z(t)
    +\widetilde A_Z(t)^\top P_n^Z
    +R_Z(t).
\end{equation}
Since the coefficient functions are deterministic and bounded on
\((\tau_n,\tau_{n+1})\), this equation admits a unique symmetric solution on each interval.

At disclosure dates, the quadratic coefficients are updated by the Gaussian jump conditions. These jump conditions preserve symmetry of \(P_n^Z\). Therefore, by solving backward from
\begin{equation}
    P_N^Z(T)=0,
\end{equation}
and applying the disclosure-date jump conditions recursively at
\(\tau_N,\ldots,\tau_1\), one obtains a unique symmetric solution on the whole interval \([0,T]\). Once \(P_n^Z\) is known on each interval, the equations for \(p_{1,n}^Z\) and \(p_{0,n}^Z\) are linear backward equations. This proves the result. Admissibility and verification are provided in Appendix~\ref{app:verification}.\\

\end{proof}

\subsection{Disclosure-date jump conditions for the Z-investor (Proposition \ref{prop:optimal-Z-jumps})}
\label{app:proof-control-Z-jumps}

\begin{proof}
At a disclosure date \(\tau_n\), \(n=1,\ldots,N\), the wealth process and the filtered cumulative
emission process are continuous. Only the filtered state jumps. The disclosure equation is
\begin{equation}
    Z_n=D_ZL_{\tau_n}^Z+\varepsilon_n,
    \qquad
    D_Z=
    \begin{pmatrix}
        0_{d\times 1} & 0_{d\times 1} & I_d
    \end{pmatrix}.
\end{equation}
The Kalman update is
\begin{equation}
    \widehat L_{\tau_n}^Z
    =
    \widehat L_{\tau_n^-}^Z
    +K_n^Z\nu_n^Z,
    \qquad
    \nu_n^Z:=Z_n-D_Z\widehat L_{\tau_n^-}^Z,
\end{equation}
where, conditionally on \(\mathcal F_{\tau_n^-}^Z\),
\begin{equation}
    \nu_n^Z\sim\mathcal N(0,S_n^Z),
    \qquad
    S_n^Z:=D_ZQ_{\tau_n^-}^ZD_Z^\top+\Sigma_n^\varepsilon.
\end{equation}
The dynamic programming principle gives
\begin{equation}
    v_{n-1}^Z(\tau_n^-,x,y,\ell)
    =
    \mathbb E\left[
        v_n^Z(\tau_n,x,y,\ell+K_n^Z\nu_n^Z)
        \,\middle|\,\mathcal F_{\tau_n^-}^Z
    \right].
\end{equation}
Since \(x\) and \(y\) do not jump, this is equivalent to
\begin{equation}
    \phi_{n-1}^Z(\tau_n^-,\ell)
    =
    -\frac1\eta
    \log
    \mathbb E\left[
        \exp\left\{
            -\eta\phi_n^Z(\tau_n,\ell+K_n^Z\nu_n^Z)
        \right\}
        \,\middle|\,\mathcal F_{\tau_n^-}^Z
    \right].
\end{equation}

Let
\begin{equation}
    p_0^+:=p_{0,n}^Z(\tau_n),
    \qquad
    p_1^+:=p_{1,n}^Z(\tau_n),
    \qquad
    P^+:=P_n^Z(\tau_n),
\end{equation}
and
\begin{equation}
    p_0^-:=p_{0,n-1}^Z(\tau_n^-),
    \qquad
    p_1^-:=p_{1,n-1}^Z(\tau_n^-),
    \qquad
    P^-:=P_{n-1}^Z(\tau_n^-).
\end{equation}
Then
\begin{equation}
\begin{split}
    \phi_n^Z(\tau_n,\ell+K_n^Z\nu)
    &=
    p_0^+
    +p_1^{+\top}\ell
    +\frac12\ell^\top P^+\ell\\
    &\quad
    +\nu^\top (K_n^Z)^\top(p_1^+ +P^+\ell)
    +\frac12\nu^\top (K_n^Z)^\top P^+K_n^Z\nu .
\end{split}
\end{equation}
Define
\begin{equation}
    A_n:=(K_n^Z)^\top P^+K_n^Z,
    \qquad
    M_n:=(S_n^Z)^{-1}+\eta A_n.
\end{equation}
We follow \cite{Gabih2024PowerUtility} and use the following identity to express the right-hand side. Provided \(M_n\) is positive definite, the Gaussian identity
\begin{equation}
\begin{split}
&-\frac1\eta
\log
\mathbb E\left[
    \exp\left\{
        -\eta b^\top\nu
        -\frac{\eta}{2}\nu^\top A\nu
    \right\}
\right] \\
&\qquad
=
\frac{1}{2\eta}\log\det(I+\eta S A)
-\frac{\eta}{2}b^\top(S^{-1}+\eta A)^{-1}b,
\qquad
\nu\sim\mathcal N(0,S),
\end{split}
\end{equation}
implies
\begin{equation}
\begin{split}
    \phi_{n-1}^Z(\tau_n^-,\ell)
    &=
    p_0^+
    +p_1^{+\top}\ell
    +\frac12\ell^\top P^+\ell
    +\frac{1}{2\eta}\log\det(I_d+\eta S_n^ZA_n)\\
    &\quad
    -\frac{\eta}{2}
    \left[
        (K_n^Z)^\top(p_1^+ +P^+\ell)
    \right]^\top
    M_n^{-1}
    \left[
        (K_n^Z)^\top(p_1^+ +P^+\ell)
    \right].
\end{split}
\end{equation}
Identifying the constant, linear and quadratic terms in \(\ell\) yields the jump conditions
\begin{equation}
    P^-
    =
    P^+
    -\eta P^+K_n^Z
    \left(
        (S_n^Z)^{-1}+\eta A_n
    \right)^{-1}
    (K_n^Z)^\top P^+,
\end{equation}
\begin{equation}
    p_1^-
    =
    p_1^+
    -\eta P^+K_n^Z
    \left(
        (S_n^Z)^{-1}+\eta A_n
    \right)^{-1}
    (K_n^Z)^\top p_1^+,
\end{equation}
and
\begin{equation}
\begin{split}
    p_0^-
    &=
    p_0^+
    +\frac{1}{2\eta}
    \log\det(I_d+\eta S_n^ZA_n)\\
    &\quad
    -\frac{\eta}{2}
    p_1^{+\top}K_n^Z
    \left(
        (S_n^Z)^{-1}+\eta A_n
    \right)^{-1}
    (K_n^Z)^\top p_1^+.
\end{split}
\end{equation}
These conditions give the terminal values for the ODE system on the preceding interval
\((\tau_{n-1},\tau_n)\). Starting from
\begin{equation}
    p_{0,N}^Z(T)=0,\qquad
    p_{1,N}^Z(T)=0,\qquad
    P_N^Z(T)=0,
\end{equation}
one solves backward on \((\tau_N,T)\), applies the jump condition at \(\tau_N\), then repeats
recursively for \(\tau_{N-1},\ldots,\tau_1\). This completes the proof.
\end{proof}

\subsection{Verification of the candidate solutions}
\label{app:verification}

The derivations in Appendices~\ref{app:proof-control-F},~\ref{app:proof-control-U} and~\ref{app:proof-control-Z} provide classical solutions of the corresponding HJB equations and feedback controls attaining the pointwise supremum. We now verify that these candidates solve the original control problems. The argument follows the standard verification approach for continuous-time portfolio problems; see, for example, \citet[Theorem~4.1]{HataSheu2018}.

\paragraph{Admissibility of the candidate optimal control.}

It remains to verify that the candidate feedback control belongs to $\Ac^H$. For $H\in\{F,U\}$, the optimal dollar exposure can be written as
\begin{equation}
    X_t\pi_t^{H,*}
    =
    \frac{1}{\eta\gamma(t)}
    \Sigma_S^{-1}
    \left[
        b_0^H(t)+B^H(t)\ell_t^H
    \right].
    \label{eq:verification-affine-control}
\end{equation}
The same representation holds intervalwise for the $Z$-investor, with
$b_{0,n}^Z$, $B_n^Z$, and $\widehat L_t^Z$. Since $\gamma$ is bounded away from zero and the deterministic coefficient functions are bounded on the finite horizon, there exists $C>0$ such that
\begin{equation}
    X_t^2
    (\pi_t^{H,*})^\top
    \Sigma_S
    \pi_t^{H,*}
    \leq
    C
    \left(
        1+\|\ell_t^H\|^2
    \right).
\end{equation}
The state and filtered-state processes satisfy linear Gaussian dynamics with bounded deterministic coefficients and therefore have finite second moments uniformly on $[0,T]$. Consequently,
\begin{equation}
    \Ebb
    \left[
        \int_0^T
        X_t^2
        (\pi_t^{H,*})^\top
        \Sigma_S
        \pi_t^{H,*}\,dt
    \right]
    <\infty.
\end{equation}
Thus the square-integrability condition defining $\Ac^H$ is satisfied.

\paragraph{Class-$D$ property of the candidate value process.}

We next verify the uniform-integrability condition. Define
\begin{equation}
    V_t^{H,*}
    :=
    v^H
    \left(
        t,X_t^*,Y_t^{H,*},\ell_t^H
    \right).
\end{equation}
Recall that
\begin{equation}
    v^H(t,x,y,\ell)
    =
    -\exp
    \left\{
        -\eta
        \left(
            \gamma(t)x-\lambda y+\phi^H(t,\ell)
        \right)
    \right\},
\end{equation}
where
\begin{equation}
    \phi^H(t,\ell)
    =
    p_0^H(t)
    +
    p_1^H(t)^\top\ell
    +
    \frac{1}{2}
    \ell^\top P^H(t)\ell.
\end{equation}
In particular,
\begin{equation}
    \nabla_\ell\phi^H(t,\ell)
    =
    p_1^H(t)+P^H(t)\ell.
\end{equation}

Because $\pi^{H,*}$ attains the supremum in the HJB equation, the drift term in the It\^o decomposition of $V^{H,*}$ vanishes. Let $W^H$ denote a Brownian motion collecting the return and state innovations, and write their martingale parts as
\begin{equation}
    dR_t^{H,m}
    =
    C_R^H(t)dW_t^H,
    \qquad
    d\ell_t^{H,m}
    =
    C_\ell^H(t)dW_t^H,
\end{equation}
where the loading matrices are deterministic and bounded on $[0,T]$. It\^o's formula then gives
\begin{equation}
    dV_t^{H,*}
    =
    V_t^{H,*}
    (\zeta_t^H)^\top dW_t^H,
\end{equation}
where
\begin{equation}
    \zeta_t^H
    =
    -\eta
    \left[
        \gamma(t)
        (C_R^H(t))^\top
        X_t\pi_t^{H,*}
        +
        (C_\ell^H(t))^\top
        \nabla_\ell\phi^H(t,\ell_t^H)
    \right].
    \label{eq:verification-zeta}
\end{equation}

By~\eqref{eq:verification-affine-control},
$\gamma(t)X_t\pi_t^{H,*}$ is affine in $\ell_t^H$, and
$\nabla_\ell\phi^H(t,\ell_t^H)$ is affine in $\ell_t^H$. Hence there exist bounded deterministic functions $a^H$ and $K^H$ such that
\begin{equation}
    \zeta_t^H
    =
    a^H(t)+K^H(t)\ell_t^H,
\end{equation}
and therefore
\begin{equation}
    \|\zeta_t^H\|
    \leq
    C\left(1+\|\ell_t^H\|\right).
    \label{eq:verification-linear-growth}
\end{equation}

The process $\ell^H$ satisfies a linear diffusion equation with bounded deterministic coefficients. The linear-growth condition~\eqref{eq:verification-linear-growth} therefore allows the Bene\v{s} criterion for stochastic exponentials to be applied \citep{Klebaner_2014}. Consequently,
\begin{equation}
    \mathcal E
    \left(
        \int_0^\cdot
        (\zeta_s^H)^\top dW_s^H
    \right)
\end{equation}
is a true martingale on $[0,T]$. Since
\begin{equation}
    \frac{V_t^{H,*}}{V_0^{H,*}}
    =
    \mathcal E
    \left(
        \int_0^t
        (\zeta_s^H)^\top dW_s^H
    \right),
\end{equation}
the candidate value process $V^{H,*}$ is a uniformly integrable martingale and is therefore of class $D$. Hence $\pi^{H,*}\in\Ac^H$.

Using the fact that the HJB inequality becomes an equality under $\pi^{H,*}$, we obtain
\begin{equation}
    \Ebb_{t,x,y,\ell}
    \left[
        -\exp
        \left\{
            -\eta
            \left(
                X_T^{*}-\lambda Y_T^{H,*}
            \right)
        \right\}
    \right]
    =
    v^H(t,x,y,\ell).
\end{equation}
Therefore,
\begin{equation}
    v^H=V^H,
    \qquad
    H\in\{F,U\},
\end{equation}
and $\pi^{H,*}$ is optimal.

\paragraph{The $Z$-investor.}

For the $Z$-investor, the same argument applies on each interval
$(\tau_n,\tau_{n+1})$. On each such interval, the filtered state follows a linear Gaussian diffusion with deterministic coefficients and the optimal exposure is affine in the filtered state. Hence the preceding Bene\v{s} argument yields the class-$D$ property intervalwise.

At a disclosure date, wealth and cumulative perceived emissions are continuous, while the filtered state is updated according to Proposition~\ref{prop:optimal-Z-jumps}. Since $P^+\succeq0$ and $S_n^Z\succ0$,
\begin{equation}
    (S_n^Z)^{-1}
    +
    \eta(K_n^Z)^\top P^+K_n^Z
    \succ0,
\end{equation}
so the Gaussian expectation entering the disclosure update is finite. Moreover,
\begin{equation}
    v_{n-1}^Z(\tau_n^-,x,y,\ell)
    =
    \Ebb
    \left[
        v_n^Z
        \left(
            \tau_n,x,y,\ell+K_n^Z\nu_n^Z
        \right)
        \middle|
        \Fbb_{\tau_n^-}^Z
    \right].
\end{equation}
Thus the martingale property of the candidate value process is preserved across each disclosure date. Since there are finitely many disclosure dates, the argument can be applied recursively backward from the final interval $(\tau_N,T]$. It follows that the candidate value process is of class $D$ on $[0,T]$, that $v_n^Z$ coincides with the value function on every interval, and that $\pi^{Z,*}$ is optimal.

\section{Calibration of $\lambda^\star$}
\label{app:lambda_calibration}

\renewcommand{\theequation}{B.\arabic{equation}}
\renewcommand{\thefigure}{B.\arabic{figure}}
\renewcommand{\thetable}{B.\arabic{table}}

\setcounter{equation}{0}
\setcounter{figure}{0}
\setcounter{table}{0}

We describe the numerical calibration of the carbon penalty $\lambda$. For each investor $H\in\{F,U,Z\}$, the optimal strategy derived in Section~4 can be written in affine form as
\begin{equation}
    \pi_t^{H,*}(\lambda)
    =
    \frac{1}{\eta\gamma(t)X_t}
    \Sigma_S^{-1}
    \left[
        b_0^H(t;\lambda)
        +
        B^H(t;\lambda)\ell_t^H
    \right],
    \label{eq:optimal_control_generic_calibration}
\end{equation}
where $\ell_t^H$ denotes the relevant state available to investor $H$. More precisely,
\begin{equation}
    \ell_t^F =
    \begin{pmatrix}
        M_t\\ U_t
    \end{pmatrix},
    \qquad
    \ell_t^U =
    \begin{pmatrix}
        \widehat M_t^U\\
        \widehat m_{\infty,t}^U\\
        U_t
    \end{pmatrix},
    \qquad
    \ell_t^Z =
    \begin{pmatrix}
        \widehat M_t^Z\\
        \widehat m_{\infty,t}^Z\\
        \widehat U_t^Z
    \end{pmatrix}.
\end{equation}
For the $Z$-investor, $b_0^Z$ and $B^Z$ are understood intervalwise as $b_{0,n}^Z$ and $B_n^Z$ for $t\in[\tau_n,\tau_{n+1})$. The functions $b_0^H$ and $B^H$ depend on $\lambda$ through both the direct carbon-penalty term and the solution of the corresponding backward system. They therefore differ across information regimes.

For a given value of $\lambda$, the realized cumulative financed
emissions generated by the optimal strategy are
\begin{equation}
    Y_T^{H,*}(\lambda)
    =
    \int_0^T E_t^{H,*}(\lambda)\,dt.
\end{equation}

Using the realized carbon-intensity process $U_t$, we write financed emissions at time $t$ as
\begin{equation}
    E_t^{H,*}(\lambda)
    =
    X_t
    \left[
        (\pi_t^{H,*}(\lambda))^\top U_t
        +
        \left(
            1-\mathbf 1^\top\pi_t^{H,*}(\lambda)
        \right)
        u_{\mathrm{rf}}(t)
    \right].
\end{equation}
Equivalently, defining
\begin{equation}
    \widetilde U_t
    :=
    U_t-u_{\mathrm{rf}}(t)\mathbf 1,
\end{equation}
we have
\begin{equation}
    E_t^{H,*}(\lambda)
    =
    X_tu_{\mathrm{rf}}(t)
    +
    X_t(\pi_t^{H,*}(\lambda))^\top\widetilde U_t.
\end{equation}
Substituting \eqref{eq:optimal_control_generic_calibration} and using
$\gamma(t)=e^{r(T-t)}- \lambda \int_{t}^{T}e^{r(s-t)}u_{rf}(s)ds$ yields
\begin{equation}
    E_t^{H,*}(\lambda)
    =
    X_tu_{\mathrm{rf}}(t)
    +
    \frac{1}{\eta \gamma(t)}
    \widetilde U_t^\top\Sigma_S^{-1}
    \left[
        b_0^H(t;\lambda)
        +
        B^H(t;\lambda)\ell_t^H
    \right].
\label{eq:emissions_affine_control}
\end{equation}

Hence,
\begin{equation}
    Y_T^{H,*}(\lambda)
    =
    \int_0^T
    \left\{
        X_tu_{\mathrm{rf}}(t)
        +
        \frac{1}{\eta\gamma(t)}
        \widetilde U_t^\top\Sigma_S^{-1}
        \left[
            b_0^H(t;\lambda)
            +
            B^H(t;\lambda)\ell_t^H
        \right]
    \right\}
    dt.
\end{equation}

For investor $H$, define the probability of exceeding the cumulative
emissions budget $\bar Y$ by
\begin{equation}
    p_H(\lambda)
    :=
    \mathbb P
    \left(
        Y_T^{H,*}(\lambda)>\bar Y
    \right).
\end{equation}
It is estimated using $N_{\mathrm{MC}}$ Monte Carlo simulations:
\begin{equation}
    \widehat p_H(\lambda)
    =
    \frac{1}{N_{\mathrm{MC}}}
    \sum_{j=1}^{N_{\mathrm{MC}}}
    \mathbf 1
    \left\{
        Y_T^{H,*,(j)}(\lambda)>\bar Y
    \right\}.
\end{equation}

An investor-specific carbon penalty could therefore be calibrated as the smallest value satisfying the target satisfaction criterion,
\begin{equation}
\lambda_H^\star
    :=
    \inf\left\{
    \lambda\in[0,\bar\lambda):
    \widehat p_H(\lambda)\leq\epsilon
    \right\},
\end{equation}
where $\bar\lambda$ is the upper bound satisfying $\inf_{t \in [0,T]} \gamma(t)>0$. In all numerical experiments, the calibrated penalty lies strictly below this bound. Moreover, throughout the numerical application, we set
\begin{equation}
    \epsilon=0.5
\end{equation}

\section{Supplementary figures}
\label{app:supplementary figures}

\renewcommand{\theequation}{C.\arabic{equation}}
\renewcommand{\thefigure}{C.\arabic{figure}}
\renewcommand{\thetable}{C.\arabic{table}}

\setcounter{equation}{0}
\setcounter{figure}{0}
\setcounter{table}{0}

\subsection{Impact of short selling}

\begin{figure}[H]
    \centering
    \includegraphics[width=0.9\linewidth]{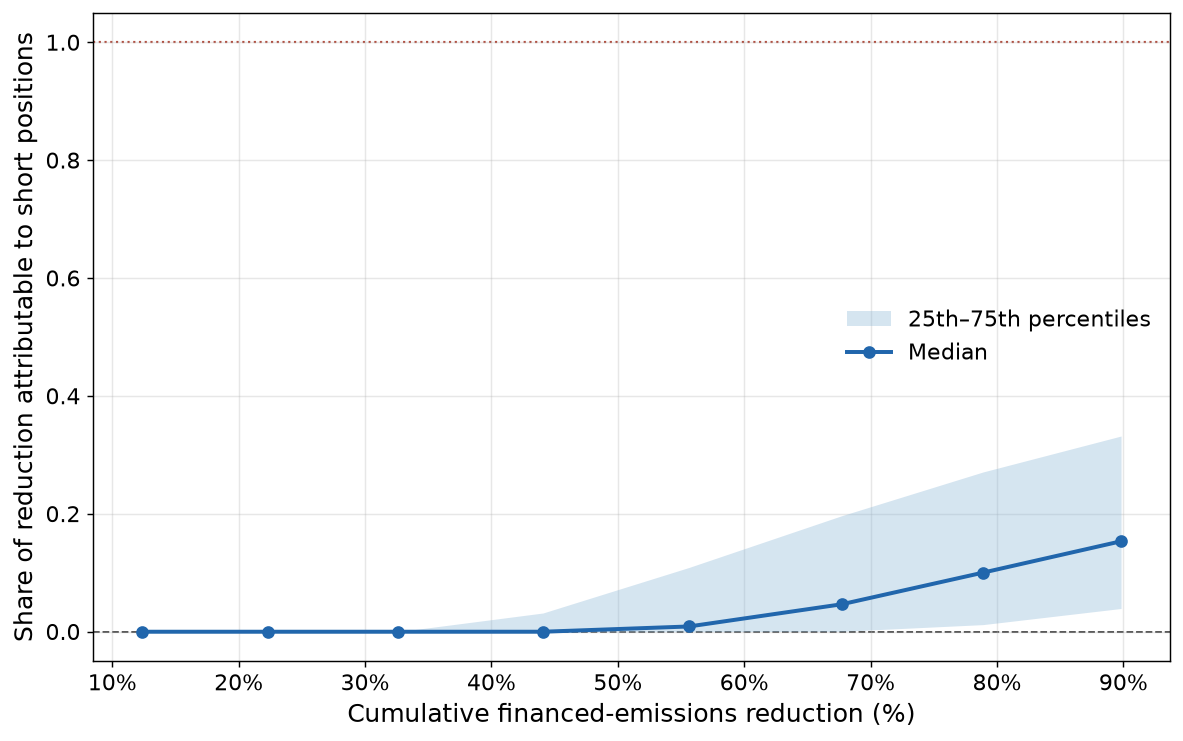}
    \caption{Contribution of short positions to the measured reduction in cumulative financed emissions.}
    \par\vspace{4pt}
    \begin{minipage}{\linewidth}
    \footnotesize
    \textit{Notes:} For each simulated path and emissions-reduction level, the figure
    reports the fraction of the achieved reduction in cumulative financed emissions that
    is attributable to negative risky-asset positions. The measure compares cumulative
    financed emissions under the unconstrained portfolio with a counterfactual in which
    the direct emissions contribution of short positions is set to zero, while leaving
    the portfolio strategy otherwise unchanged. The solid line reports the median across
    Monte Carlo paths and the shaded area the 25th--75th percentile range. A value of
    zero indicates that short positions do not contribute to the measured emissions
    reduction, while a value of one indicates that the entire measured reduction is
    attributable to negative positions. In the baseline calibration, this contribution
    remains small around the 50\% reduction used in the main analysis and becomes more
    material only for more stringent reduction levels.
    \end{minipage}
    \label{fig:negative_emissions}
\end{figure}

Because short positions enter linearly in our financed-emissions measure, they may
mechanically reduce cumulative financed emissions even though such negative
contributions would not generally be recognized under carbon-accounting conventions.
To quantify the importance of this effect, we construct a counterfactual cumulative
emissions measure in which the direct emissions contribution of negative risky-asset
positions is set to zero. For the $F$-investor, let
\begin{equation}
Y_{T,+}^{F}(\lambda)
=
\int_0^T X_t
\left[
\sum_{i=1}^d
\bigl(\pi_t^{F,i}(\lambda)\bigr)_+ U_t^i
+
\left(1-\mathbf{1}^{\top}\pi_t^{F}(\lambda)\right)
u_{\mathrm{rf}}(t)
\right]dt,
\end{equation}
where $(x)_+ := \max(x,0)$. Relative to the cumulative emissions
$Y_T^{F}(\lambda)$ generated by the original specification, the difference
$Y_{T,+}^{F}(\lambda)-Y_T^{F}(\lambda)$ measures the reduction in reported
emissions that is mechanically attributable to short positions. We therefore define
the share of the achieved emissions reduction attributable to negative positions as
\begin{equation}
\mathcal{S}_T(\lambda)
:=
\frac{
Y_{T,+}^{F}(\lambda)-Y_T^{F}(\lambda)
}{
Y_T^{0}-Y_T^{F}(\lambda)
},
\end{equation}
where $Y_T^{0}$ denotes cumulative financed emissions under the unpenalized
benchmark. A value of $\mathcal{S}_T=0$ indicates that short positions make no
contribution to the measured reduction, whereas $\mathcal{S}_T=1$ indicates that
the entire measured reduction relative to the benchmark is generated by negative
positions. Figure~\ref{fig:negative_emissions} reports the distribution of
$\mathcal{S}_T$ across Monte Carlo paths for different achieved emissions-reduction
levels.

\subsection{Realized and perceived cumulative financed emissions}

\begin{figure}[H]
    \centering
    \includegraphics[width=0.9\linewidth]{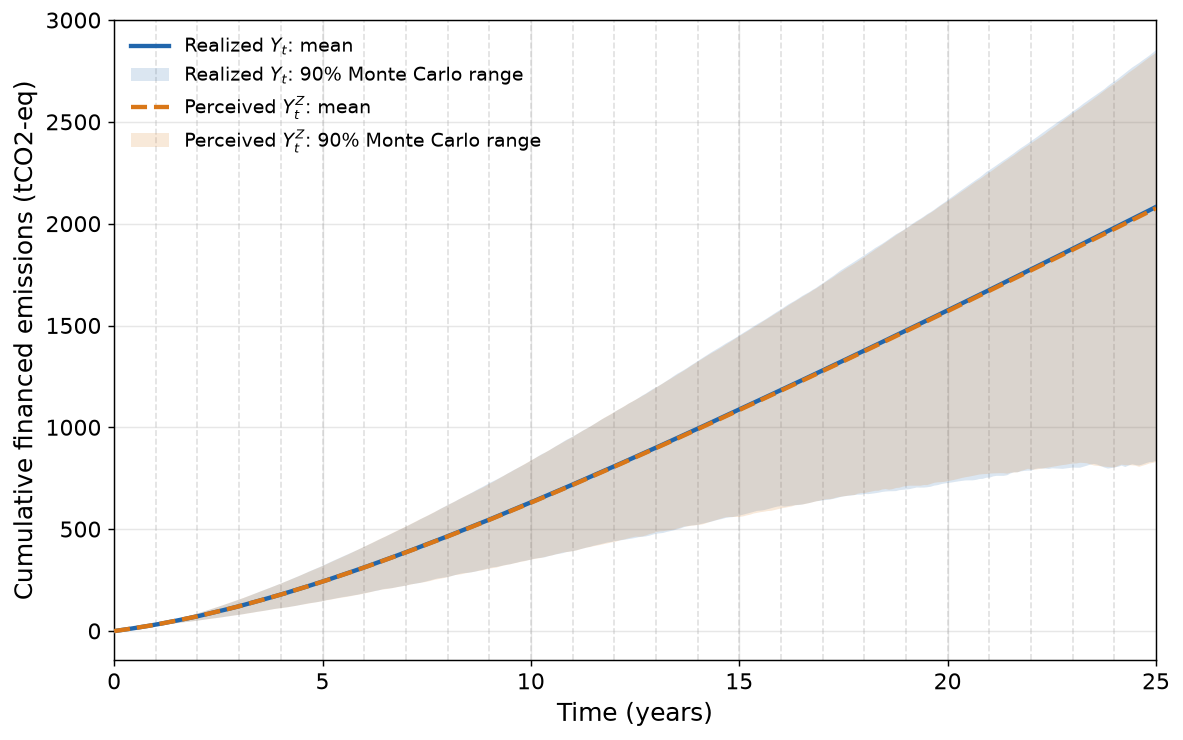}
    \caption{Perceived and realized cumulative financed emissions for the $Z$-investor.}
    \par\vspace{4pt}
    \begin{minipage}{\linewidth}
    \footnotesize
    \textit{Notes:} The figure compares cumulative financed emissions as perceived by the
    $Z$-investor, $Y_t^Z$, with realized cumulative financed emissions $Y_t$, computed
    ex post using the latent carbon-intensity process $U_t$ under the same portfolio
    strategy. Solid and dashed lines report Monte Carlo means, while shaded areas
    represent the 5th--95th percentile ranges across simulated paths. Results are shown
    for the baseline calibration with annual disclosures. The close overlap between the
    two series indicates that, in this numerical application, using filtered carbon
    intensities to construct the investor's emissions objective generates only a small
    difference between perceived and realized cumulative financed emissions.
    \end{minipage}
    \label{fig:actual_vs_perceived_emissions}
\end{figure}

For the $Z$-investor, the optimization problem is defined in terms of perceived cumulative financed emissions, $Y_t^Z$, constructed from the filtered carbon-intensity estimates available to the investor. These perceived emissions need not coincide with the cumulative financed emissions obtained ex post from the latent realized carbon intensities. To assess the quantitative importance of this distinction, Figure~\ref{fig:actual_vs_perceived_emissions} compares the perceived and realized cumulative financed emissions generated by the same $Z$-investor strategy. In our baseline calibration, the two measures remain very close over the investment horizon, suggesting that the use of filtered carbon intensities has only a limited effect on the measurement of cumulative financed emissions.

\subsection{Contribution decomposition}

\begin{figure}[H]
    \centering
    \includegraphics[width=0.9\linewidth]{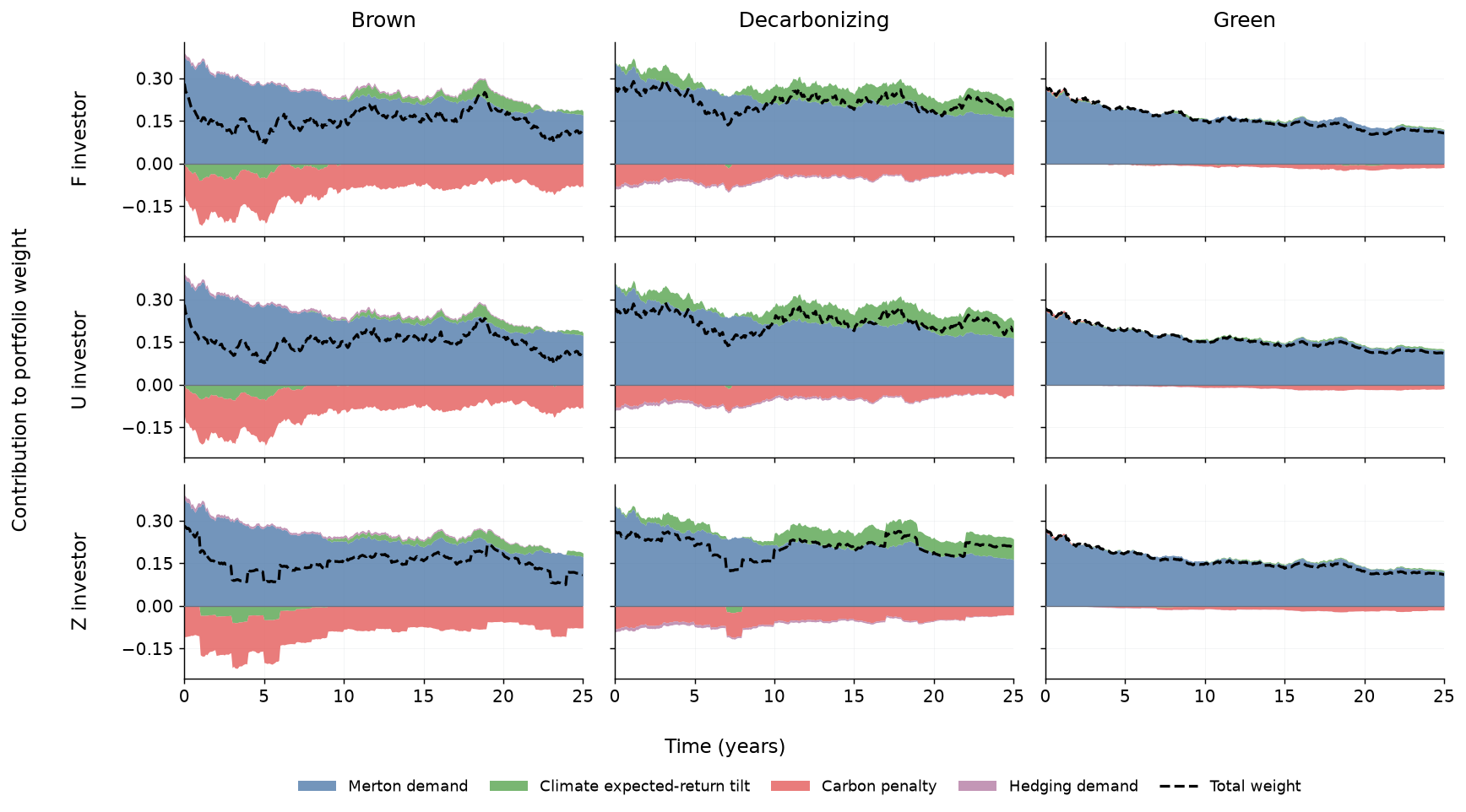}
    \caption{Decomposition of the optimal risky-asset weights by economic driver at the calibrated carbon penalty $\lambda^\star$. }
    \par\vspace{4pt}
    \begin{minipage}{\linewidth}
    \footnotesize
    \textit{Notes:} The baseline unpenalized component corresponds to the state-independent excess return $\mu-r\mathbf{1}$. The climate expected-return component contains the transition-state and carbon-intensity contributions to expected returns. The carbon component is the direct cumulative-emissions penalty, while the hedging component collects all intertemporal hedging terms. For the $F$-investor, hedging arises only from return--intensity shock correlation; for the $U$- and $Z$-investors it additionally incorporates learning about latent climate states. The dashed line reports the total optimal risky-asset weight.
    \end{minipage}
    \label{fig:decomposition}
\end{figure}

Figure \ref{fig:decomposition} presents a decomposition of the contributions of the components of the optimal strategy along the same simulated trajectory as in Figure \ref{fig:single_path}. The figure decomposes the optimal risky-asset allocation into four economic components. For each investor $H\in\{F,U,Z\}$, we write
\begin{equation}
    \pi_t^{H,*}  = \pi_t^{H,\mathrm{Unpenalized}} + \pi_t^{H,\mathrm{climate}}  +\pi_t^{H,\mathrm{carbon}}  + \pi_t^{H,\mathrm{hedge}}.
\end{equation}
The decomposition is expressed below in the original economic variables.

\medskip
\noindent
\textbf{F-investor.}
Using equation~\eqref{eq:pi_F_original_variables}, the full-information control is
\begin{align}
\pi_t^{F,*} = \frac{1}{\eta\gamma(t)X_t}\Sigma_S^{-1}\Big[&(\mu-r\mathbf{1})+
\left\{\alpha M_t+\beta(U_t-\bar U_0)\right\}\nonumber\\
&-\frac{\lambda}{\gamma(t)}
\left(U_t-u_{\mathrm{rf}}(t)\mathbf{1}\right)
-
\eta\Sigma_{SU}\nabla_u\phi^F(t,M_t,U_t)
\Big].
\end{align}
Hence the four contributions plotted in Figure~\ref{fig:decomposition}
are
\begin{equation}
\begin{aligned}
\pi_t^{F,\mathrm{Unpenalized}}
&=
\frac{1}{\eta\gamma(t)X_t}
\Sigma_S^{-1}(\mu-r\mathbf{1}),\\
\pi_t^{F,\mathrm{climate}}
&=
\frac{1}{\eta\gamma(t)X_t}
\Sigma_S^{-1}
\left[
\alpha M_t+\beta(U_t-\bar U_0)
\right],\\
\pi_t^{F,\mathrm{carbon}}
&=
-\frac{\lambda}{\eta\gamma(t)^2X_t}
\Sigma_S^{-1}
\left(U_t-u_{\mathrm{rf}}(t)\mathbf{1}\right),\\
\pi_t^{F,\mathrm{hedge}}
&=
-\frac{1}{\gamma(t)X_t}
\Sigma_S^{-1}\Sigma_{SU}
\nabla_u\phi^F(t,M_t,U_t).
\end{aligned}
\end{equation}
Under full information, the hedging component only reflects the
instantaneous covariance between return and carbon-intensity shocks.
There is no transition-learning hedging demand because $M_t$ is observed
and, under our baseline assumptions, its innovation is independent of
asset-return shocks.

\medskip
\noindent
\textbf{U-investor.}
For the $U$-investor, $U_t$ is observed while $M_t$ and $m_\infty$ are
filtered. The corresponding decomposition is
\begin{equation}
\begin{aligned}
\pi_t^{U,\mathrm{Unpenalized}}
&=
\frac{1}{\eta\gamma(t)X_t}
\Sigma_S^{-1}(\mu-r\mathbf{1}),\\
\pi_t^{U,\mathrm{climate}}
&=
\frac{1}{\eta\gamma(t)X_t}
\Sigma_S^{-1}
\left[
\alpha\widehat M_t^U+\beta(U_t-\bar U_0)
\right],\\
\pi_t^{U,\mathrm{carbon}}
&=
-\frac{\lambda}{\eta\gamma(t)^2X_t}
\Sigma_S^{-1}
\left(U_t-u_{\mathrm{rf}}(t)\mathbf{1}\right),\\
\pi_t^{U,\mathrm{hedge}}
&=
-\frac{1}{\gamma(t)X_t}\Sigma_S^{-1}
\Big[
\alpha
\left(
q_M^U(t)\partial_m\phi^U
+
q_{M,m_\infty}^U(t)\partial_\xi\phi^U
\right)
+
\Sigma_{SU}\nabla_u\phi^U
\Big],
\end{aligned}
\end{equation}

where the derivatives of $\phi^U$ are evaluated at
$(t,\widehat M_t^U,\widehat m_{\infty,t}^U,U_t)$.
The first two terms in the hedging component arise from learning about
the current and long-run transition states, whereas the last term is the
return--carbon-intensity hedging demand already present under full
information.

\medskip
\noindent
\textbf{Z-investor.}
For the disclosure-based investor, both the transition variables and the
carbon intensities are filtered. The first three components are
\begin{equation}
\begin{aligned}
\pi_t^{Z,\mathrm{Unpenalized}}
&=
\frac{1}{\eta\gamma(t)X_t}
\Sigma_S^{-1}(\mu-r\mathbf{1}),\\
\pi_t^{Z,\mathrm{climate}}
&=
\frac{1}{\eta\gamma(t)X_t}
\Sigma_S^{-1}
\left[
\alpha\widehat M_t^Z
+
\beta(\widehat U_t^Z-\bar U_0)
\right],\\
\pi_t^{Z,\mathrm{carbon}}
&=
-\frac{\lambda}{\eta\gamma(t)^2X_t}
\Sigma_S^{-1}
\left(
\widehat U_t^Z-u_{\mathrm{rf}}(t)\mathbf{1}
\right).
\end{aligned}
\end{equation}
For $t\in[\tau_n,\tau_{n+1})$, the hedging contribution is
\begin{align}
\pi_t^{Z,\mathrm{hedge}}
=
-\frac{1}{\gamma(t)X_t}\Sigma_S^{-1}
\Bigg[
&
\left(
\alpha q_M^Z(t)
+
\beta q_{MU}^Z(t)
\right)
\partial_m\phi_n^Z
\nonumber\\
&+
\left(
\alpha q_{Mm_\infty}^Z(t)
+
\beta q_{m_\infty U}^Z(t)
\right)
\partial_{m_\infty}\phi_n^Z
\nonumber\\
&+
\left(
\alpha q_{MU}^Z(t)^\top
+
\beta q_U^Z(t)
+
\Sigma_{SU}
\right)
\nabla_u\phi_n^Z
\Bigg],
\end{align}

where the derivatives are evaluated at $(t,\widehat M_t^Z,\widehat m_{\infty,t}^Z,\widehat U_t^Z)$. The hedging component therefore combines learning about the current transition state, learning about its long-run mean, uncertainty about carbon intensities, posterior cross-covariances between these latent states, and the direct return--intensity shock covariance $\Sigma_{SU}$. \\

As anticipated, the resulting effects differ substantially across the three assets. The Brown asset is primarily adversely affected by the carbon penalty, while it benefits from the intensity-hedging component. The Decarbonizing asset is likewise negatively impacted by the carbon penalty; however, this effect is largely counterbalanced by the positive contribution of the intensity-drift component, particularly after time 10. \\

For the Green asset, the optimal allocation remains comparatively stable. Initially, it benefits from the carbon penalty because its carbon intensity is lower than that of the risk-free asset; however, this advantage diminishes progressively over time. Overall, in this illustrative case, the transition state is relatively stable, implying that the associated impact on portfolio weights is marginal. Under partial observation, by contrast, the Brown and Decarbonizing assets may also benefit from the transition-parameter component, as this parameter tends to be revised downward in the estimation process, thereby underscoring the influence of partial observability on the optimal allocation. \\

\bibliographystyle{chicago}
\bibliography{biblio}

\end{document}